\documentclass[aps,prd,superscriptaddress,twocolumn]{revtex4}
\usepackage[latin9]{inputenc}
\usepackage{amsmath}
\usepackage{amssymb}
\usepackage{graphicx}
\usepackage{esint}
\usepackage{hyperref}
\usepackage{CJK}

\makeatletter

\providecommand{\tabularnewline}{\\}

\@ifundefined{textcolor}{}
{%
 \definecolor{BLACK}{gray}{0}
 \definecolor{WHITE}{gray}{1}
 \definecolor{RED}{rgb}{1,0,0}
 \definecolor{GREEN}{rgb}{0,1,0}
 \definecolor{BLUE}{rgb}{0,0,1}
 \definecolor{CYAN}{cmyk}{1,0,0,0}
 \definecolor{MAGENTA}{cmyk}{0,1,0,0}
 \definecolor{YELLOW}{cmyk}{0,0,1,0}
}

\makeatother

\usepackage[english]{babel}
\begin{document}
\begin{CJK*}{GBK}{song}

\title{Dynamical correlation functions and the related physical effects in three-dimensional Weyl/Dirac semimetals}
\begin{abstract}
We present a unified derivation of the dynamical correlation
functions including density-density, density-current and current-current,
of three-dimensional Weyl/Dirac semimetals by use of the Passarino-Veltman
reduction scheme at zero temperature. The generalized Kramers-Kronig
relations with arbitrary order of subtraction are established to verify
these correlation functions. Our results lead to the exact
chiral magnetic conductivity and directly recover the previous ones
in several limits. We also investigate the magnetic susceptibilities,
the orbital magnetization and briefly discuss the impact of electron
interactions on these physical quantities within the random phase
approximation.  Our work could provide a starting point for the investigation
of the nonlocal transport and optical properties due to the higher-order
spatial dispersion in three-dimensional Weyl/Dirac semimetals.
\end{abstract}

\author{Jianhui Zhou}
\affiliation{Department of Physics, The University of Hong Kong, Pokfulam Road,
Hong Kong, China}

\author{Hao-Ran Chang}
\email{hrchang@mail.ustc.edu.cn}
\affiliation{Department of Physics, Institute of Solid State Physics, and Center
for Computational Sciences, Sichuan Normal University, Chengdu, Sichuan
610066, China}
\affiliation{Department of Physics, McGill University, Montreal, Quebec H3A 2T8,
Canada}

\date{\today}
\maketitle

\section{introduction}

Three-dimensional (3D) Weyl semimetals are one kind of new topological
phases of matter and have a finite number of Weyl nodes characterized by the
chirality in the Brouillon zone \cite{Hosur13crp,burkov2015jpcm,WengHM2016jpcm,Armitage2017rmp}.
The Dirac semimetals usually host multiple pairs of Weyl nodes that
are protected by both time reversal symmetry and inversion symmetry.
These Weyl nodes can be seen as monopoles, sources and drains for Berry
curvature fields, in momentum space \cite{Xiao2010RMP,volovik2003oup}.
The nontrivial topology of Weyl nodes has led to a variety of exotic
electromagnetic transport phenomena, such as the chiral anomaly \cite{Nielsen1983ABJ,Goswami2013PRB,Liu2013PRB,Parameswaran2014PRX,Behrends2016PRB},
the static chiral magnetic effect (CME) \cite{VilenkinPRD1980CME,Fukushima2008PRD,Grushin2012PRD,zyuzin2012prb,zhou2013cpl,Vazifeh2013PRL,Landsteiner2014PRB,Chang2015PRBcme1,OBrien2017PRL},
the dynamical CME \cite{KharzeevPRD2009,SonDT2013PRD,Ma2015PRB,ZhongSD2016PRLGME},
the topological Fermi arc states \cite{WanPRB2011Weyl}, and negative
longitudinal magnetoresistance \cite{SonSpivakPRB2013,Burkov2014PRL,GorbarPRB2014,Gao2017PRB,Dai2017PRL}.
Recently, a series of experiments have made great efforts to characterize
the relativistic nature of 3D Weyl/Dirac fermions and detect anomalous
magnetotransport properties \cite{Kim2013PRL,Huang2015PRX,Xiong2015Science,li2015NC,LiH2016NC,LiQ2016NP}.

The correlation functions encode lots of essential information of the
systems under the perturbations of external fields and play a critical
role in investigating their physical properties \cite{forster1995hydroF}.
The dynamical correlation functions enable us to study the responses
of systems to the inhomogeneous and time-dependent external fields
and the related physical effects. The density-density correlation
function characterizes the dielectric behavior and is widely used
to study the plasmon excitations and other many-body phenomena \cite{giuliani2005qtel,MahanMPP3}.
The current-current correlation functions are directly related to
various transport properties of electrons. For example, the anomalous
Hall effect is related to the off-diagonal conductivity \cite{Nagaosa2010RMP},
while the dynamical CME or natural optical activity is obtained from
the spatially antisymmetric part of off-diagonal conductivity \cite{LandauECM}.
In fact, previous works on the dynamical CME for 3D Weyl fermions
had mainly focused on some limits: the leading order part of the chiral
magnetic conductivity \cite{KharzeevPRD2009} and the next-to-leading
order hard dense loop approximation \cite{SonDT2013PRD}. However,
the general frequency- and momentum-dependent correlation functions
for 3D Weyl/Dirac semimetals are still lacking and deserve detailed
derivations. The prime aim of this paper is to derive the dynamical
correlation functions of 3D Weyl/Dirac semimetals in a unified framework
of the Passarino-Veltman reduction scheme (PVRS) \cite{PASSARINO1979NPB},
which is capable of reducing the tensor integral over loop energy-momentum
to basic scalar integrals based on the tensor structure imposed by
Lorentz covariance.

In this paper, the dynamical correlation functions of Weyl/Dirac
semimetals are derived by use of the PVRS at zero temperature. We
establish the generalized Kramers-Kronig relations with $n$th-order
subtraction to verify these correlation functions. The gauge
invariance of these correlation functions is also clarified. We obtain
the exact analytical chiral magnetic conductivity and make a comparison
with the previous results in several limits. In addition, we explore
the magnetic susceptibilities, the orbital magnetization and briefly
discuss the impact of electron interactions on these quantities within
the random phase approximation (RPA).

This paper is organized as follows. In Sec. \ref{sec:HamCF}, we outline
the effective Hamiltonian and introduce the correlation functions.
In Sec. \ref{sec:P-VeltD}, we calculate the correlation functions
by using the PVRS. In Sec. \ref{sec:KKR}, we establish the generalized
Kramers\textendash Kronig relation and apply them to the correlation
functions. In Sec. \ref{sec:OpticCond}, the optical conductivity
is recovered. In Sec. \ref{sec:DCME}, the exact chiral magnetic conductivity
is derived and some limits are discussed. In Sec. \ref{sec:MS}, we
evaluate the magnetic susceptibilities, the orbital magnetization
and renormalizations of these physical quantities due to electron
interactions. In Sec. \ref{sec:conclusions}, the main results of
this paper are summarized. Finally, we give the detailed calculations
in the appendices.

\section{The Hamiltonian and correlation functions \label{sec:HamCF}}

We start from the effective Hamiltonian for a pair of isotropic Weyl
nodes with opposite chirality \cite{Zhang2015PRB}
\begin{align}
\mathcal{H}_{\chi} & =\chi v_{F}\hbar\sigma^{\alpha}\left(k^{\alpha}+\chi b^{\alpha}\right)-\sigma^{0}\mu_{\chi},\label{Ham}
\end{align}
where $\sigma^{\alpha}$ with $\alpha=x,y,z$ are the Pauli matrices
and $\sigma^{0}$ is the unit matrix. The velocity operator is given
as $v^{\alpha}=\partial\mathcal{H}_{\chi}/\partial\left(\hbar k^{\alpha}\right)=\chi v_{F}\sigma^{\alpha}$
with $v_{F}$ being the effective velocity. $\mu_{\chi}=\mu_{0}+\chi b_{0}$
is the chirality-dependent chemical potential, and $\chi=\pm$ denotes
the chirality of Weyl node. $\mu_{0}$ is the chemical potential at
equilibrium. $b_{0}$ refers to the chiral chemical potential induced
by the chiral anomaly or the energy difference between the two Weyl
nodes. $\boldsymbol{b}$ measures the spacing of Weyl nodes with opposite
chirality from the time reversal symmetry breaking. In particular,
for $b_{0}=0$ and $\boldsymbol{b}=0$, the effective Hamiltonian
in Eq. $\left(\ref{Ham}\right)$ reduces to a minimal model for the
Dirac semimetals. Hereafter, we shall set $\hbar=v_{F}=1$ unless
specified otherwise.

The corresponding Matsubara Green's function of Weyl fermions near
the Weyl node $\chi$, $G_{\chi}(\boldsymbol{k},i\Omega_{n})=\left(i\Omega_{n}\sigma^{0}-\mathcal{H}_{\chi}\right)^{-1}$,
is given as
\begin{equation}
G_{\chi}\left(\boldsymbol{k},i\Omega_{n}\right)=\frac{(i\Omega_{n}+\mu_{\chi})\sigma^{0}+\chi\sigma^{\alpha}\left(k^{\alpha}+\chi b^{\alpha}\right)}{\left(i\Omega_{n}+\mu_{\chi}\right)^{2}-\left(\boldsymbol{k}+\chi\boldsymbol{b}\right)^{2}},
\end{equation}
where $\Omega_{n}=\left(2n+1\right)\pi/ \beta_{T}$ is the fermionic
Matsubara frequency with \emph{$\beta_{T}=1/k_{B}T$}. In this paper,
we mainly focus on the following case~\cite{GFb}
\begin{align}
G_{\chi}(\boldsymbol{k},i\Omega_{n}) & =\frac{(i\Omega_{n}+\mu_{\chi})\sigma^{0}+\chi\sigma^{\alpha}k^{\alpha}}{\left(i\Omega_{n}+\mu_{\chi}\right)^{2}-k^{2}},
\end{align}
with $k^{2}=\left(k^{x}\right)^{2}+\left(k^{y}\right)^{2}+\left(k^{z}\right)^{2}$.

The total correlation function for 3D Weyl/Dirac semimetals is a summation
of these chirality-dependent correlation functions
\begin{align}
\Pi^{\mu\nu}(\boldsymbol{q},i\omega_{m}) & =\sum_{\chi=\pm}\Pi^{\mu\nu}(\boldsymbol{q},i\omega_{m},\chi),
\end{align}
where the indices $\mu,\nu=0$ are for the time-component, while $\mu,\nu=x,y,z$
represent the spatial components. In general, there are three kinds
of chirality-dependent correlation functions. The first kind is the
chirality-dependent density-density correlation function
\begin{align}
\Pi^{00}(\boldsymbol{q},i\omega_{m},\chi)= & \frac{1}{\mathcal{V}}\sum_{\boldsymbol{k}}\frac{1}{\beta_{T}}\sum_{i\Omega_{n}}\mathrm{Tr}\left[\sigma^{0}G_{\chi}(\boldsymbol{k},i\Omega_{n})\right.\nonumber \\
 & \left.\sigma^{0}G_{\chi}(\boldsymbol{k}+\boldsymbol{q},i\Omega_{n}+i\omega_{m})\right],
\end{align}
where $\mathrm{Tr}$ acts over the internal
degrees of freedom (pseudospin or spin). It had been detailedly derived in Refs. \cite{Abrikosov1971,LvZhangWeyl2013,ZhouJH2015PRB} and
used to extensively investigate the plasmon excitations \cite{ZhouJH2015PRB,Hofmann2015PRB,KharzeevPRL2015,PellegrinoHeliconWeyl2015,Zyuzin2015PRB,KotovPRB2016,Ferreiros2016PRB,YanZB2016PRB}
and dynamics of phonons \cite{Song2016PRB,Rinkel2017PRL,Liu2017PRL}
in 3D Weyl/Dirac semimetals. The second kind is
the chirality-dependent current-current correlation function
\begin{align}
\Pi^{\alpha\beta}(\boldsymbol{q},i\omega_{m},\chi)= & \frac{1}{\mathcal{V}}\sum_{\boldsymbol{k}}\frac{1}{\beta_{T}}\sum_{i\Omega_{n}}\mathrm{Tr}\left[\left(\chi\sigma^{\alpha}\right)G_{\chi}(\boldsymbol{k},i\Omega_{n})\right.\nonumber \\
 & \left.\left(\chi\sigma^{\beta}\right)G_{\chi}(\boldsymbol{k}+\boldsymbol{q},i\Omega_{n}+i\omega_{m})\right].
\end{align}
Note that the spatially antisymmetric part of the off-diagonal correlation
functions $\Pi_{A}^{\alpha\beta}\equiv\frac{1}{2}\left(\Pi^{\alpha\beta}-\Pi^{\beta\alpha}\right)$
has been used to study the dynamical CME in several limits \cite{KharzeevPRD2009,SonDT2013PRD}.
If $\sigma^{\alpha}$ stands for the real spin degree of freedom of
electrons, the current-current correlation functions are proportional
to the dynamical spin susceptibilities, which govern the Ruderman-Kittel-Kasuya-Yosida
interaction of magnetic impurities and the spin textures~\cite{Zhang2015PRB,Araki2016PRB}.

Finally, the chirality-dependent density-current correlation functions are given by
\begin{align}
\Pi^{0\alpha}(\boldsymbol{q},i\omega_{m},\chi)= & \frac{1}{\mathcal{V}}\sum_{\boldsymbol{k}}\frac{1}{\beta_{T}}\sum_{i\Omega_{n}}\mathrm{Tr}\left[\sigma^{0}G_{\chi}(\boldsymbol{k},i\Omega_{n})\right.\nonumber \\
 & \left.\left(\chi\sigma^{\alpha}\right)G_{\chi}(\boldsymbol{k}+\boldsymbol{q},i\Omega_{n}+i\omega_{m})\right],\\
\Pi^{\alpha0}(\boldsymbol{q},i\omega_{m},\chi)= & \frac{1}{\mathcal{V}}\sum_{\boldsymbol{k}}\frac{1}{\beta_{T}}\sum_{i\Omega_{n}}\mathrm{Tr}\left[\left(\chi\sigma^{\alpha}\right)G_{\chi}(\boldsymbol{k},i\Omega_{n})\right.\nonumber \\
 & \left.\sigma^{0}G_{\chi}(\boldsymbol{k}+\boldsymbol{q},i\Omega_{n}+i\omega_{m})\right],
\end{align}
which is critical to the gauge invariance among these correlation
functions. In the following, we evaluate all of the chirality-dependent
correlation functions $\Pi^{\mu\nu}(\boldsymbol{q},i\omega_{m},\chi)$
by utilizing the PVRS \cite{PASSARINO1979NPB}.

\section{Correlation functions from Passarino-veltman reduction scheme~\label{sec:P-VeltD}}

The essential spirit of the PVRS is to reduce the tensor integral
over loop energy-momentum to a set of basic scalar integrals by considering
the tensor structure imposed by Lorentz covariance \cite{PASSARINO1979NPB}.
This scheme is widely used in the calculations of perturbative corrections
in high energy physics. For 3D Weyl/Dirac semimetals, the Lorentz
covariance is evident such that the PVRS is applicable to calculations
of their correlation functions.

Utilizing the PVRS, summing over Matsubara frequency $\Omega_{n}$
and performing analytical continuum $i\omega_{m}\to\omega+i\eta$
with $\eta$ being a positive infinitesimal, one finally rewrites
these chirality-dependent correlation functions in terms of a set
of scalar functions (see Appendix \ref{AppDCRPVRS}). First, the chirality-dependent
density-density correlation function
\begin{align}
\Pi^{00}\left(\boldsymbol{q},\omega,\chi\right) & =\frac{1}{2}\left[A_{0}+B_{a}+q^{2}B_{1}\right].\label{Pi00}
\end{align}
Second, the chirality-dependent current-current correlation function
$\Pi^{\alpha\beta}(\boldsymbol{q},\omega,\chi)$ is given as
\begin{align}
\Pi^{\alpha\beta}\left(\boldsymbol{q},\omega,\chi\right) & =F_{T}(\boldsymbol{q},\omega,\chi)\left(\delta^{\alpha\beta}-\frac{q^{\alpha}q^{\beta}}{q^{2}}\right)+F_{L}(\boldsymbol{q},\omega,\chi)\nonumber \\
 & \times\frac{q^{\alpha}q^{\beta}}{q^{2}}+i\chi F_{A}(\boldsymbol{q},\omega,\chi)\varepsilon^{\alpha\beta\gamma}q^{\gamma},\label{PiAlphaBeta}
\end{align}
with
\begin{align}
F_{T}(\boldsymbol{q},\omega,\chi) & =\frac{1}{2}\left[A_{0}-B_{b}-q^{2}B_{1}\right],\\
F_{L}(\boldsymbol{q},\omega,\chi) & =\frac{1}{2}\left[A_{0}-B_{a}+2B_{b}+q^{2}B_{1}\right],\\
F_{A}(\boldsymbol{q},\omega,\chi) & =-\frac{1}{2}\left[C_{0}+C_{1}-D_{1}\right],
\end{align}
where $\varepsilon^{\alpha\beta\gamma}$ is the Levi-Civita symbol
and $\varepsilon^{xyz}=1$. $F_{T/L}(\boldsymbol{q},\omega,\chi)$
refers to the transverse/longitudinal component of the symmetric part
of the current-current correlation functions. Third, the chirality-dependent
density-current correlation functions $\Pi^{0\alpha/\alpha0}(\boldsymbol{q},\omega,\chi)$
are given as
\begin{align}
\Pi^{0\alpha}(\boldsymbol{q},\omega,\chi) & =\Pi^{\alpha0}(\boldsymbol{q},\omega,\chi)=F_{I}(\boldsymbol{q},\omega,\chi)q^{\alpha},\label{Pi0Alpha}
\end{align}
with
\begin{align}
 & F_{I}(\boldsymbol{q},\omega,\chi)=\frac{1}{2}\left[C_{0}+C_{1}+D_{1}\right].
\end{align}
Since $\Pi^{0\alpha}(\boldsymbol{q},\omega,\chi)$ and $\Pi^{\alpha0}(\boldsymbol{q},\omega,\chi)$
are parallel to $q^{\alpha}$, neither of them contributes to the orbital magnetization.

For convenience, we shall decompose the correlation function $\Pi^{\mu\nu}(q,\omega,\chi)$ into two parts,
$\Pi^{\mu\nu}(\boldsymbol{q},\omega,\chi)=\Pi_{-}^{\mu\nu}(\boldsymbol{q},\omega,\chi)
+\Theta(|\mu_{\chi}|)\Pi_{+}^{\mu\nu}(\boldsymbol{q},\omega,\chi)$,
where $\Pi_{-/+}^{\mu\nu}(\boldsymbol{q},\omega,\chi)$ denotes the intrinsic/extrinsic correlation functions.
After lengthy and complicated calculations, one finds the explicit
expressions of the seven scalar integrals $A_{0}^{\pm},B_{1}^{\pm},B_{a}^{\pm},B_{b}^{\pm},C_{0}^{\pm},C_{1}^{\pm}$,
and $D_{1}^{\pm}$ at zero temperature (see Appendices \ref{AppESInt} and \ref{AppEvaA0})~\cite{FitTem}.
Accordingly, one can obtain the intrinsic parts of the chirality-dependent correlation functions with $\mu_{\chi}=0$,
\begin{align}
\mathrm{Im}\Pi_{-}^{00}(\boldsymbol{q},\omega,\chi) & =-\frac{q^{2}}{24\pi}\Theta(\omega-q),\label{Pi00Im}\\
\mathrm{Re}\Pi_{-}^{00}(\boldsymbol{q},\omega,\chi) & =-\frac{q^{2}}{24\pi^{2}}\log\left|\frac{4\Lambda^{2}}{q^{2}-\omega^{2}}\right|,\label{RePi00}\\
\mathrm{Im}F_{T}^{-}(\boldsymbol{q},\omega,\chi) & =\frac{q^{2}-\omega^{2}}{24\pi}\Theta(\omega-q),\\
\mathrm{Re}F_{T}^{-}(\boldsymbol{q},\omega,\chi) & =-\frac{1}{24\pi^{2}}\left(4\Lambda^{2}-\frac{3}{5}q^{2}\right)\nonumber \\
 & +\frac{q^{2}-\omega^{2}}{24\pi^{2}}\log\left|\frac{4\Lambda^{2}}{q^{2}-\omega^{2}}\right|,\\
\mathrm{Im}F_{L}^{-}(\boldsymbol{q},\omega,\chi) & =\frac{-\omega^{2}}{24\pi}\Theta(\omega-q),\\
\mathrm{Re}F_{L}^{-}(\boldsymbol{q},\omega,\chi) & =-\frac{1}{24\pi^{2}}\left(4\Lambda^{2}-\frac{4}{5}q^{2}\right)\nonumber \\
 & -\frac{\omega^{2}}{24\pi^{2}}\log\left|\frac{4\Lambda^{2}}{q^{2}-\omega^{2}}\right|,\\
\mathrm{Im}F_{A}^{-}(\boldsymbol{q},\omega,\chi) & =0,\\
\mathrm{Re}F_{A}^{-}(\boldsymbol{q},\omega,\chi) & =\frac{\omega}{24\pi^{2}},\\
\mathrm{Im}F_{I}^{-}(\boldsymbol{q},\omega,\chi) & =-\frac{\omega}{24\pi}\Theta(\omega-q),\\
\mathrm{Re}F_{I}^{-}(\boldsymbol{q},\omega,\chi) & =-\frac{\omega}{24\pi^{2}}\log\left|\frac{4\Lambda^{2}}{q^{2}-\omega^{2}}\right|, \label{RePi0alpha}
\end{align}
and the extrinsic parts for $\mu_{\chi}>0$ (those for $\mu_{\chi}<0$ can be obtained from Appendix \ref{AppRCRpmmu})~\begin{widetext}
\begin{align}
\mathrm{Im}\Pi_{+}^{00}(\boldsymbol{q},\omega,\chi) & =-\frac{1}{64\pi}\left\{ \Theta(q-\omega)\sum_{\lambda=\pm}\lambda\Theta\left(\mu_{\chi}-\frac{q-\lambda\omega}{2}\right)G_{S}\left(q,\lambda\omega\right)\right.-\Theta(\omega-q)\nonumber \\
 & \left.\times\left[\frac{8}{3}q^{2}\Theta\left(\mu_{\chi}-\frac{\omega+q}{2}\right)+\tilde{\Theta}\left(\frac{\omega+q}{2}-\mu_{\chi}\right)\Theta\left(\mu_{\chi}-\frac{\omega-q}{2}\right)G_{S}(-q,-\omega)\right]\right\} ,\\
\mathrm{Re}\Pi_{+}^{00}(\boldsymbol{q},\omega,\chi) & =-\frac{\mu_{\chi}^{2}}{3\pi^{2}}+\frac{1}{64\pi^{2}}\sum_{\lambda,\lambda^{\prime}=\pm}G_{S}\left(\lambda q,\lambda^{\prime}\omega\right)H\left(\lambda q,\lambda^{\prime}\omega\right),\\
\mathrm{Im}F_{T}^{+}(\boldsymbol{q},\omega,\chi) & =-\frac{1}{64\pi}\left\{ \Theta(q-\omega)\sum_{\lambda=\pm}\lambda\Theta\left(\mu_{\chi}-\frac{q-\lambda\omega}{2}\right)G_{M}(q,\lambda\omega)\right.
-\Theta(\omega-q)\nonumber \\
 & \times\left.\left[-\frac{8}{3}\left(q^{2}-\omega^{2}\right)
 \Theta\left(\mu_{\chi}-\frac{\omega+q}{2}\right)
 +\tilde{\Theta}\left(\frac{\omega+q}{2}-\mu_{\chi}\right)\Theta\left(\mu_{\chi}-\frac{\omega-q}{2}\right)G_{M}(-q,-\omega)\right]\right\} ,\\
\mathrm{Re}F_{T}^{+}\left(\boldsymbol{q},\omega,\chi\right) & =\frac{\mu_{\chi}^{2}\left(q^{2}+2\omega^{2}\right)}{12\pi^{2}q^{2}}+\frac{1}{64\pi^{2}}\sum_{\lambda,\lambda^{\prime}=\pm}G_{M}\left(\lambda q,\lambda^{\prime}\omega\right)H\left(\lambda q,\lambda^{\prime}\omega\right),
\end{align}
\begin{align}
\mathrm{Im}F_{L}^{+}(\boldsymbol{q},\omega,\chi) & =-\frac{1}{64\pi}\left\{ \Theta(q-\omega)\sum_{\lambda=\pm}\lambda\Theta\left(\mu_{\chi}-\frac{q-\lambda\omega}{2}\right)G_{N}(q,\lambda\omega)\right.-\Theta(\omega-q)\nonumber \\
 & \left.\times\left[\frac{8}{3}\omega^{2}\Theta\left(\mu_{\chi}-\frac{\omega+q}{2}\right)
 +\tilde{\Theta}\left(\frac{\omega+q}{2}-\mu_{\chi}\right)\Theta\left(\mu_{\chi}-\frac{\omega-q}{2}\right)G_{N}(-q,-\omega)\right]\right\} ,\\
\mathrm{Re}F_{L}^{+}(\boldsymbol{q},\omega,\chi) & =-\frac{\mu_{\chi}^{2}\omega^{2}}{3\pi^{2}q^{2}}+\frac{1}{64\pi^{2}}\sum_{\lambda,\lambda^{\prime}=\pm}G_{N}(\lambda q,\lambda^{\prime}\omega)H(\lambda q,\lambda^{\prime}\omega),\\
\mathrm{Im}F_{A}^{+}(\boldsymbol{q},\omega,\chi) & =\frac{1}{64\pi}\left\{ \Theta(q-\omega)\sum_{\lambda=\pm}\lambda\Theta\left(\mu_{\chi}-\frac{q-\lambda\omega}{2}\right)G_{J}(q,\lambda\omega)\right.\nonumber \\
 & \left.-\Theta(\omega-q)\tilde{\Theta}\left(\frac{\omega+q}{2}-\mu_{\chi}\right)\Theta\left(\mu_{\chi}-\frac{\omega-q}{2}\right)G_{J}(-q,-\omega)\right\} ,\\
\mathrm{Re}F_{A}^{+}(\boldsymbol{q},\omega,\chi) & =\frac{(q^{2}-\omega^{2})\mu_{\chi}}{8\pi^{2}q^{2}}-\frac{1}{64\pi^{2}}\sum_{\lambda,\lambda^{\prime}=\pm}G_{J}\left(\lambda q,\lambda^{\prime}\omega\right)H\left(\lambda q,\lambda^{\prime}\omega\right),\\
\mathrm{Im}F_{I}^{+}(\boldsymbol{q},\omega,\chi) & =-\frac{1}{64\pi}\left\{ \Theta(q-\omega)\sum_{\lambda=\pm}\lambda\Theta\left(\mu-\frac{q-\lambda\omega}{2}\right)G_{I}(q,\lambda\omega)\right.-\Theta(\omega-q)\nonumber \\
 & \left.\times\left[\frac{8\omega}{3}\Theta\left(\mu_{\chi}-\frac{\omega+q}{2}\right)+\tilde{\Theta}\left(\frac{\omega+q}{2}-\mu_{\chi}\right)\Theta\left(\mu_{\chi}-\frac{\omega-q}{2}\right)G_{I}(-q,-\omega)\right]\right\} ,\\
\mathrm{Re}F_{I}^{+}(\boldsymbol{q},\omega,\chi) & =-\frac{\omega\mu_{\chi}^{2}}{3\pi^{2}q^{2}}+\frac{1}{64\pi^{2}}\sum_{\lambda,\lambda^{\prime}=\pm}G_{I}\left(\lambda q,\lambda^{\prime}\omega\right)H\left(\lambda q,\lambda^{\prime}\omega\right),\label{ReFIE}
\end{align}
\end{widetext}

where these auxiliary functions are defined as
\begin{align}
G_{f}(q,\xi) & =\frac{f(2\mu_{\chi}+\xi)-f(q)}{q},\\
H(q,\omega) & =\log\left|\frac{2\mu_{\chi}+\omega-q}{\omega-q}\right|,
\end{align}
with $f=S,M,N,J,I$ and $S(u)=2u\left(u^{2}-3q^{2}\right)/3$, $M(u)=(q^{2}-\omega^{2})\left(3q^{2}+u^{2}\right)u/3q^{2}$,
$N(u)=2\omega^{2}(u^2-3q^{2})u/3q^{2}$,
$J(u)=\left(q^{2}-\omega^{2}\right)u^{2}/q^{2}$ and $I(u)=2\omega u\left(u^{2}-3q^{2}\right)/3q^{2}$.
$\Theta\left(x\right)$ is the Heaviside step function and the tilted
Heaviside step function implies that $\tilde{\Theta}\left(x\right)=1$
for $x\geq1$, and vanishes otherwise. $\Lambda$ is a cut-off wave vector
relative to each Weyl node. Note that the density-density correlation
function via the PVRS is the same as the one via other methods in Ref.
\cite{ZhouJH2015PRB}. Equations $\left(\ref{Pi00Im}\right)$-$\left(\ref{ReFIE}\right)$
are the central result of this paper.

The general correlation functions should obey several fundamental
relations, such as the gauge invariance and the Kramers\textendash Kronig
relations. The gauge invariance (Ward identity) of the chirality-dependent correlation functions reads
\begin{align}
q_{\mu}\Pi^{\mu\nu}(\boldsymbol{q},\omega,\chi) & =0.
\end{align}
Inserting the expressions of correlation functions and using the relations
$q_{0}=q^{0}=\omega$, $q_{\alpha}=-q^{\alpha}$, $q^{\alpha}q^{\alpha}=q_{\alpha}q_{\alpha}=q^{2}$,
$\delta_{\alpha\beta}=\delta^{\alpha\beta}$, one finds
\begin{align}
\omega\Pi^{00}(\boldsymbol{q},\omega,\chi)-q^{2}F_{I}(\boldsymbol{q},\omega,\chi) & =0,\label{GIA}\\
\left[\omega F_{I}(\boldsymbol{q},\omega,\chi)-F_{L}(\boldsymbol{q},\omega,\chi)\right]q^{\beta} & =0.\label{GIB}
\end{align}
From Eq. $\left(\ref{GIA}\right)$, one immediately finds that the
static density-current correlation functions vanish $\Pi^{0\alpha}\left(\boldsymbol{q},0,\chi\right)=\Pi^{\alpha0}\left(\boldsymbol{q},0,\chi\right)=0$.
From Eq. $\left(\ref{GIB}\right)$, one has the relation $F_{L}(\boldsymbol{q},0,\chi)=0$.

It is straightforward to verify that the extrinsic part and the imaginary
part of the intrinsic correlation functions satisfy the Ward identity.
However, the nature of the cutoff scheme makes the intrinsic correlation
function always contain some gauge-violating terms \cite{peskin1995qft}.
The intrinsic correlation function is usually written as
\begin{equation}
\Pi_{-}^{\mu\nu}\left(\boldsymbol{q},\omega,\chi\right)=\left[\left(\omega^{2}-q^{2}\right)g^{\mu\nu}-q^{\mu}q^{\nu}\right]\Pi\left(\boldsymbol{q},\omega,\chi\right),
\end{equation}
where $g^{\mu\nu}=\mathrm{diag}\left(1,-1,-1,-1\right)$ is the metric
tensor, and the overall scalar function $\Pi(\boldsymbol{q},\omega,\chi)$
needs to be determined. From either the
density-density correlation function via the PVRS in Eq. $\left(\ref{RePi00}\right)$ or the results in Ref.~\cite{ZhouJH2015PRB}, one finds
\begin{align}
\Pi_{-}^{00}(\boldsymbol{q},\omega,\chi) & =-\frac{q^{2}}{24\pi^{2}}\log\frac{4\Lambda^{2}}{q^{2}-\omega^{2}},
\end{align}
which leads to the overall scalar function as
\begin{align}
\Pi\left(\boldsymbol{q},\omega,\chi\right) & =\frac{1}{24\pi^{2}}\log\frac{4\Lambda^{2}}{q^{2}-\omega^{2}}.\label{Pisf0}
\end{align}
In fact, this scalar function can be also extracted from the density-current correlations functions in Eqs. $(\ref{Pi0Alpha})$ and $(\ref{RePi0alpha})$.
To restore the gauge invariance of the intrinsic correlation function, we would like to subtract the non-logarithmic terms in $\mathrm{Re}F_{T}^{-}$
and $\mathrm{Re}F_{L}^{-}$ and get
\begin{align}
\mathrm{Re}F_{T}^{-}(\boldsymbol{q},\omega,\chi) & =\frac{\left(q^{2}-\omega^{2}\right)}{24\pi^{2}}\log\left|\frac{4\Lambda^{2}}{q^{2}-\omega^{2}}\right|,\\
\mathrm{Re}F_{L}^{-}(\boldsymbol{q},\omega,\chi) & =-\frac{\omega^{2}}{24\pi^{2}}\log\left|\frac{4\Lambda^{2}}{q^{2}-\omega^{2}}\right|.
\end{align}

\begin{widetext}

\begin{figure}
\includegraphics[scale=0.7]{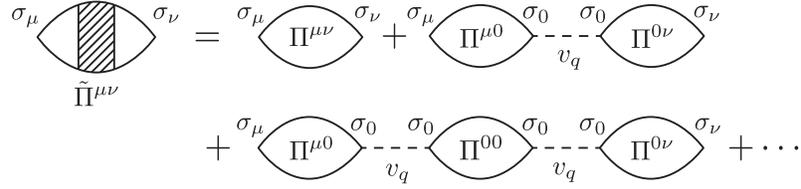}\caption{Diagrammatic relation between the RPA correlation function $\tilde{\Pi}^{\mu\nu}\left(\boldsymbol{q},\omega,\chi\right)$
and the noninteracting one $\Pi^{\mu\nu}\left(\boldsymbol{q},\omega,\chi\right)$.
The dashed line refers to the bare Coulomb interaction $v_{q}$. }
\label{rpa}
\end{figure}
\end{widetext}

Let us consider the correlation functions for the interacting electrons.
The simple way to encode the electron-electron interactions is the
RPA. Within the RPA as illustrated in Fig. \ref{rpa}, the correlation
functions of the interacting 3D Weyl/Dirac semimetals can be expressed as
\begin{align}
\tilde{\Pi}^{\mu\nu}\left(\boldsymbol{q},\omega,\chi\right) & =\Pi^{\mu\nu}\left(\boldsymbol{q},\omega,\chi\right)\nonumber \\
 & +\frac{\Pi^{\mu0}\left(\boldsymbol{q},\omega,\chi\right)v_{q}\Pi^{0\nu}\left(\boldsymbol{q},\omega,\chi\right)}{1-v_{q}\Pi^{00}\left(\boldsymbol{q},\omega,\chi\right)},
\end{align}
where $v_{q}=4\pi e^{2}/\kappa q^{2}$ is the Fourier transform of
3D Coulomb interaction, $\kappa$ is the effective dielectric constant.

For example, the interacting density-density response function with
$\mu=\nu=0$ becomes
\begin{equation}
\tilde{\Pi}^{00}\left(\boldsymbol{q},\omega,\chi\right)=\frac{\Pi^{00}\left(\boldsymbol{q},\omega,\chi\right)}{1-v_{q}\Pi^{00}\left(\boldsymbol{q},\omega,\chi\right)},
\end{equation}
which recovers the widely-used RPA density-density correlation function.
On the other hand, the current-current correlation functions within
the RPA are given as
\begin{align}
\tilde{\Pi}^{\alpha\beta}\left(\boldsymbol{q},\omega,\chi\right) & =\Pi^{\alpha\beta}\left(\boldsymbol{q},\omega,\chi\right)\nonumber \\
 & +\frac{\Pi^{\alpha0}\left(\boldsymbol{q},\omega,\chi\right)v_{q}\Pi^{0\beta}\left(\boldsymbol{q},\omega,\chi\right)}{1-v_{q}\Pi^{00}\left(\boldsymbol{q},\omega,\chi\right)}.
\end{align}
It is clear that the nonvanishing density-current response functions
give rise to corrections to the current-current correlation functions.
Since $\Pi^{\alpha0/0\alpha}\left(\boldsymbol{q},\omega,\chi\right)$
is parallel to the wave vector $q^{\alpha}$, there is no contribution to the antisymmetric
part or the transverse part of the current-current correlation functions from the electron
interactions. Note that, for the 3D conventional electron gases in the absence of magnetic
fields, the density-current response functions usually vanish.
Thus the electron interaction does not renormalize
the corresponding current-current correlation function within the RPA~\cite{giuliani2005qtel}.

\section{Generalized Kramers-Kronig relations \label{sec:KKR}}

The Kramers-Kronig relation establishes a connection between the real and imaginary
parts of the correlation functions and also
allows us to verify the correctness of the correlation functions. Before
generalizing the conventional Kramers-Kronig relation to the one with
$n$th-order subtraction, let us first examine the causality relations
among the correlation functions that enable us to find the negative-frequency
part by taking Hermitian conjugate of the positive-frequency part.

From the causality relations for chirality-dependent correlation functions
(see Appendix \ref{AppCR}), one could write down the relations between
the positive- and negative-frequency parts of each correlation function as follows
\begin{align}
\mathrm{Re}\Pi^{00}(\boldsymbol{q},-\omega,\chi) & =\mathrm{Re}\Pi^{00}(\boldsymbol{q},\omega,\chi),\nonumber \\
\mathrm{Im}\Pi^{00}(\boldsymbol{q},-\omega,\chi) & =-\mathrm{Im}\Pi^{00}(\boldsymbol{q},\omega,\chi),\nonumber \\
\mathrm{Re}F_{I}(\boldsymbol{q},-\omega,\chi) & =-\mathrm{Re}F_{I}(\boldsymbol{q},\omega,\chi),\nonumber \\
\mathrm{Im}F_{I}(\boldsymbol{q},-\omega,\chi) & =\mathrm{Im}F_{I}(\boldsymbol{q},\omega,\chi),\\
\mathrm{Re}F_{X}(\boldsymbol{q},-\omega,\chi) & =\mathrm{Re}F_{X}(\boldsymbol{q},\omega,\chi),\nonumber \\
\mathrm{Im}F_{X}(\boldsymbol{q},-\omega,\chi) & =-\mathrm{Im}F_{X}(\boldsymbol{q},\omega,\chi),\nonumber
\end{align}
where $X=T,L,A$. Based on all of the explicit expressions of chirality-dependent
correlation functions in previous section, it is instructive to crosscheck
the real part from the imaginary part via the Kramers-Kronig relations
or vice versa.

The conventional Kramers-Kronig relation for a complex function $f\left(\xi\right)$,
\begin{equation}
f\left(\omega\right)=\frac{1}{\pi i}\mathcal{P}\int_{-\infty}^{\infty}d\xi\frac{f(\xi)}{\xi-\omega} +\mathcal{C}_{\infty} \label{CKKR}
\end{equation}
requires that the contribution from the semicircle at infinity $\mathcal{C}_{\infty} = C_{\infty} + i C_{\infty}^{'}$ tends to vanish, i.e.,~$C_{\infty}=0$ and $C_{\infty}^{'}=0$.
Here $\mathcal{P}$ denotes the principal value of the integral. It is applicable to many
causal response functions, such as the dielectric functions for the
conventional electron gases~\cite{giuliani2005qtel} and 2D Dirac
fermions~\cite{Kotov2012RMP}. Historically, Bjorken and Drell~\cite{bjorken1965RQF}
had generalized Eq. $\left(\ref{CKKR}\right)$ to a case with $\mathcal{C}_{\infty}$ being a complex constant that needs a 1st-order subtraction.
Their generalization had been used to derive the correct density-density
correlation function of 3D Weyl semimetals \cite{ZhouJH2015PRB}.
However, neither the conventional Kramers-Kronig relation nor Bjorken
and Drell's generalization is adequate for the calculations of correlation
functions of our interest. In this paper, we would like to provide
a more general Kramers-Kronig relation with $n$th-order
subtraction, which is not only useful for our present calculations
but also of great interest to many other physical problems.

If $f\left(\xi\right)$ does not diverge more than $\xi^{n-1}$ as $\xi\to\infty$, the generalized
Kramers-Kronig relation is given as (The detailed proof is presented
in Appendix \ref{AppKKR})
\begin{align}
f(\omega)\prod_{m=1}^{n}\frac{1}{(\omega-\omega_{m})} & =\sum_{l=1}^{n}\frac{f(\omega_{l})}{(\omega-\omega_{l})}\prod_{m=1,m\neq l}^{n}\frac{1}{(\omega_{l}-\omega_{m})}\nonumber \\
 & +\frac{1}{\pi i}\mathcal{P}\int_{-\infty}^{+\infty}d\xi\frac{f(\xi)}{g(\xi)}+\mathcal{C}_{\infty},
\end{align}
where $g(\xi)=\left(\xi-\omega\right)\prod_{m=1}^{n}(\xi-\omega_{m})$.
The corresponding real and imaginary parts are given as
\begin{align}
\mathrm{Re}[f(\omega)]\prod_{m=1}^{n}\frac{1}{(\omega-\omega_{m})} & =\sum_{l=1}^{n}\frac{\mathrm{Re}f(\omega_{l})}{(\omega-\omega_{l})}\prod_{m=1,m\neq l}^{n}\frac{1}{(\omega_{l}-\omega_{m})}\nonumber \\
+\frac{1}{\pi}\mathcal{P} & \int_{-\infty}^{+\infty}d\xi\frac{\mathrm{Im}f(\xi)}{g(\xi)}+C_{\infty},\\
\mathrm{Im}[f(\omega)]\prod_{m=1}^{n}\frac{1}{(\omega-\omega_{m})} & =\sum_{l=1}^{n}\frac{\mathrm{Im}f(\omega_{l})}{(\omega-\omega_{l})}\prod_{m=1,m\neq l}^{n}\frac{1}{(\omega_{l}-\omega_{m})}\nonumber \\
-\frac{1}{\pi}\mathcal{P} & \int_{-\infty}^{+\infty}d\xi\frac{\mathrm{Re}f(\xi)}{g(\xi)}+C_{\infty}^{'},
\end{align}
which ensure $\mathcal{C}_{\infty}$ vanishes and are named the Kramers-Kronig relations with $n$th-order
subtraction. It is one of the main results in this paper. In principle, the
quantities of $\omega_{1}$, $\omega_{2}$, $\omega_{3}$, $\cdots$,
and $\omega_{n}$ can be arbitrarily chosen as if it is not equal
to $q$ in the real calculations. For the sake of simplicity, one could
choose $\omega_{i}=\alpha_{i}q$ for $i=1,2,3,\cdots,n$ with $\alpha_{1}\neq\alpha_{2}\neq\alpha_{3}\neq\cdots\neq\alpha_{n}\neq1$.
Note that the final result is independent of the specific values of $\omega_{i}$.

Let us explicitly verify the correlation functions by using the generalized
Kramers-Kronig relations. The extrinsic part $F^{+}(\boldsymbol{q},\xi,\chi)\to0$ as $\xi\to\infty$, so that one
only needs to use the conventional Kramers-Kronig relations in Eq.
$\left(\ref{CKKR}\right)$. However, since the intrinsic part $F^{-}(\boldsymbol{q},\xi,\chi)$
does not vanish as $\xi\to\infty$, one must utilize the generalized
Kramers-Kronig relations with $n$th-order subtraction. The least orders
of subtraction of the chirality-dependent correlation functions are
tabulated in Table \ref{Tab1}. If the least order of subtraction
is $m$, a higher-order subtraction $n>m$ would yield the same results
except for more tedious mathematical manipulations.

\begin{table}
\begin{tabular}{|c|c|c|c|c|}
\hline
correlation functions & $\Pi_{+}^{\mu\nu}$ & $\Pi_{-}^{00}$ & $F_{A}^{-}$,$\:$$F_{I}^{-}$ & $F_{L}^{-}$,$\:$$F_{T}^{-}$\tabularnewline
\hline
\hline
$n$ & 0 & 1 & 2 & 3\tabularnewline
\hline
\end{tabular}
\caption{The least order of subtraction, $n$, in the generalized Kramers-Kronig
relations for the correlation functions of 3D Weyl/Dirac semimetals.
The index $+\left(-\right)$ refers to the extrinsic (intrinsic) part. }
\label{Tab1}
\end{table}

To be specific, we take the intrinsic part of the density-current
correlation function as an example to demonstrate the application
of the generalized Kramers-Kronig relations. We set $f(\xi)=F_{I}^{-}\left(\boldsymbol{q},\xi,\chi\right)$
and have
\begin{equation}
\mathrm{Im}f(\xi)=\frac{-\xi}{24\pi}\Theta(\xi-q).
\end{equation}
Since $\mathrm{Im}F_{I}^{-}\left(\boldsymbol{q},\xi,\chi\right)$
does not diverge more than $\xi$ as $\xi\to\infty$, the generalized
Kramers-Kronig relations with at least 2nd-order subtraction are needed to
calculate $\mathrm{Re}F_{I}^{-}(\boldsymbol{q},\omega,\chi)$ from
$\mathrm{Im}F_{I}^{-}(\boldsymbol{q},\omega,\chi)$. The
Kramers-Kronig relations with 2nd-order subtraction has the form
\begin{align}
\mathrm{Re}f(\omega) & =\frac{\left(\omega-\omega_{2}\right)\mathrm{Re}f(\omega_{1})}{\left(\omega_{1}-\omega_{2}\right)}+\frac{\left(\omega-\omega_{1}\right)\mathrm{Re}f(\omega_{2})}{\left(\omega_{2}-\omega_{1}\right)}\nonumber \\
 & +\frac{1}{\pi}\mathcal{P}\int_{-\infty}^{+\infty}d\xi\frac{\left(\omega-\omega_{1}\right)\left(\omega-\omega_{2}\right)\mathrm{Im}f(\xi)}{\left(\xi-\omega\right)\left(\xi-\omega_{1}\right)\left(\xi-\omega_{2}\right)}.
\end{align}
Without loss of generality, we choose $\omega_{1}=\frac{1}{4}q$ and
$\omega_{2}=\frac{1}{2}q$. After some straightforward calculations, we obtain
\begin{equation}
\mathrm{Re}F_{I}^{-}(\boldsymbol{q},\omega,\chi)=-\frac{\omega}{24\pi^{2}}\log\left|\frac{4\Lambda^{2}}{q^{2}-\omega^{2}}\right|,
\end{equation}
which is identical to the one in Eq. $\left(\ref{RePi0alpha}\right)$.
Additionally, other functions involved in the correlation functions
can be calculated in a similar way.

\section{the optical conductivity \label{sec:OpticCond}}

In this section, the optical conductivity is obtained from the current-current
correlation function in Weyl/Dirac semimetals. The general formula
for the optical conductivity tensor is given as \cite{MahanMPP3}
\begin{align}
\sigma^{\alpha\beta}(\omega) & =\frac{i}{\omega+i\eta}\lim_{\boldsymbol{q}\rightarrow0}\left[\Pi^{\alpha\beta}(\boldsymbol{q},\omega)-\Pi^{\alpha\beta}(\boldsymbol{q},0)\right].
\end{align}
After taking the limit $\boldsymbol{q}\rightarrow0$, the second term
in the square bracket vanishes identically. The vanishing of the second
term implies no diamagnetic current, similar to 2D Dirac fermions in
graphene \cite{Stauber2015PRB}. The real part of the optical conductivity
$\sigma^{\alpha\beta}(\omega)$ is responsible to optical absorption
and has the form
\begin{align}
\mathrm{Re}\sigma^{\alpha\beta}(\omega) & =\sum_{\chi=\pm}\left[\frac{e^{2}\mu_{\chi}^{2}}{6\pi v_{F}\hbar^{3}}\delta\left(\omega\right)\right.\nonumber \\
 & +\left.\frac{e^{2}\omega}{24\pi v_{F}\hbar}\Theta\left(\hbar\omega-2\left|\mu_{\chi}\right|\right)\right]\delta^{\alpha\beta}.
\end{align}
It is clear that the first term corresponds to the intraband part,
while the second term is the interband part which onsets only above $2\left|\mu_{\chi}\right|$
with $\left(\mu_{\chi}\neq0\right)$. When the Fermi level crosses
the Weyl nodes $\mu_{\chi}=0$, the diagonal optical conductivity
reduces to $\mathrm{Re}\sigma^{\alpha\alpha}(\omega)=\frac{e^{2}\omega}{24\pi\hbar v_{F}}\Theta\left(\omega\right)$
\cite{Tabert2016PRB,Roy2016SR,roy2017prb}. Note that we have restored the
factors of $\hbar$, $v_{F}$, and $e$ to make the physical units clear in the final equalities.

\section{dynamical chiral magnetic effect \label{sec:DCME}}

\begin{figure}
\includegraphics[scale=0.6]{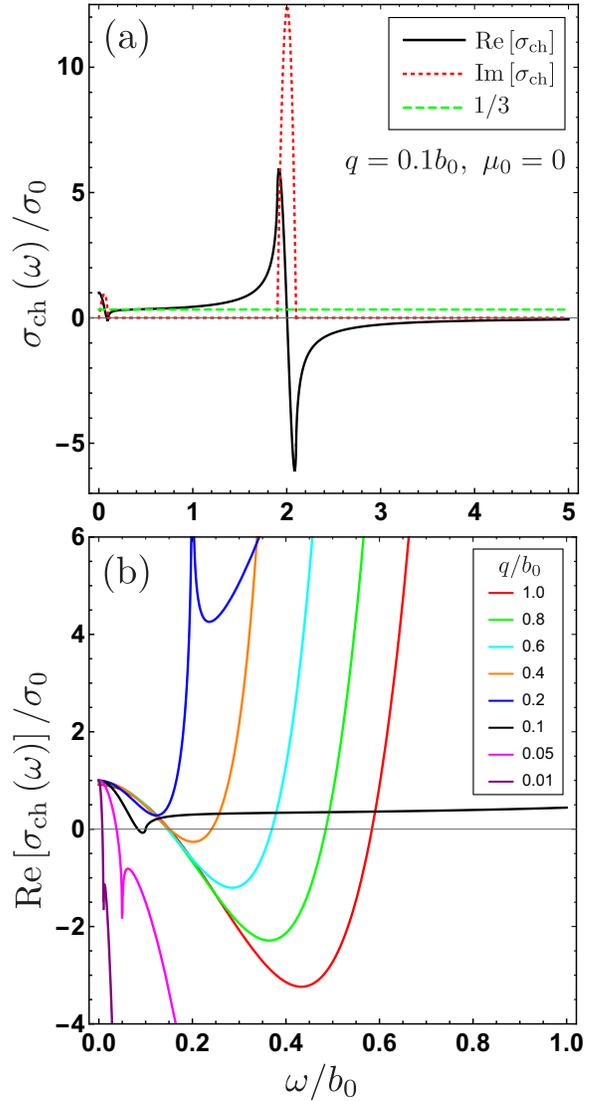}
\caption{(color online) The chiral magnetic conductivity as a function of frequency
at zero temperature. (a) real (black, solid) and imaginary (red, dotted)
part of the chiral magnetic conductivity at $q=0.1b_{0}$, $\mu_{0}=0$.
The green dashed line indicates $1/3$. (b) the real part of the chiral
magnetic conductivity for a set of values $q$. All the conductivities
are scaled by $\sigma_{0}=e^{2}b_{0}/2\pi^{2}$. }
\label{cmcQ}
\end{figure}

The chiral magnetic conductivity, the antisymmetric part of off-diagonal
electronic conductivity tensor, of 3D Weyl fermions has been studied
in some limits, such as the leading order part \cite{KharzeevPRD2009}
and the next-to-leading order hard dense loop approximation ($\omega,q\ll |\mu_{\chi}|$)
\cite{SonDT2013PRD}. Very recently, the dynamical CME or natural
optical activity was revisited in the context of Weyl semimetals and
metals without inversion symmetry from the semiclassical kinetic theory
\cite{Ma2015PRB,ZhongSD2016PRLGME}. It has been pointed out that
the dynamical CME has a geometric origin of Bloch bands and is directly
related to the intrinsic magnetic moment of Bloch electrons on the
Fermi surface. However, the semiclassical kinetic theory \cite{Xiao2010RMP,zhou2013cpl,SonDT2013PRD,Ma2015PRB,Chang2015PRBcme1,ZhongSD2016PRLGME}
does not work well when the Fermi level is very close to the Weyl
nodes, where the Berry curvature proportional to $1/k^{2}$ tends
to diverge as $k\rightarrow0$ \cite{Xiao2010RMP}. Meanwhile, the
strong interband correlation at a tiny $k$ makes the single-band
approximation in the semiclassical kinetic theory invalid. To comprehensively
understand both the static and dynamical CMEs, one needs the exact current-current
correlation functions, which allows us not only to reproduce the static
and dynamical CMEs in some limits but also to obtain the corrections
due to the higher-order spatial dispersion.

Within the linear response theory, the total current density induced
by the optical field $\boldsymbol{A}\left(\mathbf{r},t\right)=\boldsymbol{A}\left(\boldsymbol{q},\omega\right)e^{i\boldsymbol{q}\cdot\mathbf{r}-i\omega t}$
can be written as
\begin{equation}
j^{\alpha}\left(\boldsymbol{q},\omega\right)=\Pi^{\alpha\beta}\left(\boldsymbol{q},\omega\right)A^{\beta}\left(\boldsymbol{q},\omega\right).
\end{equation}
The Faraday's law, $\boldsymbol{B}\left(\boldsymbol{q},\omega\right)=\boldsymbol{q}\times\boldsymbol{E}\left(\boldsymbol{q},\omega\right)/\omega$
implies that a time-dependent magnetic field always comes together
with a perpendicular electric field. Since the electric field is a
vector and the magnetic field is a pseudovector, the CME coefficient
is parity-odd. Due to the rotational and gauge invariance,
one could adopt the chiral magnetic conductivity as \cite{KharzeevPRD2009}
\begin{equation}
\sigma_{\mathrm{ch}}\left(\boldsymbol{q},\omega\right)=\frac{1}{2iq^{\alpha}}\varepsilon^{\alpha\beta\gamma}\Pi^{\beta\gamma}\left(\boldsymbol{q},\omega\right),
\end{equation}
which only involves the antisymmetric part (or the parity-odd part)
of the current-current correlation function. Inserting the specific
expression of current-current correlation function, $\Pi_{A}^{\alpha\beta}(\boldsymbol{q},\omega)=\sum_{\chi=\pm}i\chi F_{A}(\boldsymbol{q},\omega,\chi)\varepsilon^{\alpha\beta\gamma}q^{\gamma}$,
immediately leads to
\begin{equation}
\sigma_{\mathrm{ch}}\left(\boldsymbol{q},\omega\right)=\sum_{\chi=\pm}\chi F_{A}\left(\boldsymbol{q},\omega,\chi\right).\label{SigCh}
\end{equation}
It is nothing else but the exact chiral magnetic conductivity of 3D Weyl fermions.

Several remarks are in order here. First, $\sigma_{\mathrm{ch}}\left(\boldsymbol{q},\omega\right)$
does not rely on those approximations made in Refs. \cite{KharzeevPRD2009,SonDT2013PRD,Ma2015PRB,ZhongSD2016PRLGME}.
Second, it should be emphasized that, in the large chemical potential
limit $|\mu_{\chi}|\gg q,\omega$, the chiral magnetic conductivity
in Eq. $\left(\ref{SigCh}\right)$ exactly recovers the previous result
\cite{SonDT2013PRD}. Third, $F_{A}\left(\boldsymbol{q},\omega,\chi\right)$
is an odd function of the chirality-dependent chemical potential $\mu_{\chi}$
(see Appendix \ref{AppRCRpmmu}). Finally, from the RPA procedure
in Fig. \ref{rpa}, the electron interaction does not modify the antisymmetric
part of current-current correlation functions such that the dynamical
CME remains unchanged. Therefore, our results provide a more throughout
understanding of the dynamical CME at nonzero frequencies and momentum.
It is a second main result in this paper.

\begin{figure}
\includegraphics[scale=0.75]{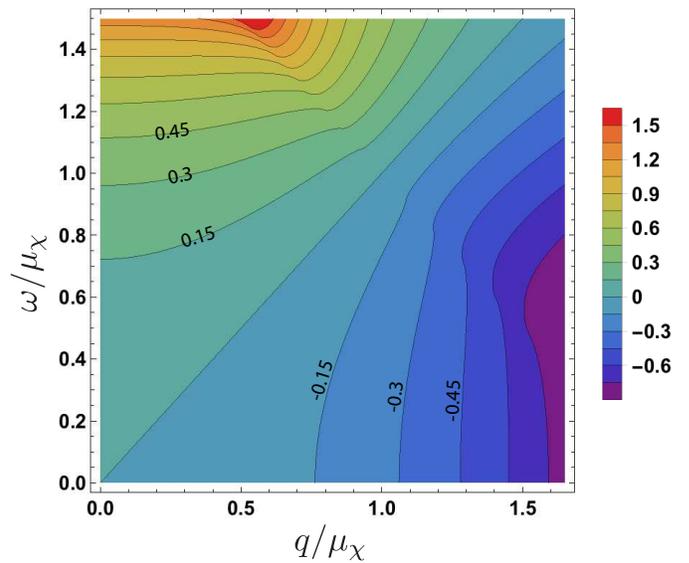}
\caption{(color online) Contour plot of the difference between the real part
of the chiral magnetic conductivity of one Weyl node in Eq. $\left(\ref{SigCh}\right)$
and that of Eq. (84) in Ref. \cite{SonDT2013PRD} at zero temperature.
The results are normalized to $e^{2}\mu_{\chi}/12\pi^{2}$. }
\label{cmcDif}
\end{figure}

Fig. \ref{cmcQ}(a) depicts the real and imaginary parts of the exact
chiral magnetic conductivity at $q=0.1b_{0}$ in Eq. $\left(\ref{SigCh}\right)$
and captures all essential features of the chiral magnetic conductivity
in Fig. 1 in Ref. \cite{KharzeevPRD2009}. First, one can clearly
see a typical resonance behavior with a peak at $\omega=2b_{0}$.
Second, the chiral magnetic conductivity (black solid line in (a))
drops from $\sigma_{0}$ at $\omega=0$ to $\sigma_{0}/3$. As shown
in Fig. \ref{cmcQ}(b), the behavior of the chiral magnetic conductivity
strongly depends on the magnitude of the wave vector $q$. The difference
between the real part of the chiral magnetic conductivity in Eq. $\left(\ref{SigCh}\right)$
and that in the hard dense loop approximation \cite{SonDT2013PRD}
is illustrated in Fig. \ref{cmcDif}. The approximate result in Ref.
\cite{SonDT2013PRD} is in a good agreement with ours when $\omega/\left|\mu_{\chi}\right|$
or $q/\left|\mu_{\chi}\right|$ is less than 0.4. However, the approximate
chiral magnetic conductivity shows a noticeable deviation from our
exact one when the ratio of $\omega/\left|\mu_{\chi}\right|$ or $q/\left|\mu_{\chi}\right|$
starts to exceed 0.6.

There are two distinct limits for the chiral magnetic conductivity:
the static limit ($\omega=0$ before $\boldsymbol{q}\rightarrow0$)
and the uniform limit ($\boldsymbol{q}=0$ before $\omega\rightarrow0$)
\cite{MahanMPP3}. Let us first examine the static limit
\begin{equation}
\lim_{\boldsymbol{q}\rightarrow0}\lim_{\omega\rightarrow0}\mathrm{Re}F_{A}\left(\boldsymbol{q},\omega,\chi\right)=\frac{e^{2}\mu_{\chi}}{4\pi^{2}}.
\end{equation}
Thus the chiral magnetic conductivity for a pair of Weyl nodes in the static limit becomes
\begin{equation}
\sigma_{\mathrm{ch}}\left(\boldsymbol{q},\omega\right)=\frac{e^{2}b_{0}}{2\pi^{2}\hbar^{2}c}.
\end{equation}
If $b_{0}$ is regarded as the chiral chemical potential induced by
the parallel electric and magnetic fields via the chiral anomaly \cite{ZhouJH2015PRB},
the static chiral magnetic conductivity vanishes identically and agrees
with the general semiclassical analysis \cite{zhou2013cpl} and the
numerical simulations in lattice models \cite{Vazifeh2013PRL,Chang2015PRBcme1}.
On the other hand, if $b_{0}$ is the energy difference of Weyl nodes
due to the inversion symmetry breaking, one may naively expect a nonzero
electric current induced by a static magnetic field, which is unfortunately
inconsistent with the fact that there is no equilibrium current in
solids in the static limit \cite{MahanMPP3}. The controversy of the
chiral magnetic conductivity might be resolved by introducing the Bardeen-Zumino
Chern-Simons term~\cite{Bardeen1969PR,BARDEEN1984NPB,Landsteiner2014PRB,GorbarPRL2017,Huang2017PRB}.

Similarly, one evaluates the uniform limit and obtains
\begin{equation}
\lim_{\omega\rightarrow0}\lim_{\boldsymbol{q}\rightarrow0}\mathrm{Re}F_{A}\left(\boldsymbol{q},\omega,\chi\right)=\frac{e^{2}\mu_{\chi}}{12\pi^{2}},
\end{equation}
which gives rise to the corresponding chiral magnetic conductivity
in the uniform limit
\begin{equation}
\sigma_{\mathrm{ch}}\left(\boldsymbol{q},\omega\right)=\frac{e^{2}b_{0}}{6\pi^{2}\hbar^{2}c},
\end{equation}
where $b_{0}$ refers to the energy difference of Weyl nodes with
opposite chirality. Our result here is consistent with the ones in
Refs. \cite{KharzeevPRD2009,SonDT2013PRD,Ma2015PRB,ZhongSD2016PRLGME}.

To gain more insights into the dynamical CME, we consider the limit
that the Fermi energy is far from the Weyl nodes $\left|\mu_{\chi}\right|\gg\omega,q$.
Expanding $\mathrm{Re}F_{A}\left(\boldsymbol{q},\omega,\chi\right)$
in power of $\boldsymbol{q}$ or $\omega$ and keeping the correction
up to $\mathcal{O}\left(q^{3}\right)$ or $\mathcal{O}\left(\omega^{3}\right)$,
one finds the real part
\begin{eqnarray}
\mathrm{Re}F_{A}\left(\boldsymbol{q},\omega,\chi\right) & \approx & \begin{cases}
\frac{e^{2}\mu_{\chi}}{12\pi^{2}}\left(1-\frac{2q^{2}}{5\omega^{2}}\right), & \omega\gg q;\\
\frac{e^{2}\mu_{\chi}}{4\pi^{2}}\left(1-\frac{2\omega^{2}}{q^{2}}\right), & q\gg\omega.
\end{cases}\label{dcme2nd}
\end{eqnarray}
The first term in each line on the right hand side corresponds to
the chiral magnetic conductivities in two different limits (uniform
and static), while the second term is the new leading order correction.
In sum, the exact chiral magnetic conductivity in Eq. $\left(\ref{SigCh}\right)$
possesses more rich features than the two distinct limits.

\section{magnetic susceptibilities~\label{sec:MS}}

In this section, we first consider the Pauli susceptibility and the
orbital magnetic susceptibility of the noninteracting 3D Weyl/Dirac
semimetals in the weak magnetic field limit thus neglecting the Landau
level structure of Weyl nodes. We then briefly discuss the impact
of electron interactions on these magnetic susceptibilities within
the RPA.

The Pauli spin susceptibility of noninteracting electrons follows
from the limit
\begin{equation}
\chi_{P}^{0}=-\left(\frac{g\mu_{B}}{2}\right)^{2}\lim_{q\rightarrow0}\sum_{\chi=\pm}\Pi^{zz}\left(q\hat{z},0,\chi\right),\label{chiP0}
\end{equation}
where $\mu_{B}$ is the Bohr magneton of free electrons and $g$ is
the material-dependent $g$ factor.

If the Pauli matrices $\sigma^{i}$ in Eq. $\left(\ref{Ham}\right)$
refer to the pseudospin degree of freedom (3D analogs of graphene),
the spin response function of a noninteracting system equals the density-density susceptibility,
\begin{align}
\chi_{P}^{0} & =-\left(\frac{g\mu_{B}}{2}\right)^{2}\lim_{q\rightarrow0}\sum_{\chi=\pm}\Pi^{00}\left(q,0,\chi\right)\nonumber \\
 & =\left(\frac{g\mu_{B}}{2}\right)^{2}\sum_{\chi=\pm}N\left(\mu_{\chi}\right),
\end{align}
which is proportional to the sum of the density of states near the Fermi level
$N\left(\mu_{\chi}\right)=\mu_{\chi}^{2}/\left(2\pi^{2}v_{F}^{3}\hbar^{3}\right)$,
similar to the conventional electron gases and graphene \cite{giuliani2005qtel,Kotov2012RMP}.

On the other hand, if the Pauli matrices $\sigma^{i}$ in Eq. $\left(\ref{Ham}\right)$
refer to the real spin of electrons, the spin response function of
a noninteracting system vanishes identically,
\begin{align}
\chi_{P}^{0} & =-\left(\frac{g\mu_{B}}{2}\right)^{2}\lim_{q\rightarrow0}\sum_{\chi=\pm}F_{L}\left(q,0,\chi\right)=0,
\end{align}
which can be traced back to the fact that the spin-momentum locking
causes the average Zeeman energy over the Fermi surface near each
Weyl node to vanish. It is worth noting that the vanishing of the
Pauli spin susceptibilities had been calculated in the Landau level
basis \cite{Koshino2016PRB} and possibly observed in NbAs, a candidate
of Weyl semimetal, at the quantum limit \cite{moll2016NC}.

For 3D Weyl/Dirac semimetals, a magnetic field usually produce both
the orbital diamagnetism and the splitting of Weyl nodes with opposite
chirality in momentum space $\boldsymbol{b}\neq0$ though the orbital
motion of electrons and the Zeeman interaction, respectively. The
former corresponds to the orbital magnetic susceptibility, while the
latter leads to a finite orbital magnetization \cite{Xiao2010RMP}.
The transverse current-current correlation function allows us to calculate
the noninteracting orbital magnetic susceptibility induced by a static magnetic field
\begin{align}
\chi_{\mathrm{orb}}^{0} & =-\frac{e^{2}v_{F}^2}{c^{2}}\sum_{\chi=\pm}\lim_{q\rightarrow0}\frac{\Pi^{zz}\left(q\hat{x},0,\chi\right)}{q^{2}}\nonumber \\
 & =\frac{-e^{2}v_{F}}{12\pi^{2}\hbar c^{2}}\log\frac{\varepsilon_c^{2}}{\left|\mu_{+} \mu_{-}\right|}+\frac{5e^{2}v_{F}}{36\pi^{2}\hbar c^{2}}.\label{chiOrb0}
\end{align}
where $\varepsilon_c = v_{F}\hbar\Lambda$ is the cut-off energy measured from each Weyl node.
The first term in the second line exactly agrees with the one obtained from the thermodynamic potential in the quantum limit
\cite{Koshino2016PRB}. However, the second term was absent there.
Since the cut-off energy $\varepsilon_c$ is far greater than the chirality-dependent
chemical potential $\left|\mu_{\chi}\right|$, the second term becomes negligible,
leading to an orbital diamagnetism. This logarithmic divergence of
the orbital diamagnetism was attributed to the relativistic nature
of 3D Weyl/Dirac fermions.

According to the semiclassical formula of the orbital magnetization
\cite{Xiao2010RMP}, the orbital magnetization for a pair of Weyl
nodes with a finite momentum spacing $\boldsymbol{b}=\left(0,0,b\right)$
is given as
\begin{equation}
M_{z}=-\frac{\mu_{0}}{ec}\sigma_{xy}=\frac{e\mu_{0}b}{2\pi^{2}\hbar c},
\end{equation}
where $\sigma_{xy}=-\frac{e^{2}b}{2\pi^{2}\hbar}$ is the corresponding
anomalous Hall conductivity at $\mu_{0}=0$ \cite{AHEmu} and $b=g\mu_{B}B$.
Our semiclassical estimation of the orbital magnetization is consistent
with the spin-orbit component from the direct derivative of the thermodynamic
potential with respect to the orbital part and the spin part of magnetic
fields \cite{Koshino2016PRB}. This part of orbital magnetization
is related to the Fermi arc states that connect the two Weyl nodes.
It is well known that the Pauli susceptibility and the orbital magnetic
susceptibility of the 3D noninteracting electron gases with a single-parabolic
band satisfy the relation $\chi_{P}^{0}=-3\chi_{\mathrm{orb}}^{0}$
\cite{giuliani2005qtel}. It is clear that the ratio between the orbital
and Pauli susceptibilities of 3D Weyl fermions greatly deviates from
$-1/3$ of the 3D electron gases.

Before closing this section, let us briefly discuss the impact of
electron interactions on the magnetic susceptibilities. Following
the RPA procedure in Fig. \ref{rpa}, one finds that there is no correction
to either the Pauli susceptibility or the orbital magnetic susceptibility
from electron interactions. In addition, another RPA procedure was
proposed to compute the orbital magnetic susceptibility in the context of graphene \cite{Principi2009PRB,Scholz2011PRB}
\begin{align}
\tilde{\chi}_{\mathrm{orb}} & =-\frac{e^{2}v_{F}^2}{c^{2}}\lim_{q\rightarrow0}\frac{\Pi^{zz}\left(q\hat{x},0\right)}{q^{2}\left[1-v_{q}\Pi^{zz}\left(q\hat{x},0\right)\right]}.
\end{align}
Within this RPA procedure, the orbital magnetic susceptibility of
3D interacting Weyl fermions becomes
\begin{align}
\tilde{\chi}_{\mathrm{orb}} & =\sum_{\chi=\pm}\frac{\chi_{\mathrm{orb}}^{0}\left(\chi\right)}{\left[1-\frac{4\pi e^{2}}{\kappa}\cdot\frac{1}{12\pi^{2}v_{F}\hbar}\left(\log\frac{\varepsilon_c}{\left| \mu_{\chi}\right|}-\frac{5}{6}\right)\right]},
\end{align}
where $\chi_{\mathrm{orb}}^{0}\left(\chi\right)$ refers to the orbital
magnetic susceptibilities of noninteracting Weyl fermions of chirality
$\chi$ in Eq. $\left(\ref{chiOrb0}\right)$. One can see that the
interaction effect gives rise to a $\Lambda$-dependent renormalization
of the orbital magnetic susceptibilities. For a sufficiently large
ratio $\varepsilon_c/\left| \mu_{\chi} \right|$, the interacting orbital magnetic
susceptibility $\tilde{\chi}_{\mathrm{orb}}$ might change a sign.
In other words, a transition from the orbital diamagnetism to the
orbital paramagnetism could occur through tuning the Fermi energy
or the carrier concentration. The interacting susceptibility for the
orbital paramagnetism becomes
\begin{align}
\tilde{\chi}_{\mathrm{orb}} & =\frac{\kappa v_{F}^{2}}{2\pi c^{2}},
\end{align}
which turns out to be independent of the cutoff $\Lambda$. One should
bear in mind that when the Fermi level approaches the Weyl nodes,
the carrier density becomes very low such that the RPA might become
invalid. Thus, the interaction-driven transition of orbital magnetism
requires a more sophisticated treatment beyond the simple RPA, such
as the higher-order perturbation theory and the GW approximation~\cite{giuliani2005qtel}.

\section{conclusions and discussions \label{sec:conclusions}}

In summary, the dynamical correlation functions of 3D Weyl/Dirac
semimetals have been derived analytically via the PVRS at zero temperature.
The gauge invariance and Kramers-Kronig relations among these correlation
functions have been verified in details. We have obtained the exact
chiral magnetic conductivity and deepened the understanding of the
CME. We have calculated the magnetic susceptibilities as well as
the orbital magnetization. The impacts of electron interactions within
the RPA on the magnetic susceptibilities are also discussed. In addition,
the dynamical correlation functions might be useful to explore
the nonlocal transport and optical properties due to the higher-order
spatial dispersion of 3D Weyl/Dirac semimetals in the presence of
time- and spatially-varying external fields, such as Lorentz birefringence
and Jones birefringence \cite{raab2005MTEM}.

\section{ACKNOWLEDGMENTS}

We are grateful to Wen-Yu Shan for useful discussions and to Shi-Xiong
Wang for preparing the figures and a careful reading of the manuscript.
J.Z. was supported by the Research Grant Council, University Grants
Committee, Hong Kong under Grant No. 17301116 and C6026-16W. H.-R.C.
was supported by the National Natural Science Foundation of China
under Grant No. 11547200, the China Scholarship Council, the NSERC of Canada, and FQRNT of
Quebec (Hong Guo). J.Z. also acknowledges the hospitality of Department of Physics in Southern University of Science and Technology.

\textit{Note added.}\textendash While we were finalizing the paper,
an independent study \cite{Thakur2017} appeared which derives the
diagonal current-current correlation functions by a different approach
and computes some related quantities.

\appendix

\section{Decomposition of the correlation functions via PVRS \label{AppDCRPVRS}}

To apply the PVRS to evaluate the correlation functions contains two
steps. The first step is to reduce the correlation functions to six
basic tensor integrals (including two basic scalar integrals). The
second is to further decompose them into seven basic scalar integrals
by utilizing PVRS.

By utilizing the traces of products of Pauli matrices
\begin{align}
\mathrm{Tr}[\sigma^{\mu}] & =2\delta^{\mu0},\nonumber \\
\mathrm{Tr}[\sigma^{\alpha}\sigma^{\beta}] & =2\delta^{\alpha\beta},\nonumber \\
\mathrm{Tr}[\sigma^{\alpha}\sigma^{\beta}\sigma^{\tau}] & =2i\varepsilon^{\alpha\beta\tau},\\
\mathrm{Tr}[\sigma^{\alpha}\sigma^{\tau}\sigma^{\beta}\sigma^{\rho}] & =2(\delta^{\alpha\tau}\delta^{\beta\rho}-\delta^{\alpha\beta}\delta^{\tau\rho}+\delta^{\alpha\rho}\delta^{\beta\tau}),\nonumber
\end{align}
we decompose the four chirality-dependent correlation functions as follows \begin{widetext}
\begin{align}
\Pi^{00}(\boldsymbol{q},i\omega_{m},\chi) & =\frac{1}{2}\left[A_{0}+\delta^{\tau\rho}\left(B^{\tau\rho}+B^{\tau}q^{\rho}\right)\right],\\
\Pi^{\alpha\beta}(\boldsymbol{q},i\omega_{m},\chi) & =\frac{1}{2}\left[\delta^{\alpha\beta}A_{0}+\left(\delta^{\alpha\tau}\delta^{\beta\rho}-\delta^{\alpha\beta}\delta^{\tau\rho}+\delta^{\alpha\rho}\delta^{\beta\tau}\right)\left(B^{\tau\rho}+B^{\tau}q^{\rho}\right)+i\chi\varepsilon^{\alpha\beta\tau}\left(C^{\tau}+C_{0}q^{\tau}-D^{\tau}\right)\right],\\
\Pi^{0\alpha}(\boldsymbol{q},i\omega_{m},\chi) & =\frac{1}{2}\left[\delta^{\alpha\tau}\left(C^{\tau}+C_{0}q^{\tau}\right)+\delta^{\alpha\tau}D_{\tau}+\chi i\varepsilon^{\tau\alpha\rho}\left(B^{\tau\rho}+B^{\tau}q^{\rho}\right)\right],\\
\Pi^{\alpha0}(\boldsymbol{q},i\omega_{m},\chi) & =\frac{1}{2}\left[\delta^{\alpha\tau}\left(C^{\tau}+C_{0}q^{\tau}\right)+\delta^{\alpha\tau}D^{\tau}+\chi i\varepsilon^{\alpha\tau\rho}\left(B^{\tau\rho}+B^{\tau}q^{\rho}\right)\right],
\end{align}
which clearly show that the four chirality-dependent correlation functions are reduced to six integrals
\begin{align}
A_{0}(\boldsymbol{q},i\omega_{m},\chi) & \equiv\frac{1}{\mathcal{V}}\sum_{\boldsymbol{k}}\frac{1}{\beta_{T}}\sum_{i\Omega_{n}}\sum_{\lambda,\lambda^{\prime}=\pm}\frac{1}{i\Omega_{n}+\mu_{\chi}-\varepsilon_{\lambda}(\boldsymbol{k})}\frac{1}{i\Omega_{n}+i\omega_{m}+\mu_{\chi}-\varepsilon_{\lambda^{\prime}}(\boldsymbol{k}+\boldsymbol{q})},\\
B^{\tau\rho}(\boldsymbol{q},i\omega_{m},\chi) & \equiv\frac{1}{\mathcal{V}}\sum_{\boldsymbol{k}}\frac{1}{\beta_{T}}\sum_{i\Omega_{n}}\sum_{\lambda,\lambda^{\prime}=\pm}\lambda\lambda^{\prime}\frac{k^{\tau}k^{\rho}}{|\boldsymbol{k}||\boldsymbol{k}+\boldsymbol{q}|}\frac{1}{i\Omega_{n}+\mu_{\chi}-\varepsilon_{\lambda}(\boldsymbol{k})}\frac{1}{i\Omega_{n}+i\omega_{m}+\mu_{\chi}-\varepsilon_{\lambda^{\prime}}(\boldsymbol{k}+\boldsymbol{q})},\\
B^{\tau}\left(\boldsymbol{q},i\omega_{m},\chi\right) & \equiv\frac{1}{\mathcal{V}}\sum_{\boldsymbol{k}}\frac{1}{\beta_{T}}\sum_{i\Omega_{n}}\sum_{\lambda,\lambda^{\prime}=\pm}\lambda\lambda^{\prime}\frac{k^{\tau}}{|\boldsymbol{k}||\boldsymbol{k}+\boldsymbol{q}|}\frac{1}{i\Omega_{n}+\mu_{\chi}-\varepsilon_{\lambda}(\boldsymbol{k})}\frac{1}{i\Omega_{n}+i\omega_{m}+\mu_{\chi}-\varepsilon_{\lambda^{\prime}}(\boldsymbol{k}+\boldsymbol{q})},
\end{align}

\begin{align}
C^{\tau}(\boldsymbol{q},i\omega_{m},\chi) & \equiv\frac{1}{\mathcal{V}}\sum_{\boldsymbol{k}}\frac{1}{\beta_{T}}\sum_{i\Omega_{n}}\sum_{\lambda,\lambda^{\prime}=\pm}\lambda^{\prime}\frac{k^{\tau}}{|\boldsymbol{k}+\boldsymbol{q}|}\frac{1}{i\Omega_{n}+\mu_{\chi}-\varepsilon_{\lambda}(\boldsymbol{k})}\frac{1}{i\Omega_{n}+i\omega_{m}+\mu_{\chi}-\varepsilon_{\lambda^{\prime}}(\boldsymbol{k}+\boldsymbol{q})},\\
C_{0}(\boldsymbol{q},i\omega_{m},\chi) & \equiv\frac{1}{\mathcal{V}}\sum_{\boldsymbol{k}}\frac{1}{\beta_{T}}\sum_{i\Omega_{n}}\sum_{\lambda,\lambda^{\prime}=\pm}\lambda^{\prime}\frac{1}{|\boldsymbol{k}+\boldsymbol{q}|}\frac{1}{i\Omega_{n}+\mu_{\chi}-\varepsilon_{\lambda}(\boldsymbol{k})}\frac{1}{i\Omega_{n}+i\omega_{m}+\mu_{\chi}-\varepsilon_{\lambda^{\prime}}(\boldsymbol{k}+\boldsymbol{q})},\\
D^{\tau}(\boldsymbol{q},i\omega_{m},\chi) & \equiv\frac{1}{\mathcal{V}}\sum_{\boldsymbol{k}}\frac{1}{\beta_{T}}\sum_{i\Omega_{n}}\sum_{\lambda,\lambda^{\prime}=\pm}\lambda\frac{k^{\tau}}{|\boldsymbol{k}|}\frac{1}{i\Omega_{n}+\mu_{\chi}-\varepsilon_{\lambda}(\boldsymbol{k})}\frac{1}{i\Omega_{n}+i\omega_{m}+\mu_{\chi}-\varepsilon_{\lambda^{\prime}}(\boldsymbol{k}+\boldsymbol{q})}.
\end{align}
Here $A_{0}(\boldsymbol{q},i\omega_{m},\chi)$ and $C_{0}(\boldsymbol{q},i\omega_{m},\chi)$
are scalar integrals and relatively easy to be evaluated. Our main
task is to compute the complicated tensor integrals $B^{\tau\rho}(\boldsymbol{q},i\omega_{m},\chi)$,
$B^{\tau}(\boldsymbol{q},i\omega_{m},\chi)$, $C^{\tau}(\boldsymbol{q},i\omega_{m},\chi)$,
and $D^{\tau}(\boldsymbol{q},i\omega_{m},\chi)$ via PVRS. Using the Ansatz
\begin{align}
B^{\tau}\left(\boldsymbol{q},i\omega_{m},\chi\right) & \equiv B_{1}(\boldsymbol{q},i\omega_{m},\chi)q^{\tau},\\
B^{\tau\rho}(\boldsymbol{q},i\omega_{m},\chi) & \equiv B_{00}(\boldsymbol{q},i\omega_{m},\chi)\delta^{\tau\rho}+B_{11}(\boldsymbol{q},i\omega_{m},\chi)q^{\tau}q^{\rho},\\
C^{\tau}(\boldsymbol{q},i\omega_{m},\chi) & \equiv C_{1}(\boldsymbol{q},i\omega_{m},\chi)q^{\tau},\\
D^{\tau}(\boldsymbol{q},i\omega_{m},\chi) & \equiv D_{1}(\boldsymbol{q},i\omega_{m},\chi)q^{\tau},
\end{align}
we have
\begin{align}
B^{\tau}(\boldsymbol{q},i\omega_{m},\chi)q^{\tau} & \equiv B_{1}(\boldsymbol{q},i\omega_{m},\chi)q^{\tau}q^{\tau}=B_{1}(\boldsymbol{q},i\omega_{m},\chi)q^{2},\\
C^{\tau}(\boldsymbol{q},i\omega_{m},\chi)q^{\tau} & \equiv C_{1}(\boldsymbol{q},i\omega_{m},\chi)q^{\tau}q^{\tau}=C_{1}(\boldsymbol{q},i\omega_{m},\chi)q^{2},\\
D^{\tau}(\boldsymbol{q},i\omega_{m},\chi)q^{\tau} & \equiv D_{1}(\boldsymbol{q},i\omega_{m},\chi)q^{\tau}q^{\tau}=D_{1}(\boldsymbol{q},i\omega_{m},\chi)q^{2},\\
B^{\tau\rho}(\boldsymbol{q},i\omega_{m},\chi)\delta^{\tau\rho} & \equiv B_{00}(\boldsymbol{q},i\omega_{m},\chi)\delta^{\tau\rho}\delta^{\tau\rho}+B_{11}(\boldsymbol{q},i\omega_{m},\chi)q^{\tau}q^{\rho}\delta^{\tau\rho}\nonumber \\
 & =3B_{00}(\boldsymbol{q},i\omega_{m},\chi)+q^{2}B_{11}(\boldsymbol{q},i\omega_{m},\chi)\equiv B_{a}(\boldsymbol{q},i\omega_{m},\chi),\\
B^{\tau\rho}(\boldsymbol{q},i\omega_{m},\chi)q^{\tau}q^{\rho} & \equiv B_{00}(\boldsymbol{q},i\omega_{m},\chi)\delta^{\tau\rho}q^{\tau}q^{\rho}+B_{11}(\boldsymbol{q},i\omega_{m},\chi)q^{\tau}q^{\rho}q^{\tau}q^{\rho}\nonumber \\
 & =B_{00}(\boldsymbol{q},i\omega_{m},\chi)q^{2}+B_{11}(\boldsymbol{q},i\omega_{m},\chi)(q^{2})^{2}\equiv q^{2}B_{b}(\boldsymbol{q},i\omega_{m},\chi),
\end{align}
where we have summed over the repeated indices and applied the relations
$\delta^{\tau\rho}\delta^{\tau\rho}=3$ and $q^{\tau}q^{\tau}=\boldsymbol{q}^{2}=q^{2}$.
Solving $B_{00}(\boldsymbol{q},i\omega_{m},\chi)$ and $B_{11}(\boldsymbol{q},i\omega_{m},\chi)$
in the last two equations in terms of $B_{a}(\boldsymbol{q},i\omega_{m},\chi)$
and $B_{b}(\boldsymbol{q},i\omega_{m},\chi)$ leads to
\begin{align}
B_{00}(\boldsymbol{q},i\omega_{m},\chi) & \equiv\frac{B_{a}(\boldsymbol{q},i\omega_{m},\chi)-B_{b}(\boldsymbol{q},i\omega_{m},\chi)}{2},\\
B_{11}(\boldsymbol{q},i\omega_{m},\chi) & \equiv\frac{3B_{b}(\boldsymbol{q},i\omega_{m},\chi)-B_{a}(\boldsymbol{q},i\omega_{m},\chi)}{2q^{2}}.
\end{align}
The tensor integrals $B^{\tau\rho}(\boldsymbol{q},i\omega_{m},\chi)$,
$B^{\tau}(\boldsymbol{q},i\omega_{m},\chi)$, $C^{\tau}(\boldsymbol{q},i\omega_{m},\chi)$,
and $D^{\tau}(\boldsymbol{q},i\omega_{m},\chi)$ can be expressed
in terms of the following five scalar integrals $B_{a}$, $B_{b}$,
$B_{1}$, $C_{1}$, and $D_{1}$
\begin{align}
B_{a}(\boldsymbol{q},i\omega_{m},\chi) & \equiv\frac{1}{\mathcal{V}}\sum_{\boldsymbol{k}}\frac{1}{\beta_{T}}\sum_{i\Omega_{n}}\sum_{\lambda,\lambda^{\prime}=\pm}\lambda\lambda^{\prime}\frac{|\boldsymbol{k}|}{|\boldsymbol{k}+\boldsymbol{q}|}\frac{1}{i\Omega_{n}+\mu_{\chi}-\varepsilon_{\lambda}(\boldsymbol{k})}\frac{1}{i\Omega_{n}+i\omega_{m}+\mu_{\chi}-\varepsilon_{\lambda^{\prime}}(\boldsymbol{k}+\boldsymbol{q})},\\
q^{2}B_{b}(\boldsymbol{q},i\omega_{m},\chi) & \equiv\frac{1}{\mathcal{V}}\sum_{\boldsymbol{k}}\frac{1}{\beta_{T}}\sum_{i\Omega_{n}}\sum_{\lambda,\lambda^{\prime}=\pm}\lambda\lambda^{\prime}\frac{(\boldsymbol{k}\cdot\boldsymbol{q})^{2}}{|\boldsymbol{k}||\boldsymbol{k}+\boldsymbol{q}|}\frac{1}{i\Omega_{n}+\mu_{\chi}-\varepsilon_{\lambda}(\boldsymbol{k})}\frac{1}{i\Omega_{n}+i\omega_{m}+\mu_{\chi}-\varepsilon_{\lambda^{\prime}}(\boldsymbol{k}+\boldsymbol{q})},\\
q^{2}B_{1}(\boldsymbol{q},i\omega_{m},\chi) & \equiv\frac{1}{\mathcal{V}}\sum_{\boldsymbol{k}}\frac{1}{\beta_{T}}\sum_{i\Omega_{n}}\sum_{\lambda,\lambda^{\prime}=\pm}\lambda\lambda^{\prime}\frac{\boldsymbol{k}\cdot\boldsymbol{q}}{|\boldsymbol{k}||\boldsymbol{k}+\boldsymbol{q}|}\frac{1}{i\Omega_{n}+\mu_{\chi}-\varepsilon_{\lambda}(\boldsymbol{k})}\frac{1}{i\Omega_{n}+i\omega_{m}+\mu_{\chi}-\varepsilon_{\lambda^{\prime}}(\boldsymbol{k}+\boldsymbol{q})},\\
q^{2}C_{1}(\boldsymbol{q},i\omega_{m},\chi) & \equiv\frac{1}{\mathcal{V}}\sum_{\boldsymbol{k}}\frac{1}{\beta_{T}}\sum_{i\Omega_{n}}\sum_{\lambda,\lambda^{\prime}=\pm}\lambda^{\prime}\frac{k_{\tau}}{|\boldsymbol{k}+\boldsymbol{q}|}\frac{1}{i\Omega_{n}+\mu_{\chi}-\varepsilon_{\lambda}(\boldsymbol{k})}\frac{1}{i\Omega_{n}+i\omega_{m}+\mu_{\chi}-\varepsilon_{\lambda^{\prime}}(\boldsymbol{k}+\boldsymbol{q})},\\
q^{2}D_{1}(\boldsymbol{q},i\omega_{m},\chi) & \equiv\frac{1}{\mathcal{V}}\sum_{\boldsymbol{k}}\frac{1}{\beta_{T}}\sum_{i\Omega_{n}}\sum_{\lambda,\lambda^{\prime}=\pm}\lambda\frac{k_{\tau}}{|\boldsymbol{k}|}\frac{1}{i\Omega_{n}+\mu_{\chi}-\varepsilon_{\lambda}(\boldsymbol{k})}\frac{1}{i\Omega_{n}+i\omega_{m}+\mu_{\chi}-\varepsilon_{\lambda^{\prime}}(\boldsymbol{k}+\boldsymbol{q})}.
\end{align}
Summing over Matsubara frequency $\Omega_{n}$, and performing analytical
continuum $i\omega_{m}\to\omega+i\eta$, one can obtain the following
seven scalar functions
\begin{align}
A_{0}(\boldsymbol{q},\omega,\chi) & =\frac{1}{\mathcal{V}}\sum_{\boldsymbol{k}}\sum_{\lambda,\lambda^{\prime}=\pm}\mathcal{F}_{\lambda\lambda^{\prime}}\left(\boldsymbol{k},\boldsymbol{k}+\boldsymbol{q},\omega\right),\\
q^{2}B_{1}(\boldsymbol{q},\omega,\chi) & =\frac{1}{\mathcal{V}}\sum_{\boldsymbol{k}}\frac{(\boldsymbol{k}+\boldsymbol{q})^{2}-\boldsymbol{k}^{2}-q^{2}}{2|\boldsymbol{k}||\boldsymbol{k}+\boldsymbol{q}|}\sum_{\lambda,\lambda^{\prime}=\pm}\lambda\lambda^{\prime}\mathcal{F}_{\lambda\lambda^{\prime}}\left(\boldsymbol{k},\boldsymbol{k}+\boldsymbol{q},\omega\right),\\
B_{a}(\boldsymbol{q},\omega,\chi) & =\frac{1}{\mathcal{V}}\sum_{\boldsymbol{k}}\frac{|\boldsymbol{k}|}{|\boldsymbol{k}+\boldsymbol{q}|}\sum_{\lambda,\lambda^{\prime}=\pm}\lambda\lambda^{\prime}\mathcal{F}_{\lambda\lambda^{\prime}}\left(\boldsymbol{k},\boldsymbol{k}+\boldsymbol{q},\omega\right),\\
q^{2}B_{b}(\boldsymbol{q},\omega,\chi) & =\frac{1}{\mathcal{V}}\sum_{\boldsymbol{k}}\frac{\left((\boldsymbol{k}+\boldsymbol{q})^{2}-\boldsymbol{k}^{2}-q^{2}\right)^{2}}{4|\boldsymbol{k}||\boldsymbol{k}+\boldsymbol{q}|}\sum_{\lambda,\lambda^{\prime}=\pm}\lambda\lambda^{\prime}\mathcal{F}_{\lambda\lambda^{\prime}}\left(\boldsymbol{k},\boldsymbol{k}+\boldsymbol{q},\omega\right),\\
C_{0}(\boldsymbol{q},\omega,\chi) & =\frac{1}{\mathcal{V}}\sum_{\boldsymbol{k}}\frac{1}{|\boldsymbol{k}+\boldsymbol{q}|}\sum_{\lambda,\lambda^{\prime}=\pm}\lambda^{\prime}\mathcal{F}_{\lambda\lambda^{\prime}}\left(\boldsymbol{k},\boldsymbol{k}+\boldsymbol{q},\omega\right),\\
q^{2}C_{1}(\boldsymbol{q},\omega,\chi) & =\frac{1}{\mathcal{V}}\sum_{\boldsymbol{k}}\frac{(\boldsymbol{k}+\boldsymbol{q})^{2}-\boldsymbol{k}^{2}-q^{2}}{2|\boldsymbol{k}||\boldsymbol{k}+\boldsymbol{q}|}\sum_{\lambda,\lambda^{\prime}=\pm}\lambda^{\prime}\mathcal{F}_{\lambda\lambda^{\prime}}\left(\boldsymbol{k},\boldsymbol{k}+\boldsymbol{q},\omega\right),\\
q^{2}D_{1}(\boldsymbol{q},\omega,\chi) & =\frac{1}{\mathcal{V}}\sum_{\boldsymbol{k}}\frac{(\boldsymbol{k}+\boldsymbol{q})^{2}-\boldsymbol{k}^{2}-q^{2}}{2|\boldsymbol{k}||\boldsymbol{k}+\boldsymbol{q}|}\sum_{\lambda,\lambda^{\prime}=\pm}\lambda\mathcal{F}_{\lambda\lambda^{\prime}}\left(\boldsymbol{k},\boldsymbol{k}+\boldsymbol{q},\omega\right),
\end{align}
where
\begin{equation}
\mathcal{F}_{\lambda\lambda^{\prime}}\left(\boldsymbol{k},\boldsymbol{k}+\boldsymbol{q},\omega\right)=\frac{n_{F}[\varepsilon_{\lambda}(\boldsymbol{k})]-n_{F}\big[\varepsilon_{\lambda^{\prime}}\left(\boldsymbol{k}+\boldsymbol{q}\right)]}{\omega+\varepsilon_{\lambda}(\boldsymbol{k})-\varepsilon_{\lambda^{\prime}}\left(\boldsymbol{k}+\boldsymbol{q}\right)+i\eta},
\end{equation}
with $n_{F}(x)=1/\left[\exp\left\{ \beta_{T}\left(x-\mu_{\chi}\right)\right\} +1\right]$
being the Fermi distribution function and $\varepsilon_{\lambda}(\boldsymbol{k})=\lambda\left|\boldsymbol{k}\right|$.

\section{Expressions of seven scalar integrals \label{AppESInt}}

In this appendix, we list the final expressions of the seven scalar integrals at zero temperature where the Fermi distribution function $n_{F}[\varepsilon_{\lambda}(\boldsymbol{k})]$ reduce to be the Heaviside step function $\Theta(\mu_{\chi}-\lambda|\boldsymbol{k}|)$. In order to express the analytical result in a compact form, we introduce the following auxiliary functions
\begin{align}
T(u)&=\frac{1}{3}u^{3}-\omega^{2}u,\\
U(u)&=\omega u^{2}-2q^{2}u,\\
V(u)&=\frac{1}{3}u^{3}-\omega u^{2}+\omega^{2}u,\\
W(u)&=\frac{1}{3}\omega^{2}u^{3}-q^{2}\omega u^{2}+q^{4}u,\\
X(u)&=u^{2}-2\omega u,\\
Y(u)&=\frac{1}{3}\omega u^{3}-\frac{(q^{2}+\omega^{2})}{2}u^{2}+q^{2}\omega u,\\
Z(u)&=\frac{1}{3}\omega u^{3}-\frac{(q^{2}-\omega^{2})}{2}u^{2}-q^{2}\omega u,\\
H(q,\omega)&=\log\left|\frac{2\mu_{\chi}+\omega-q}{\omega-q}\right|.
\end{align}

After lengthy and complicated calculations, one finds the explicit
expressions of the seven scalar integrals $A_{0}^{\pm},B_{1}^{\pm},B_{a}^{\pm},B_{b}^{\pm},C_{0}^{\pm},C_{1}^{\pm}$,
and $D_{1}^{\pm}$ as follows.

\subsection{Expression of $A_{0}(\boldsymbol{q},\omega,\chi)$}

The scalar integral $A_{0}(\boldsymbol{q},\omega,\chi)$ can de decomposed
into the intrinsic and extrinsic part as follows
\begin{align}
A_{0} & =\frac{1}{\mathcal{V}}\sum_{\boldsymbol{k}}\sum_{\lambda=\pm}\sum_{\lambda^{\prime}=\pm}\frac{n_{F}[\varepsilon_{\lambda}(\boldsymbol{k})]-n_{F}\big[\varepsilon_{\lambda^{\prime}}(\boldsymbol{k}+\boldsymbol{q})]}{\omega+\varepsilon_{\lambda}(\boldsymbol{k})-\varepsilon_{\lambda^{\prime}}(\boldsymbol{k}+\boldsymbol{q})+i\eta}=A_{0}^{-}+A_{0}^{+},
\end{align}
with the intrinsic part
\begin{align}
\mathrm{Im}A_{0}^{-} & =\frac{(q^{2}-3\omega^{2})}{48\pi}\Theta(\omega-q),\label{ImA0I}\\
\mathrm{Re}A_{0}^{-} & =\frac{(q^{2}-6\Lambda^{2})}{24\pi^{2}}
+\frac{(q^{2}-3\omega^{2})}{48\pi^{2}}\log\left|\frac{4\Lambda^{2}}{q^{2}-\omega^{2}}\right|,\label{ReA0I}
\end{align}
and the extrinsic part
\begin{align}
\mathrm{Im}A_{0}^{+} & =\frac{1}{32\pi q}\left\{ -\Theta(q-\omega)\sum_{\lambda=\pm}\lambda\Theta\left(\mu_{\chi}-\frac{q-\lambda\omega}{2}\right)
\left[T(2\mu_{\chi}+\lambda\omega)-T(q)\right]\right.-\Theta(\omega-q)\nonumber \\
 & \left.\times\left[\Theta\left(\mu_{\chi}-\frac{\omega+q}{2}\right)\left[T(q)-T(-q)\right]
 +\tilde{\Theta}\left(\frac{\omega+q}{2}-\mu_{\chi}\right)\Theta\left(\mu_{\chi}-\frac{\omega-q}{2}\right)\left[T(2\mu_{\chi}-\omega)-T(-q)\right]\right]\right\} ,\label{ImA0E}\\
\mathrm{Re}A_{0}^{+} & =\frac{1}{32\pi^{2}q}\left\{ \sum_{\lambda=\pm}\sum_{\lambda^{\prime}=\pm}\lambda\left[T(2\mu_{\chi}+\lambda^{\prime}\omega)-T(\lambda q)\right]H(\lambda q,\lambda^{\prime}\omega)-\frac{8q\mu_{\chi}^{2}}{3}\right\} .\label{ReA0E}
\end{align}

\subsection{Expression of $B_{1}(\boldsymbol{q},\omega,\chi)$}

The scalar integral $B_{1}(\boldsymbol{q},\omega,\chi)$ is given as
\begin{align}
q^{2}B_{1} & =\frac{1}{\mathcal{V}}\sum_{\boldsymbol{k}}\sum_{\lambda=\pm}\sum_{\lambda^{\prime}=\pm}\lambda\lambda^{\prime}\frac{(\boldsymbol{k}+\boldsymbol{q})^{2}-\boldsymbol{k}^{2}-q^{2}}{2|\boldsymbol{k}||\boldsymbol{k}+\boldsymbol{q}|}\frac{n_{F}[\varepsilon_{\lambda}(\boldsymbol{k})]-n_{F}\big[\varepsilon_{\lambda^{\prime}}(\boldsymbol{k}+\boldsymbol{q})]}{\omega+\varepsilon_{\lambda}(\boldsymbol{k})-\varepsilon_{\lambda^{\prime}}(\boldsymbol{k}+\boldsymbol{q})+i\eta}=q^{2}B_{1}^{-}+q^{2}B_{1}^{+},
\end{align}
with the intrinsic part
\begin{align}
q^{2}\mathrm{Im}B_{1}^{-} & =-\frac{q^{2}}{8\pi}\Theta(\omega-q),\\
q^{2}\mathrm{Re}B_{1}^{-} & =\frac{q^{2}}{12\pi^{2}}-\frac{q^{2}}{8\pi^{2}}\log\left|\frac{4\Lambda^{2}}{q^{2}-\omega^{2}}\right|,
\end{align}
and the extrinsic part
\begin{align}
q^{2}\mathrm{Im}B_{1}^{+} & =\frac{1}{32\pi q}\left\{ -\Theta(q-\omega)\sum_{\lambda=\pm}\lambda\Theta\left(\mu_{\chi}-\frac{q-\lambda\omega}{2}\right)\left[U(2\mu_{\chi}+\lambda\omega)-U(q)\right]\right.-\Theta(\omega-q)\nonumber \\
 & \left.\times\left[\Theta\left(\mu_{\chi}-\frac{\omega+q}{2}\right)\left[U(q)-U(-q)\right]+\tilde{\Theta}\left(\frac{\omega+q}{2}-\mu_{\chi}\right)\Theta\left(\mu_{\chi}-\frac{\omega-q}{2}\right)\left[U(2\mu_{\chi}-\omega)-U(-q)\right]\right]\right\} ,\\
q^{2}\mathrm{Re}B_{1}^{+} & =\frac{1}{32\pi^{2}q}\left\{ \sum_{\lambda=\pm}\sum_{\lambda^{\prime}=\pm}\lambda\left[U(2\mu_{\chi}+\lambda^{\prime}\omega)-U(\lambda q)\right]H(\lambda q,\lambda^{\prime}\omega)-8q\text{\ensuremath{\omega}}\mu_{\chi}\right\} .
\end{align}

\subsection{Expression of $B_{a}(\boldsymbol{q},\omega,\chi)$}

The scalar integral $B_{a}(\boldsymbol{q},\omega,\chi)$ is given as
\begin{align}
B_{a} & =\frac{1}{\mathcal{V}}\sum_{\boldsymbol{k}}\sum_{\lambda=\pm}\sum_{\lambda^{\prime}=\pm}\lambda\lambda^{\prime}\frac{|\boldsymbol{k}|}{|\boldsymbol{k}+\boldsymbol{q}|}\frac{n_{F}[\varepsilon_{\lambda}(\boldsymbol{k})]-n_{F}\big[\varepsilon_{\lambda^{\prime}}(\boldsymbol{k}+\boldsymbol{q})]}{\omega+\varepsilon_{\lambda}(\boldsymbol{k})-\varepsilon_{\lambda^{\prime}}(\boldsymbol{k}+\boldsymbol{q})+i\eta}=q^{2}B_{a}^{-}+q^{2}B_{a}^{+},
\end{align}
with the intrinsic part
\begin{align}
\mathrm{Im}B_{a}^{-} & =\frac{\left(q^{2}+3\omega^{2}\right)}{48\pi}\Theta(\omega-q),\\
\mathrm{Re}B_{a}^{-} & =\frac{\left(2\Lambda^{2}-q^{2}\right)}{8\pi^{2}}
+\frac{\left(q^{2}+3\omega^{2}\right)}{48\pi^{2}}
\log\left|\frac{4\Lambda^{2}}{q^{2}-\omega^{2}}\right|,
\end{align}
and the extrinsic part
\begin{align}
\mathrm{Im}B_{a}^{+} & =\frac{1}{32\pi q}\left\{ -\Theta(q-\omega)\sum_{\lambda=\pm}\lambda\Theta\left(\mu_{\chi}-\frac{q-\lambda\omega}{2}\right)\left[V(2\mu_{\chi}+\lambda\omega)-V(q)\right]\right.-\Theta(\omega-q)\nonumber \\
 & \left.\times\left[\Theta\left(\mu_{\chi}-\frac{\omega+q}{2}\right)\left[V(q)-V(-q)\right]+\tilde{\Theta}\left(\frac{\omega+q}{2}-\mu_{\chi}\right)\Theta\left(\mu_{\chi}-\frac{\omega-q}{2}\right)\left[V(2\mu_{\chi}-\omega)-V(-q)\right]\right]\right\} ,\\
\mathrm{Re}B_{a}^{+} & =\frac{1}{32\pi^{2}q}\left\{ \sum_{\lambda=\pm}\sum_{\lambda^{\prime}=\pm}\lambda\left[V(2\mu_{\chi}+\lambda^{\prime}\omega)-V(\lambda q)\right]H(\lambda q,\lambda^{\prime}\omega)-\frac{8}{3}q\mu_{\chi}\left(7\mu_{\chi}-3\omega\right)\right\} .
\end{align}

\subsection{Expression of $B_{b}(\boldsymbol{q},\omega,\chi)$}

The scalar integral $B_{b}(\boldsymbol{q},\omega,\chi)$ is given as
\begin{align}
q^{2}B_{b} & =\frac{1}{\mathcal{V}}\sum_{\boldsymbol{k}}\sum_{\lambda=\pm}\sum_{\lambda^{\prime}=\pm}\lambda\lambda^{\prime}\frac{\left[(\boldsymbol{k}+\boldsymbol{q})^{2}-\boldsymbol{k}^{2}-q^{2}\right]^{2}}{4|\boldsymbol{k}||\boldsymbol{k}+\boldsymbol{q}|}\frac{n_{F}[\varepsilon_{\lambda}(\boldsymbol{k})]-n_{F}\big[\varepsilon_{\lambda^{\prime}}(\boldsymbol{k}+\boldsymbol{q})]}{\omega+\varepsilon_{\lambda}(\boldsymbol{k})-\varepsilon_{\lambda^{\prime}}(\boldsymbol{k}+\boldsymbol{q})+i\eta}=q^{2}B_{b}^{-}+q^{2}B_{b}^{+},
\end{align}
with the intrinsic part
\begin{align}
q^{2}\mathrm{Im}B_{b}^{-} & =\frac{q^{2}\left(3q^{2}+\omega^{2}\right)}{48\pi}
\Theta(\omega-q),\\
q^{2}\mathrm{Re}B_{b}^{-} & =\frac{q^{2}\left(10\Lambda^{2}-11q^{2}\right)}{120\pi^2}
+\frac{q^{2}\left(3q^{2}+\omega^{2}\right)}{48\pi^2}
\log\left|\frac{4\Lambda^{2}}{q^{2}-\omega^{2}}\right|,
\end{align}
and the extrinsic part
\begin{align}
q^{2}\mathrm{Im}B_{b}^{+} & =\frac{1}{32\pi q}\left\{ -\Theta(q-\omega)\sum_{\lambda=\pm}\lambda\Theta\left(\mu_{\chi}-\frac{q-\lambda\omega}{2}\right)\left[W(2\mu_{\chi}+\lambda\omega)-W(q)\right]\right.-\Theta(\omega-q)\nonumber \\
 & \left.\times\left[\Theta\left(\mu_{\chi}-\frac{\omega+q}{2}\right)\left[W(q)-W(-q)\right]+\tilde{\Theta}\left(\frac{\omega+q}{2}-\mu_{\chi}\right)\Theta\left(\mu_{\chi}-\frac{\omega-q}{2}\right)\left[W(2\mu_{\chi}-\omega)-W(-q)\right]\right]\right\} ,\\
q^{2}\mathrm{Re}B_{b}^{+} & =\frac{1}{32\pi^{2}q}\left\{ \sum_{\lambda=\pm}\sum_{\lambda^{\prime}=\pm}\lambda\left[W(2\mu_{\chi}+\lambda^{\prime}\omega)W(\lambda q)\right]H(\lambda q,\lambda^{\prime}\omega)-\frac{32}{3}q\omega^{2}\mu_{\chi}^{2}-8q^{3}\mu_{\chi}(\mu_{\chi}-\omega)\right\} .
\end{align}

\subsection{Expression of $C_{0}(\boldsymbol{q},\omega,\chi)$}

The scalar integral $C_{0}(\boldsymbol{q},\omega,\chi)$ is given as
\begin{align}
C_{0} & =\frac{1}{\mathcal{V}}\sum_{\boldsymbol{k}}\sum_{\lambda=\pm}\sum_{\lambda^{\prime}=\pm}\lambda^{\prime}\frac{1}{|\boldsymbol{k}+\boldsymbol{q}|}\frac{n_{F}[\varepsilon_{\lambda}(\boldsymbol{k})]-n_{F}\big[\varepsilon_{\lambda^{\prime}}(\boldsymbol{k}+\boldsymbol{q})]}{\omega+\varepsilon_{\lambda}(\boldsymbol{k})-\varepsilon_{\lambda^{\prime}}(\boldsymbol{k}+\boldsymbol{q})+i\eta}=C_{0}^{-}+C_{0}^{+},
\end{align}
with the intrinsic part
\begin{align}
\mathrm{Im}C_{0}^{-} & =-\frac{\omega}{8\pi}\Theta(\omega-q),\\
\mathrm{Re}C_{0}^{-} & =-\frac{\omega}{8\pi^{2}}\log\left|\frac{4\Lambda^{2}}{q^{2}-\omega^{2}}\right|,
\end{align}
and the extrinsic part
\begin{align}
\mathrm{Im}C_{0}^{+} & =\frac{1}{32\pi q}\left\{ -\Theta(q-\omega)\sum_{\lambda=\pm}\lambda\Theta\left(\mu_{\chi}-\frac{q-\lambda\omega}{2}\right)\left[X(2\mu_{\chi}+\lambda\omega)-X(q)\right]\right.-\Theta(\omega-q)\nonumber \\
 & \left.\times\left[\Theta\left(\mu_{\chi}-\frac{\omega+q}{2}\right)\left[X(q)-X(-q)\right]+\tilde{\Theta}\left(\frac{\omega+q}{2}-\mu_{\chi}\right)\Theta\left(\mu_{\chi}-\frac{\omega-q}{2}\right)\left[X(2\mu_{\chi}-\omega)-X(-q)\right]\right]\right\} ,\\
\mathrm{Re}C_{0}^{+} & =\frac{1}{32\pi^{2}q}\left\{ \sum_{\lambda=\pm}\sum_{\lambda^{\prime}=\pm}\lambda\left[X(2\mu_{\chi}+\lambda^{\prime}\omega)-X(\lambda q)\right]H(\lambda q,\lambda^{\prime}\omega)-8q\mu_{\chi}\right\} .
\end{align}

\subsection{Expression of $C_{1}(\boldsymbol{q},\omega,\chi)$}

The scalar integral $C_{1}(\boldsymbol{q},\omega,\chi)$ is given as
\begin{align}
q^{2}C_{1} & =\frac{1}{\mathcal{V}}\sum_{\boldsymbol{k}}\sum_{\lambda=\pm}\sum_{\lambda^{\prime}=\pm}\lambda^{\prime}\frac{(\boldsymbol{k}+\boldsymbol{q})^{2}-\boldsymbol{k}^{2}-q^{2}}{2|\boldsymbol{k}+\boldsymbol{q}|}\frac{n_{F}[\varepsilon_{\lambda}(\boldsymbol{k})]-n_{F}\big[\varepsilon_{\lambda^{\prime}}(\boldsymbol{k}+\boldsymbol{q})]}{\omega+\varepsilon_{\lambda}(\boldsymbol{k})-\varepsilon_{\lambda^{\prime}}(\boldsymbol{k}+\boldsymbol{q})+i\eta}=q^{2}C_{1}^{-}+q^{2}C_{1}^{+},
\end{align}
with the intrinsic part
\begin{align}
q^{2}\mathrm{Im}C_{1}^{-} & =\frac{q^{2}\omega}{12\pi}\Theta(\omega-q),\\
q^{2}\mathrm{Re}C_{1}^{-} & =-\frac{q^{2}\omega}{24\pi^2}+
\frac{q^{2}\omega}{12\pi^2}\log\left|\frac{4\Lambda^{2}}{q^{2}-\omega^{2}}\right|,
\end{align}
and the extrinsic part
\begin{align}
q^{2}\mathrm{Im}C_{1}^{+} & =\frac{1}{32\pi q}\left\{ -\Theta(q-\omega)\sum_{\lambda=\pm}\lambda\Theta\left(\mu_{\chi}-\frac{q-\lambda\omega}{2}\right)\left[Y(2\mu_{\chi}+\lambda\omega)-Y(q)\right]\right.-\Theta(\omega-q)\nonumber \\
 & \left.\times\left[\Theta\left(\mu_{\chi}-\frac{\omega+q}{2}\right)\left[Y(q)-Y(-q)\right]+\tilde{\Theta}\left(\frac{\omega+q}{2}-\mu_{\chi}\right)\Theta\left(\mu_{\chi}-\frac{\omega-q}{2}\right)\left[Y(2\mu_{\chi}-\omega)-Y(-q)\right]\right]\right\} ,\\
q^{2}\mathrm{Re}C_{1}^{+} & =\frac{1}{32\pi^{2}q}\left\{ \sum_{\lambda=\pm}\sum_{\lambda^{\prime}=\pm}\lambda\left[Y(2\mu_{\chi}+\lambda^{\prime}\omega)-Y(\lambda q)\right]H(\lambda q,\lambda^{\prime}\omega)-\frac{32}{3}q\omega\mu_{\chi}^{2}+4q(q^{2}+\omega^{2})\mu_{\chi}\right\} .
\end{align}

\subsection{Expression of $D_{1}(\boldsymbol{q},\omega,\chi)$}

The scalar integral $D_{1}(\boldsymbol{q},\omega,\chi)$ is given as
\begin{align}
q^{2}D_{1} & =\frac{1}{\mathcal{V}}\sum_{\boldsymbol{k}}\sum_{\lambda=\pm}\sum_{\lambda^{\prime}=\pm}\lambda\frac{(\boldsymbol{k}+\boldsymbol{q})^{2}-\boldsymbol{k}^{2}-q^{2}}{2|\boldsymbol{k}|}\frac{n_{F}[\varepsilon_{\lambda}(\boldsymbol{k})]-n_{F}\big[\varepsilon_{\lambda^{\prime}}(\boldsymbol{k}+\boldsymbol{q})]}{\omega+\varepsilon_{\lambda}(\boldsymbol{k})-\varepsilon_{\lambda^{\prime}}(\boldsymbol{k}+\boldsymbol{q})+i\eta}=q^{2}D_{1}^{-}+q^{2}D_{1}^{+},
\end{align}
with the intrinsic part
\begin{align}
q^{2}\mathrm{Im}D_{1}^{-} & =-\frac{q^{2}\omega}{24\pi}\Theta(\omega-q),\\
q^{2}\mathrm{Re}D_{1}^{-} & =\frac{q^{2}\omega}{24\pi^{2}}
-\frac{q^{2}\omega}{24\pi^{2}}\log\left|\frac{4\Lambda^{2}}{q^{2}-\omega^{2}}\right|,
\end{align}
and the extrinsic part
\begin{align}
q^{2}\mathrm{Im}D_{1}^{+} & =\frac{1}{32\pi q}\left\{ -\Theta(q-\omega)\sum_{\lambda=\pm}\lambda\Theta\left(\mu_{\chi}-\frac{q-\lambda\omega}{2}\right)\left[Z(2\mu_{\chi}+\lambda\omega)-Z(q)\right]\right.-\Theta(\omega-q)\nonumber \\
 & \left.\times\left[\Theta\left(\mu_{\chi}-\frac{\omega+q}{2}\right)\left[Z(q)-Z(-q)\right]+\tilde{\Theta}\left(\frac{\omega+q}{2}-\mu_{\chi}\right)\Theta\left(\mu_{\chi}-\frac{\omega-q}{2}\right)\left[Z(2\mu_{\chi}-\omega)-Z(-q)\right]\right]\right\} ,\\
q^{2}\mathrm{Re}D_{1}^{+} & =\frac{1}{32\pi^{2}q}\left\{ \sum_{\lambda=\pm}\sum_{\lambda^{\prime}=\pm}\lambda\left[Z(2\mu_{\chi}+\lambda^{\prime}\omega)-Z(\lambda q)\right]H(\lambda q,\lambda^{\prime}\omega)-\frac{32}{3}q\omega\mu_{\chi}^{2}+4q(q^{2}-\omega^{2})\mu_{\chi}\right\} .
\end{align}

\section{Evaluation of $A_{0}(\boldsymbol{q},\omega,\chi)$ \label{AppEvaA0}}

In this appendix, we take $A_{0}(\boldsymbol{q},\omega,\chi)$ as an
example to outline the main procedure and tricks to evaluate the seven
scalar integrals at zero temperature.

\subsection{Intrinsic case}

For the intrinsic case $\mu_{\chi}=0$, at zero temperature the Fermi distribution
function $n_{F}[x]=\Theta(\mu_{\chi}-x)$, leading to $A_{0}^{+}=0$
and $A_{0}^{-}\neq0$.
\begin{align}
A_{0}^{-} & =\frac{1}{\mathcal{V}}\sum_{\boldsymbol{k}}\left\{ \sum_{\lambda^{\prime}=\pm}\frac{n_{F}[\varepsilon_{-}(\boldsymbol{k})]}{\omega+\varepsilon_{-}(\boldsymbol{k})-\varepsilon_{\lambda^{\prime}}(\boldsymbol{k}+\boldsymbol{q})+i\eta}-\sum_{\lambda=\pm}\frac{n_{F}\big[\varepsilon_{-}(\boldsymbol{k}+\boldsymbol{q})]}{\omega+\varepsilon_{\lambda}(\boldsymbol{k})-\varepsilon_{-}(\boldsymbol{k}+\boldsymbol{q})+i\eta}\right\} =A_{0a}^{-}+A_{0b}^{-}.
\end{align}
It is noted that the two terms in the curly brackets cancel each other
line in $A_{0}^{-}$ vanishes. The left two terms can be evaluated as
\begin{align}
A_{0a}^{-} & =\mbox{\ensuremath{\mathrm{Re}}}A_{0a}^{-}+i\,\mathrm{Im}A_{0a}^{-}=\frac{1}{\mathcal{V}}\sum_{\boldsymbol{k}}\frac{n_{F}[\varepsilon_{-}(\boldsymbol{k})]}{\omega+\varepsilon_{-}(\boldsymbol{k})-\varepsilon_{+}(\boldsymbol{k}+\boldsymbol{q})+i\eta}=\frac{1}{\mathcal{V}}\sum_{\boldsymbol{k}}\frac{1}{\omega+\varepsilon_{-}(\boldsymbol{k})-\varepsilon_{+}(\boldsymbol{k}+\boldsymbol{q})+i\eta}\nonumber \\
 & =\frac{1}{4\pi^{2}q}\int_{0}^{\Lambda}kdk\int_{\left|k-q\right|}^{k+q}ydy\frac{1}{\omega-k-y+i\eta}=\frac{1}{4\pi^{2}q}\int_{0}^{\Lambda}kdk\int_{\left|k-q\right|}^{k+q}ydy\left\{ \mathcal{P}\frac{1}{\omega-k-y}-i\pi\delta(\omega-k-y)\right\} ,
\end{align}
similarly
\begin{align}
A_{0b}^{-} & =\mbox{\ensuremath{\mathrm{Re}}}A_{0b}^{-}+i\,\mathrm{Im}A_{0b}^{-}=\frac{1}{\mathcal{V}}\sum_{\boldsymbol{k}}\frac{-n_{F}\big[\varepsilon_{-}(\boldsymbol{k}+\boldsymbol{q})]}{\omega+\varepsilon_{+}(\boldsymbol{k})-\varepsilon_{-}(\boldsymbol{k}+\boldsymbol{q})+i\eta}=\frac{1}{\mathcal{V}}\sum_{\boldsymbol{k}}\frac{-1}{\omega+\varepsilon_{+}(\boldsymbol{k})-\varepsilon_{-}(\boldsymbol{k}+\boldsymbol{q})+i\eta}\nonumber \\
 & =\frac{-1}{4\pi^{2}q}\int_{0}^{\Lambda}kdk\int_{\left|k-q\right|}^{k+q}ydy\frac{1}{\omega+k+y+i\eta}=\frac{-1}{4\pi^{2}q}\int_{0}^{\Lambda}kdk\int_{\left|k-q\right|}^{k+q}ydy\left\{ \mathcal{P}\frac{1}{\omega+k+y}-i\pi\delta(\omega+k+y)\right\} .
\end{align}
Integrating over $y$ and $k$ leads to $A_{0}^{-}$ for the intrinsic
case in Eqs. $\left(\mathrm{\ref{ImA0I}}\right)$ and $\left(\mathrm{\ref{ReA0I}}\right)$.

\subsection{Extrinsic case}

Let us calculate the extrinsic case $\mu_{\chi}>0$ as follows
\begin{align}
A_{0}^{+} & =\frac{1}{\mathcal{V}}\sum_{\boldsymbol{k}}\left\{ \sum_{\lambda^{\prime}=\pm}\frac{n_{F}[\varepsilon_{+}(\boldsymbol{k})]}{\omega+\varepsilon_{+}(\boldsymbol{k})-\varepsilon_{\lambda^{\prime}}(\boldsymbol{k}+\boldsymbol{q})+i\eta}-\sum_{\lambda=\pm}\frac{n_{F}\big[\varepsilon_{+}(\boldsymbol{k}+\boldsymbol{q})]}{\omega+\varepsilon_{\lambda}(\boldsymbol{k})-\varepsilon_{+}(\boldsymbol{k}+\boldsymbol{q})+i\eta}\right\} =A_{0a}^{+}+A_{0b}^{+},
\end{align}
where $A_{0a}^{+}$ and $A_{0b}^{+}$ are given as
\begin{align}
A_{0a}^{+} & =\frac{1}{\mathcal{V}}\sum_{\boldsymbol{k}}n_{F}[\varepsilon_{+}(\boldsymbol{k})]\left\{ \frac{1}{\omega+\varepsilon_{+}(\boldsymbol{k})-\varepsilon_{-}(\boldsymbol{k}+\boldsymbol{q})+i\eta}+\frac{1}{\omega+\varepsilon_{+}(\boldsymbol{k})-\varepsilon_{+}(\boldsymbol{k}+\boldsymbol{q})+i\eta}\right\} \nonumber \\
 & =\frac{1}{4\pi^{2}q}\int_{0}^{\Lambda}kdk\int_{\left|k-q\right|}^{k+q}ydy\Theta(\mu_{\chi}-k)\left\{ \frac{1}{\omega+k+y+i\eta}+\frac{1}{\omega+k-y+i\eta}\right\} \nonumber \\
 & =\frac{1}{4\pi^{2}q}\int_{0}^{\mu_{\chi}}kdk\int_{\left|k-q\right|}^{k+q}ydy\left\{ \mathcal{P}\frac{2(\omega+k)}{(\omega+k)^{2}-y^{2}}-2i\pi\left|\omega+k\right|\delta\left[(\omega+k)^{2}-y^{2}\right]\right\} ,
\end{align}
and
\begin{align}
A_{0b}^{+} & =\frac{1}{\mathcal{V}}\sum_{\boldsymbol{k}}n_{F}\big[\varepsilon_{+}(\boldsymbol{k})]\left\{ \frac{1}{\omega-\varepsilon_{+}(\boldsymbol{k})+\varepsilon_{-}(\boldsymbol{k}+\boldsymbol{q})+i\eta}+\frac{1}{\omega-\varepsilon_{+}(\boldsymbol{k})+\varepsilon_{+}(\boldsymbol{k}+\boldsymbol{q})+i\eta}\right\} \nonumber \\
 & =\frac{-1}{4\pi^{2}q}\int_{0}^{\Lambda}kdk\int_{\left|k-q\right|}^{k+q}ydy\Theta(\mu_{\chi}-k)\left\{ \frac{1}{\omega-k-y+i\eta}+\frac{1}{\omega-k+y+i\eta}\right\} \nonumber \\
 & =\frac{-1}{4\pi^{2}q}\int_{0}^{\mu_{\chi}}kdk\int_{\left|k-q\right|}^{k+q}ydy\left\{ \mathcal{P}\left[\frac{2(\omega-k)}{(\omega-k)^{2}-y^{2}}\right]-2i\pi\left|\omega-k\right|\delta\left[(\omega-k)^{2}-y^{2}\right]\right\} .
\end{align}
After integrating over $y$ and $k$, we obtain $A_{0}^{+}$ for the
extrinsic case in Eqs. $\left(\mathrm{\ref{ImA0E}}\right)$ and $\left(\mathrm{\ref{ReA0E}}\right)$.

\section{Causality relations \label{AppCR}}

Let us rewrite the chirality-dependent correlation functions $\Pi^{00}(\boldsymbol{q},\omega,\chi)$,
$\Pi^{\alpha\beta}(\boldsymbol{q},\omega,\chi)$, and $\Pi^{0\alpha/\alpha0}(\boldsymbol{q},\omega,\chi)$ as
\begin{align}
\Pi^{00}(\boldsymbol{q},\omega,\chi) & \equiv\frac{1}{2}\left[A_{0}(\boldsymbol{q},\omega,\chi)+B_{a}(\boldsymbol{q},\omega,\chi)+q^{2}B_{1}(\boldsymbol{q},\omega,\chi)\right],\\
\Pi^{\alpha\beta}(\boldsymbol{q},\omega,\chi) & \equiv F_{T}(\boldsymbol{q},\omega,\chi)\left(\delta^{\alpha\beta}-\frac{q^{\alpha}q^{\beta}}{q^{2}}\right)+F{}_{L}(\boldsymbol{q},\omega,\chi)\frac{q^{\alpha}q^{\beta}}{q^{2}}+i\chi F_{A}(\boldsymbol{q},\omega,\chi)\varepsilon^{\alpha\beta\tau}q^{\tau},\\
\Pi^{\alpha0}(\boldsymbol{q},\omega,\chi) & =\Pi^{0\alpha}(\boldsymbol{q},\omega,\chi)\equiv F_{I}(\boldsymbol{q},\omega,\chi)q^{\alpha},
\end{align}
where the four auxiliary functions read
\begin{align}
F_{T}(\boldsymbol{q},\omega,\chi) & =\frac{1}{2}\left[A_{0}(\boldsymbol{q},\omega,\chi)-B_{b}(\boldsymbol{q},\omega,\chi)-q^{2}B_{1}(\boldsymbol{q},\omega,\chi)\right],\\
F_{L}(\boldsymbol{q},\omega,\chi) & =\frac{1}{2}\left[A_{0}(\boldsymbol{q},\omega,\chi)-B_{a}(\boldsymbol{q},\omega,\chi)+2B_{b}(\boldsymbol{q},\omega,\chi)+q^{2}B_{1}(\boldsymbol{q},\omega,\chi)\right],\\
F_{A}(\boldsymbol{q},\omega,\chi) & =-\frac{1}{2}\left[C_{1}(\boldsymbol{q},\omega,\chi)+C_{0}(\boldsymbol{q},\omega,\chi)-D_{1}(\boldsymbol{q},\omega,\chi)\right],\\
F_{I}(\boldsymbol{q},\omega,\chi) & =\frac{1}{2}\left[C_{1}(\boldsymbol{q},\omega,\chi)+C_{0}(\boldsymbol{q},\omega,\chi)+D_{1}(\boldsymbol{q},\omega,\chi)\right].
\end{align}
 Next we take $\Pi^{0\alpha/\alpha0}(\boldsymbol{q},\omega,\chi)$
as an example to derive the causality relation of each chirality-dependent
correlation function. Let us first take the Hermitian conjugate of
$\Pi^{0\alpha}(\boldsymbol{q},\omega,\chi)$ as
\begin{align}
\left[\Pi^{0\alpha}(\boldsymbol{q},\omega,\chi)\right]^{\dagger} & =\left[\Pi^{\alpha0}(\boldsymbol{q},\omega,\chi)\right]^{\ast}=\left[\Pi^{0\alpha}(\boldsymbol{q},\omega,\chi)\right]^{\ast}=\left[F_{I}(\boldsymbol{q},\omega,\chi)\right]^{\ast}q^{\alpha},
\end{align}
where the complex conjugate of $F_{I}(\boldsymbol{q},\omega,\chi)$ reads
\begin{align}
\left[F_{I}(\boldsymbol{q},\omega,\chi)\right]^{\ast} & =\frac{1}{2\mathcal{V}}\sum_{\boldsymbol{k}}\sum_{\lambda,\lambda^{\prime}=\pm}\left(\lambda^{\prime}\frac{(\boldsymbol{k}+\boldsymbol{q})^{2}-\boldsymbol{k}^{2}+q^{2}}{2q^{2}|\boldsymbol{k}+\boldsymbol{q}|}+\lambda\frac{(\boldsymbol{k}+\boldsymbol{q})^{2}-\boldsymbol{k}^{2}-q^{2}}{2q^{2}|\boldsymbol{k}|}\right)\frac{n_{F}[\varepsilon_{\lambda}(\boldsymbol{k})]-n_{F}\big[\varepsilon_{\lambda^{\prime}}(\boldsymbol{k}+\boldsymbol{q})]}{\omega+\varepsilon_{\lambda}(\boldsymbol{k})-\varepsilon_{\lambda^{\prime}}(\boldsymbol{k}+\boldsymbol{q})-i\eta}.
\end{align}
Relabeling $\left(\boldsymbol{k},\lambda\right)\leftrightarrow\left(-\boldsymbol{k}-\boldsymbol{q},\lambda^{\prime}\right)$ leads to
\begin{align}
\left[F_{I}(\boldsymbol{q},\omega,\chi)\right]^{\ast} & =-\frac{1}{2\mathcal{V}}\sum_{\boldsymbol{k}}\sum_{\lambda,\lambda^{\prime}=\pm}\left(\lambda^{\prime}\frac{(\boldsymbol{k}+\boldsymbol{q})^{2}-\boldsymbol{k}^{2}+q^{2}}{2q^{2}|\boldsymbol{k}+\boldsymbol{q}|}+\lambda\frac{(\boldsymbol{k}+\boldsymbol{q})^{2}-\boldsymbol{k}^{2}-q^{2}}{2q^{2}|\boldsymbol{k}|}\right)\frac{n_{F}[\varepsilon_{\lambda}(\boldsymbol{k})]-n_{F}\big[\varepsilon_{\lambda^{\prime}}(\boldsymbol{k}+\boldsymbol{q})]}{-\omega+\varepsilon_{\lambda}(\boldsymbol{k})-\varepsilon_{\lambda^{\prime}}(\boldsymbol{k}+\boldsymbol{q})+i\eta}\nonumber \\
 & =-F_{I}(\boldsymbol{q},-\omega,\chi),
\end{align}
which implies
\begin{align}
\left[\Pi^{0\alpha}(\boldsymbol{q},\omega,\chi)\right]^{\dagger} & =\left[\Pi^{\alpha0}(\boldsymbol{q},\omega,\chi)\right]^{\ast}=-F{}_{I}(\boldsymbol{q},-\omega,\chi)q^{\alpha}=-\Pi^{0\alpha}(\boldsymbol{q},-\omega,\chi).
\end{align}
Other correlation function can be calculated in a similar way. Finally, we have
\begin{align}
\mathrm{Re}\Pi^{00}(\boldsymbol{q},-\omega,\chi) & =\mathrm{Re}\Pi^{00}(\boldsymbol{q},\omega,\chi),\\
\mathrm{Im}\Pi^{00}(\boldsymbol{q},-\omega,\chi) & =-\mathrm{Im}\Pi^{00}(\boldsymbol{q},\omega,\chi),\\
\mathrm{Re}F{}_{X}(\boldsymbol{q},-\omega,\chi) & =\mathrm{Re}F{}_{X}(\boldsymbol{q},\omega,\chi),\\
\mathrm{Im}F{}_{X}(\boldsymbol{q},-\omega,\chi) & =-\mathrm{Im}F{}_{X}(\boldsymbol{q},\omega,\chi),\\
\mathrm{Re}F{}_{I}(\boldsymbol{q},-\omega,\chi) & =-\mathrm{Re}F{}_{I}(\boldsymbol{q},\omega,\chi),\\
\mathrm{Im}F{}_{I}(\boldsymbol{q},-\omega,\chi) & =\mathrm{Im}F{}_{I}(\boldsymbol{q},\omega,\chi),
\end{align}
which helps us to obtain the negative-frequency part of $\Pi^{\mu\nu}(\boldsymbol{q},\omega,\chi)$
by taking the Hermitian conjugate of that with
the positive frequency. Here $X=T,L,A$. Thus, we only need to consider
the positive frequency part $\omega>0$.

\section{Kramers-Kronig relation with $n$th-order subtraction \label{AppKKR}}

\begin{figure}
\includegraphics[scale=0.7]{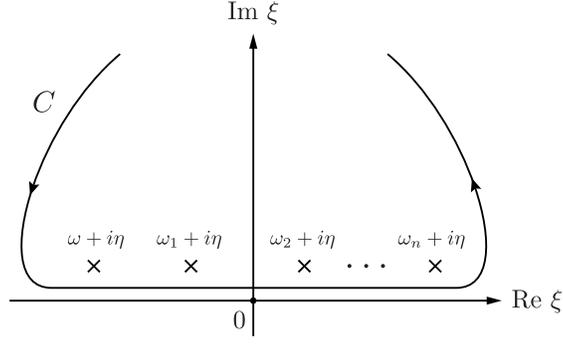}
\caption{Contour in the upper half $\xi$ plane for Cauchy integral with $n+1$
poles in Eq. $\left(\mathrm{\ref{Jn-1}}\right)$.}
\label{KKRfig}
\end{figure}
In this appendix, we give a proof of the dispersion relation with $n$th-order subtraction. For an analytic
function $f(\xi)$ in the upper half plane, if $f\left(\xi\right)$ does not diverge more than $\xi^{n-1}$
as $\xi\to\infty$, we construct a Cauchy integral
\begin{align}
\mathcal{J}_{n-1} & =\lim_{\eta\to0^{+}}\frac{1}{2\pi i}\oint_{C}d\xi\frac{f(\xi)}{[\xi-(\omega+i\eta)]}\prod_{m=1}^{n}\frac{1}{[\xi-(\omega_{m}+i\eta)]},\label{Jn-1}
\end{align}
which has $n+1$ poles at $\xi=\omega+i\eta$, $\xi=\omega_{m}+i\eta$
with $\omega\neq\omega_{m}\neq\omega_{l}$, $m,l=1,2,3,\cdots,n$
and $m\neq l$. The auxiliary function $\mathcal{J}_{n-1}$ can be
calculated in two ways. The first one is
\begin{align}
 & \mathcal{J}_{n-1}=\lim_{\eta\to0^{+}}\frac{1}{2\pi i}\oint_{C}d\xi\frac{f(\xi)}{(\xi-\omega-i\eta)}\prod_{m=1}^{n}\frac{1}{[\xi-(\omega_{m}+i\eta)]}\nonumber \\
 & =\lim_{\eta\to0^{+}}\left[\frac{(\xi-\omega-i\eta)f(\xi)}{(\xi-\omega-i\eta)}\prod_{m=1}^{n}\frac{1}{[\xi-(\omega_{m}+i\eta)]}\right]_{\xi=\omega+i\eta}+\sum_{l=1}^{n}\lim_{\eta\to0^{+}}\left[\frac{(\xi-\omega_{l}-i\eta)f(\xi)}{(\xi-\omega-i\eta)}\prod_{m=1}^{n}\frac{1}{[\xi-(\omega_{m}+i\eta)]}\right]_{\xi=\omega_{l}+i\eta}\nonumber \\
 & =f(\omega)\prod_{m=1}^{n}\frac{1}{(\omega-\omega_{m})}+\sum_{l=1}^{n}\frac{f(\omega_{l})}{(\omega_{l}-\omega)}\prod_{m=1,m\neq l}^{n}\frac{1}{(\omega_{l}-\omega_{m})}.
\end{align}
The second one is
\begin{align}
 & \mathcal{J}_{n-1}=\lim_{\eta\to0^{+}}\frac{1}{2\pi i}\oint_{C}d\xi\frac{f(\xi)}{[\xi-(\omega+i\eta)]}\prod_{m=1}^{n}\frac{1}{[\xi-(\omega_{m}+i\eta)]}=\lim_{\eta\to0^{+}}\frac{1}{2\pi i}\left(\int_{-\infty}^{+\infty}d\xi\frac{f(\xi)}{g(\xi)-i\eta g^{\prime}(\xi)}+i\pi \mathcal{C}_{\infty}\right)\nonumber \\
 & =\frac{1}{2\pi i}\left(\mathcal{P}\int_{-\infty}^{+\infty}d\xi\frac{f(\xi)}{g(\xi)}+i\pi\int_{-\infty}^{+\infty}d\xi f(\xi)\mathrm{Sgn}[g^{\prime}(\xi)]\delta\left[g(\xi)\right]+i\pi\mathcal{C}_{\infty}\right)\nonumber \\
 & =\frac{1}{2\pi i}\mathcal{P}\int_{-\infty}^{+\infty}d\xi\frac{f(\xi)}{g(\xi)}+\frac{1}{2}\int_{-\infty}^{+\infty}d\xi f(\xi)\mathrm{Sgn}[g^{\prime}(\xi)]\sum_{j=1}^{n+1}\frac{\delta(\xi-\xi_{0,j})}{\big|g^{\prime}(\xi)\big|_{\xi=\xi_{0,j}}}+\frac{1}{2}\mathcal{C}_{\infty}\nonumber \\
 & =\frac{1}{2\pi i}\mathcal{P}\int_{-\infty}^{+\infty}d\xi\frac{f(\xi)}{g(\xi)}+\frac{1}{2}\sum_{j=1}^{n+1}\int_{-\infty}^{+\infty}d\xi f(\xi)\frac{\delta(\xi-\xi_{0,j})}{\big[g^{\prime}(\xi)\big]_{\xi=\xi_{0,j}}}+\frac{1}{2}\mathcal{C}_{\infty}\nonumber \\
 & =\frac{1}{2\pi i}\mathcal{P}\int_{-\infty}^{+\infty}d\xi\frac{f(\xi)}{g(\xi)}+\frac{1}{2}f(\omega)\prod_{m=1}^{n}\frac{1}{(\omega-\omega_{m})}+\frac{1}{2}\sum_{l=1}^{n}\frac{f(\omega_{l})}{(\omega_{l}-\omega)}\prod_{m=1,m\neq l}^{n}\frac{1}{(\omega_{l}-\omega_{m})}+\frac{1}{2}\mathcal{C}_{\infty},
\end{align}
where $g(\xi)=(\xi-\omega)\prod_{m=1}^{n}(\xi-\omega_{m})$, $g^{\prime}(\xi)=\frac{dg(\xi)}{d\xi}$,
$\mathcal{P}$ denotes the principal value of the integral along the
real axis from $-\infty$ to $+\infty$, and $C$ denotes the contour
drawn in Fig. \ref{KKRfig}. The contribution from the infinite semicircle
is a complex quantity $\mathcal{C}_{\infty}=C_{\infty}+iC_{\infty}^{'}$.
Note that we have also utilized the Dirac identity $\int dx\frac{f(x)}{x-x_{0}+i\eta}=\mathcal{P}\int dx\frac{f(x)}{x-x_{0}}-i\pi\int dxf(x)\delta(x-x_{0})$
and $\mathrm{Sgn}(x)/|x|=1/x$.

From these two equations, we thus arrive at
\begin{align}
f(\omega)\prod_{m=1}^{n}\frac{1}{(\omega-\omega_{m})} & =\sum_{l=1}^{n}\frac{f(\omega_{l})}{(\omega-\omega_{l})}\prod_{m=1,m\neq l}^{n}\frac{1}{(\omega_{l}-\omega_{m})}+\frac{1}{\pi i}\mathcal{P}\int_{-\infty}^{+\infty}d\xi\frac{f(\xi)}{g(\xi)}+\mathcal{C}_{\infty},
\end{align}
whose real and imaginary parts are given as
\begin{align}
\mathrm{Re}[f(\omega)]\prod_{m=1}^{n}\frac{1}{(\omega-\omega_{m})} & =\sum_{l=1}^{n}\frac{\mathrm{Re}f(\omega_{l})}{(\omega-\omega_{l})}\prod_{m=1,m\neq l}^{n}\frac{1}{(\omega_{l}-\omega_{m})}+\frac{1}{\pi}\mathcal{P}\int_{-\infty}^{+\infty}d\xi\frac{\mathrm{Im}f(\xi)}{g(\xi)}+C_{\infty},\\
\mathrm{Im}[f(\omega)]\prod_{m=1}^{n}\frac{1}{(\omega-\omega_{m})} & =\sum_{l=1}^{n}\frac{\mathrm{Im}f(\omega_{l})}{(\omega-\omega_{l})}\prod_{m=1,m\neq l}^{n}\frac{1}{(\omega_{l}-\omega_{m})}-\frac{1}{\pi}\mathcal{P}\int_{-\infty}^{+\infty}d\xi\frac{\mathrm{Re}f(\xi)}{g(\xi)}+C_{\infty}^{'},
\end{align}
which are named the Kramers-Kronig relation with $n$th-order
subtraction. It is one of the main results in this paper.
If $f\left(\xi\right)$ does not diverge more than $\xi^{n-1}$ as $\xi\to\infty$,
which leads to $\mathcal{C}_{\infty}=C_{\infty}+iC_{\infty}^{'}=0$.

Several remarks are in order here. First, for $n=0,$ and $n=1$,
our result recovers the conventional Kramers-Kronig relation and the
one with 1st-order subtraction \cite{bjorken1965RQF},
respectively. Second, the quantities of $\omega_{1}$, $\omega_{2}$,
$\omega_{3}$, $\cdots$, and $\omega_{n}$ can be arbitrarily chosen
as if it is not equal to $q$. For the sake of simplicity, we choose $\omega_{j}=\alpha_{j}q$
with $\alpha_{i}\neq\alpha_{j}\neq1$. Third, $\mathcal{J}_{m}$ can
be used to calculate all the cases for $f(\xi)\sim\xi^{n}$ with $n\le m$,
$\cdots$. Finally, if a function $f(\xi)$ is $\sim\xi^{1}$, either
$\mathcal{J}_{2}$ or $\mathcal{J}_{1}$ is applicable, but the latter
is much convenient than the former, especially for the extrinsic parts.

\section{Relations between the correlation functions with opposite chemical potentials
\label{AppRCRpmmu}}

In this Appendix, we prove relations between chirality-dependent correlation
functions with opposite chemical potentials. Next we take the antisymmetric
part of $\Pi^{\alpha\beta}(\boldsymbol{q},\omega,\chi)$ as an example
to derive the relation of chirality-dependent correlation functions
with opposite chemical potentials. For the positive chemical potential
$+\mu_{\chi}$, we have
\begin{align}
F_{A}(\boldsymbol{q},\omega,\chi,\mu_{\chi}) & =\frac{1}{2\mathcal{V}}\sum_{\boldsymbol{k}}\sum_{\lambda,\lambda^{\prime}=\pm}\left(\lambda\frac{(\boldsymbol{k}+\boldsymbol{q})^{2}-\boldsymbol{k}^{2}-q^{2}}{2q^{2}|\boldsymbol{k}|}-\lambda^{\prime}\frac{(\boldsymbol{k}+\boldsymbol{q})^{2}-\boldsymbol{k}^{2}+q^{2}}{2q^{2}|\boldsymbol{k}+\boldsymbol{q}|}\right)\frac{n_{F}[\varepsilon_{\lambda}(\boldsymbol{k})]-n_{F}\big[\varepsilon_{\lambda^{\prime}}(\boldsymbol{k}+\boldsymbol{q})]}{\omega+\varepsilon_{\lambda}(\boldsymbol{k})-\varepsilon_{\lambda^{\prime}}(\boldsymbol{k}+\boldsymbol{q})+i\eta}.
\end{align}
while for the negative chemical potential $-\mu_{\chi}$,
\begin{align}
 & F_{A}(\boldsymbol{q},\omega,\chi,-\mu_{\chi})=\frac{1}{2\mathcal{V}}\sum_{\boldsymbol{k}}\sum_{\lambda,\lambda^{\prime}=\pm}\left(\lambda\frac{(\boldsymbol{k}+\boldsymbol{q})^{2}-\boldsymbol{k}^{2}-q^{2}}{2q^{2}|\boldsymbol{k}|}-\lambda^{\prime}\frac{(\boldsymbol{k}+\boldsymbol{q})^{2}-\boldsymbol{k}^{2}+q^{2}}{2q^{2}|\boldsymbol{k}+\boldsymbol{q}|}\right)\frac{\tilde{n}_{F}[\varepsilon_{\lambda}(\boldsymbol{k})]-\tilde{n}_{F}\big[\varepsilon_{\lambda^{\prime}}(\boldsymbol{k}+\boldsymbol{q})]}{\omega+\varepsilon_{\lambda}(\boldsymbol{k})-\varepsilon_{\lambda^{\prime}}(\boldsymbol{k}+\boldsymbol{q})+i\eta}\nonumber \\
 & =\frac{1}{2\mathcal{V}}\sum_{\boldsymbol{k}}\sum_{\lambda,\lambda^{\prime}=\pm}\left(\lambda\frac{(\boldsymbol{k}+\boldsymbol{q})^{2}-\boldsymbol{k}^{2}-q^{2}}{2q^{2}|\boldsymbol{k}|}-\lambda^{\prime}\frac{(\boldsymbol{k}+\boldsymbol{q})^{2}-\boldsymbol{k}^{2}+q^{2}}{2q^{2}|\boldsymbol{k}+\boldsymbol{q}|}\right)\frac{n_{F}\big[-\varepsilon_{\lambda^{\prime}}(\boldsymbol{k}+\boldsymbol{q})]-n_{F}[-\varepsilon_{\lambda}(\boldsymbol{k})]}{\omega-\varepsilon_{\lambda^{\prime}}(\boldsymbol{k}+\boldsymbol{q})+\varepsilon_{\lambda}(\boldsymbol{k})+i\eta}.
\end{align}
where $\tilde{n}_{F}[x]=1/\left[\exp\left\{ \beta_T \left(x+\mu_{\chi}\right)\right\} +1\right]$
and $\tilde{n}_{F}[x]+n_{F}[-x]=1$. Because of the particle-hole
symmetry for each Weyl node, the energy dispersion relation satisfies
$\varepsilon_{\lambda}(\boldsymbol{k})=-\varepsilon_{-\lambda}(\boldsymbol{k})$
such that
\begin{align}
 & F_{A}(\boldsymbol{q},\omega,\chi,-\mu_{\chi})=\frac{1}{2\mathcal{V}}\sum_{\boldsymbol{k}}\sum_{\lambda,\lambda^{\prime}=\pm}\left(\lambda\frac{(\boldsymbol{k}+\boldsymbol{q})^{2}-\boldsymbol{k}^{2}-q^{2}}{2q^{2}|\boldsymbol{k}|}-\lambda^{\prime}\frac{(\boldsymbol{k}+\boldsymbol{q})^{2}-\boldsymbol{k}^{2}+q^{2}}{2q^{2}|\boldsymbol{k}+\boldsymbol{q}|}\right)\frac{n_{F}\big[\varepsilon_{-\lambda^{\prime}}(\boldsymbol{k}+\boldsymbol{q})]-n_{F}[\varepsilon_{-\lambda}(\boldsymbol{k})]}{\omega+\varepsilon_{-\lambda^{\prime}}(\boldsymbol{k}+\boldsymbol{q})-\varepsilon_{-\lambda}(\boldsymbol{k})+i\eta}.
\end{align}
Relabeling $\left(\boldsymbol{k},\lambda\right)\leftrightarrow\left(-\boldsymbol{k}-\boldsymbol{q},-\lambda^{\prime}\right)$
leads to
\begin{align}
 & F_{A}(\boldsymbol{q},\omega,\chi,-\mu_{\chi})=\frac{1}{2\mathcal{V}}\sum_{\boldsymbol{k}}\sum_{\lambda,\lambda^{\prime}=\pm}\left(-\lambda^{\prime}\frac{\boldsymbol{k}^{2}-(\boldsymbol{k}+\boldsymbol{q})^{2}-q^{2}}{2q^{2}|\boldsymbol{k}+\boldsymbol{q}|}+\lambda\frac{\boldsymbol{k}^{2}-(\boldsymbol{k}+\boldsymbol{q})^{2}+q^{2}}{2q^{2}|\boldsymbol{k}|}\right)\frac{n_{F}[\varepsilon_{\lambda}(\boldsymbol{k})]-n_{F}\big[\varepsilon_{\lambda^{\prime}}(\boldsymbol{k}+\boldsymbol{q})]}{\omega+\varepsilon_{\lambda}(\boldsymbol{k})-\varepsilon_{\lambda^{\prime}}(\boldsymbol{k}+\boldsymbol{q})+i\eta}.
\end{align}
Clearly, $F_{A}(\boldsymbol{q},\omega,\chi,-\mu_{\chi})$ is nothing
but $-F_{A}(\boldsymbol{q},\omega,\chi,+\mu_{\chi})$. Other chirality-dependent
correlation functions can be calculated in a parallel way. Finally,
we have
\begin{align}
\Pi^{00}\left(\boldsymbol{q},\omega,\chi,-\mu_{\chi}\right) & =\Pi^{00}\left(\boldsymbol{q},\omega,\chi,\mu_{\chi}\right),\\
F_{X}\left(\boldsymbol{q},\omega,\chi,-\mu_{\chi}\right) & =F_{X}\left(\boldsymbol{q},\omega,\chi,\mu_{\chi}\right),\\
F_{A}\left(\boldsymbol{q},\omega,\chi,-\mu_{\chi}\right) & =-F_{A}\left(\boldsymbol{q},\omega,\chi,\mu_{\chi}\right),
\end{align}
where $X=T,L,I$. Note that the relation of $\Pi^{00}$ is consistent
with our previous result \cite{ZhouJH2015PRB}. \end{widetext}

\bibliographystyle{apsrev4-1}

\begin{thebibliography}{72}%
\makeatletter
\providecommand \@ifxundefined [1]{%
 \@ifx{#1\undefined}
}%
\providecommand \@ifnum [1]{%
 \ifnum #1\expandafter \@firstoftwo
 \else \expandafter \@secondoftwo
 \fi
}%
\providecommand \@ifx [1]{%
 \ifx #1\expandafter \@firstoftwo
 \else \expandafter \@secondoftwo
 \fi
}%
\providecommand \natexlab [1]{#1}%
\providecommand \enquote  [1]{``#1''}%
\providecommand \bibnamefont  [1]{#1}%
\providecommand \bibfnamefont [1]{#1}%
\providecommand \citenamefont [1]{#1}%
\providecommand \href@noop [0]{\@secondoftwo}%
\providecommand \href [0]{\begingroup \@sanitize@url \@href}%
\providecommand \@href[1]{\@@startlink{#1}\@@href}%
\providecommand \@@href[1]{\endgroup#1\@@endlink}%
\providecommand \@sanitize@url [0]{\catcode `\\12\catcode `\$12\catcode
  `\&12\catcode `\#12\catcode `\^12\catcode `\_12\catcode `\%12\relax}%
\providecommand \@@startlink[1]{}%
\providecommand \@@endlink[0]{}%
\providecommand \url  [0]{\begingroup\@sanitize@url \@url }%
\providecommand \@url [1]{\endgroup\@href {#1}{\urlprefix }}%
\providecommand \urlprefix  [0]{URL }%
\providecommand \Eprint [0]{\href }%
\providecommand \doibase [0]{http://dx.doi.org/}%
\providecommand \selectlanguage [0]{\@gobble}%
\providecommand \bibinfo  [0]{\@secondoftwo}%
\providecommand \bibfield  [0]{\@secondoftwo}%
\providecommand \translation [1]{[#1]}%
\providecommand \BibitemOpen [0]{}%
\providecommand \bibitemStop [0]{}%
\providecommand \bibitemNoStop [0]{.\EOS\space}%
\providecommand \EOS [0]{\spacefactor3000\relax}%
\providecommand \BibitemShut  [1]{\csname bibitem#1\endcsname}%
\let\auto@bib@innerbib\@empty
\bibitem [{\citenamefont {Hosur}\ and\ \citenamefont {Qi}(2013)}]{Hosur13crp}%
  \BibitemOpen
  \bibfield  {author} {\bibinfo {author} {\bibfnamefont {P.}~\bibnamefont
  {Hosur}}\ and\ \bibinfo {author} {\bibfnamefont {X.}~\bibnamefont {Qi}},\
  }\href {\doibase 10.1016/j.crhy.2013.10.010} {\bibfield  {journal} {\bibinfo
  {journal} {C. R. Physique}\ }\textbf {\bibinfo {volume} {14}},\ \bibinfo
  {pages} {857} (\bibinfo {year} {2013})}\BibitemShut {NoStop}%
\bibitem [{\citenamefont {Burkov}(2015)}]{burkov2015jpcm}%
  \BibitemOpen
  \bibfield  {author} {\bibinfo {author} {\bibfnamefont {A.}~\bibnamefont
  {Burkov}},\ }\href@noop {} {\bibfield  {journal} {\bibinfo  {journal}
  {Journal of Physics: Condensed Matter}\ }\textbf {\bibinfo {volume} {27}},\
  \bibinfo {pages} {113201} (\bibinfo {year} {2015})}\BibitemShut {NoStop}%
\bibitem [{\citenamefont {Weng}\ \emph {et~al.}(2016)\citenamefont {Weng},
  \citenamefont {Dai},\ and\ \citenamefont {Fang}}]{WengHM2016jpcm}%
  \BibitemOpen
  \bibfield  {author} {\bibinfo {author} {\bibfnamefont {H.}~\bibnamefont
  {Weng}}, \bibinfo {author} {\bibfnamefont {X.}~\bibnamefont {Dai}}, \ and\
  \bibinfo {author} {\bibfnamefont {Z.}~\bibnamefont {Fang}},\ }\href
  {http://stacks.iop.org/0953-8984/28/i=30/a=303001} {\bibfield  {journal}
  {\bibinfo  {journal} {Journal of Physics: Condensed Matter}\ }\textbf
  {\bibinfo {volume} {28}},\ \bibinfo {pages} {303001} (\bibinfo {year}
  {2016})}\BibitemShut {NoStop}%
%
\bibitem [{\citenamefont {Armitage}\ \emph {et~al.}()\citenamefont {Armitage},
  \citenamefont {Mele},\ and\ \citenamefont {Vishwanath}}]{Armitage2017rmp}%
  \BibitemOpen
  \bibfield  {author} {\bibinfo {author} {\bibfnamefont {N.~P.}\ \bibnamefont
  {Armitage}}, \bibinfo {author} {\bibfnamefont {E.~J.}\ \bibnamefont {Mele}},
  \ and\ \bibinfo {author} {\bibfnamefont {A.}~\bibnamefont {Vishwanath}},\
  }\href@noop {} {\bibinfo  {journal} {arXiv:1705.01111 [cond-mat.str-el]}\
  }\BibitemShut {NoStop}%
%
\bibitem [{\citenamefont {Xiao}\ \emph {et~al.}(2010)\citenamefont {Xiao},
  \citenamefont {Chang},\ and\ \citenamefont {Niu}}]{Xiao2010RMP}%
  \BibitemOpen
\bibfield  {journal} {  }\bibfield  {author} {\bibinfo {author} {\bibfnamefont
  {D.}~\bibnamefont {Xiao}}, \bibinfo {author} {\bibfnamefont {M.-C.}\
  \bibnamefont {Chang}}, \ and\ \bibinfo {author} {\bibfnamefont
  {Q.}~\bibnamefont {Niu}},\ }\href {\doibase 10.1103/RevModPhys.82.1959}
  {\bibfield  {journal} {\bibinfo  {journal} {Rev. Mod. Phys.}\ }\textbf
  {\bibinfo {volume} {82}},\ \bibinfo {pages} {1959} (\bibinfo {year}
  {2010})}\BibitemShut {NoStop}%
\bibitem [{\citenamefont {Volovik}(2003)}]{volovik2003oup}%
  \BibitemOpen
  \bibfield  {author} {\bibinfo {author} {\bibfnamefont {G.~E.}\ \bibnamefont
  {Volovik}},\ }\href@noop {} {\emph {\bibinfo {title} {The universe in a
  helium droplet}}}\ (\bibinfo  {publisher} {Oxford University Press},\
  \bibinfo {address} {Oxford, UK},\ \bibinfo {year} {2003})\BibitemShut
  {NoStop}%
\bibitem [{\citenamefont {Nielsen}\ and\ \citenamefont
  {Ninomiya}(1983)}]{Nielsen1983ABJ}%
  \BibitemOpen
  \bibfield  {author} {\bibinfo {author} {\bibfnamefont {H.}~\bibnamefont
  {Nielsen}}\ and\ \bibinfo {author} {\bibfnamefont {M.}~\bibnamefont
  {Ninomiya}},\ }\href {\doibase
  http://dx.doi.org/10.1016/0370-2693(83)91529-0} {\bibfield  {journal}
  {\bibinfo  {journal} {Phys. Lett. B}\ }\textbf {\bibinfo {volume} {130}},\
  \bibinfo {pages} {389 } (\bibinfo {year} {1983})}\BibitemShut {NoStop}%
\bibitem [{\citenamefont {Goswami}\ and\ \citenamefont
  {Tewari}(2013)}]{Goswami2013PRB}%
  \BibitemOpen
  \bibfield  {author} {\bibinfo {author} {\bibfnamefont {P.}~\bibnamefont
  {Goswami}}\ and\ \bibinfo {author} {\bibfnamefont {S.}~\bibnamefont
  {Tewari}},\ }\href {\doibase 10.1103/PhysRevB.88.245107} {\bibfield
  {journal} {\bibinfo  {journal} {Phys. Rev. B}\ }\textbf {\bibinfo {volume}
  {88}},\ \bibinfo {pages} {245107} (\bibinfo {year} {2013})}\BibitemShut
  {NoStop}%
\bibitem [{\citenamefont {Liu}\ \emph {et~al.}(2013)\citenamefont {Liu},
  \citenamefont {Ye},\ and\ \citenamefont {Qi}}]{Liu2013PRB}%
  \BibitemOpen
  \bibfield  {author} {\bibinfo {author} {\bibfnamefont {C.-X.}\ \bibnamefont
  {Liu}}, \bibinfo {author} {\bibfnamefont {P.}~\bibnamefont {Ye}}, \ and\
  \bibinfo {author} {\bibfnamefont {X.-L.}\ \bibnamefont {Qi}},\ }\href
  {\doibase 10.1103/PhysRevB.87.235306} {\bibfield  {journal} {\bibinfo
  {journal} {Phys. Rev. B}\ }\textbf {\bibinfo {volume} {87}},\ \bibinfo
  {pages} {235306} (\bibinfo {year} {2013})}\BibitemShut {NoStop}%
\bibitem [{\citenamefont {Parameswaran}\ \emph {et~al.}(2014)\citenamefont
  {Parameswaran}, \citenamefont {Grover}, \citenamefont {Abanin}, \citenamefont
  {Pesin},\ and\ \citenamefont {Vishwanath}}]{Parameswaran2014PRX}%
  \BibitemOpen
  \bibfield  {author} {\bibinfo {author} {\bibfnamefont {S.~A.}\ \bibnamefont
  {Parameswaran}}, \bibinfo {author} {\bibfnamefont {T.}~\bibnamefont
  {Grover}}, \bibinfo {author} {\bibfnamefont {D.~A.}\ \bibnamefont {Abanin}},
  \bibinfo {author} {\bibfnamefont {D.~A.}\ \bibnamefont {Pesin}}, \ and\
  \bibinfo {author} {\bibfnamefont {A.}~\bibnamefont {Vishwanath}},\ }\href
  {\doibase 10.1103/PhysRevX.4.031035} {\bibfield  {journal} {\bibinfo
  {journal} {Phys. Rev. X}\ }\textbf {\bibinfo {volume} {4}},\ \bibinfo {pages}
  {031035} (\bibinfo {year} {2014})}\BibitemShut {NoStop}%
\bibitem [{\citenamefont {Behrends}\ \emph {et~al.}(2016)\citenamefont
  {Behrends}, \citenamefont {Grushin}, \citenamefont {Ojanen},\ and\
  \citenamefont {Bardarson}}]{Behrends2016PRB}%
  \BibitemOpen
  \bibfield  {author} {\bibinfo {author} {\bibfnamefont {J.}~\bibnamefont
  {Behrends}}, \bibinfo {author} {\bibfnamefont {A.~G.}\ \bibnamefont
  {Grushin}}, \bibinfo {author} {\bibfnamefont {T.}~\bibnamefont {Ojanen}}, \
  and\ \bibinfo {author} {\bibfnamefont {J.~H.}\ \bibnamefont {Bardarson}},\
  }\href {\doibase 10.1103/PhysRevB.93.075114} {\bibfield  {journal} {\bibinfo
  {journal} {Phys. Rev. B}\ }\textbf {\bibinfo {volume} {93}},\ \bibinfo
  {pages} {075114} (\bibinfo {year} {2016})}\BibitemShut {NoStop}%
\bibitem [{\citenamefont {Vilenkin}(1980)}]{VilenkinPRD1980CME}%
  \BibitemOpen
  \bibfield  {author} {\bibinfo {author} {\bibfnamefont {A.}~\bibnamefont
  {Vilenkin}},\ }\href {\doibase 10.1103/PhysRevD.22.3080} {\bibfield
  {journal} {\bibinfo  {journal} {Phys. Rev. D}\ }\textbf {\bibinfo {volume}
  {22}},\ \bibinfo {pages} {3080} (\bibinfo {year} {1980})}\BibitemShut
  {NoStop}%
\bibitem [{\citenamefont {Fukushima}\ \emph {et~al.}(2008)\citenamefont
  {Fukushima}, \citenamefont {Kharzeev},\ and\ \citenamefont
  {Warringa}}]{Fukushima2008PRD}%
  \BibitemOpen
  \bibfield  {author} {\bibinfo {author} {\bibfnamefont {K.}~\bibnamefont
  {Fukushima}}, \bibinfo {author} {\bibfnamefont {D.~E.}\ \bibnamefont
  {Kharzeev}}, \ and\ \bibinfo {author} {\bibfnamefont {H.~J.}\ \bibnamefont
  {Warringa}},\ }\href {\doibase 10.1103/PhysRevD.78.074033} {\bibfield
  {journal} {\bibinfo  {journal} {Phys. Rev. D}\ }\textbf {\bibinfo {volume}
  {78}},\ \bibinfo {pages} {074033} (\bibinfo {year} {2008})}\BibitemShut
  {NoStop}%
\bibitem [{\citenamefont {Grushin}(2012)}]{Grushin2012PRD}%
  \BibitemOpen
  \bibfield  {author} {\bibinfo {author} {\bibfnamefont {A.~G.}\ \bibnamefont
  {Grushin}},\ }\href {\doibase 10.1103/PhysRevD.86.045001} {\bibfield
  {journal} {\bibinfo  {journal} {Phys. Rev. D}\ }\textbf {\bibinfo {volume}
  {86}},\ \bibinfo {pages} {045001} (\bibinfo {year} {2012})}\BibitemShut
  {NoStop}%
\bibitem [{\citenamefont {Zyuzin}\ and\ \citenamefont
  {Burkov}(2012)}]{zyuzin2012prb}%
  \BibitemOpen
  \bibfield  {author} {\bibinfo {author} {\bibfnamefont {A.~A.}\ \bibnamefont
  {Zyuzin}}\ and\ \bibinfo {author} {\bibfnamefont {A.~A.}\ \bibnamefont
  {Burkov}},\ }\href {\doibase 10.1103/PhysRevB.86.115133} {\bibfield
  {journal} {\bibinfo  {journal} {Phys. Rev. B}\ }\textbf {\bibinfo {volume}
  {86}},\ \bibinfo {pages} {115133} (\bibinfo {year} {2012})}\BibitemShut
  {NoStop}%
\bibitem [{\citenamefont {Zhou}\ \emph {et~al.}(2013)\citenamefont {Zhou},
  \citenamefont {Jiang}, \citenamefont {Niu},\ and\ \citenamefont
  {Shi}}]{zhou2013cpl}%
  \BibitemOpen
  \bibfield  {author} {\bibinfo {author} {\bibfnamefont {J.}~\bibnamefont
  {Zhou}}, \bibinfo {author} {\bibfnamefont {H.}~\bibnamefont {Jiang}},
  \bibinfo {author} {\bibfnamefont {Q.}~\bibnamefont {Niu}}, \ and\ \bibinfo
  {author} {\bibfnamefont {J.}~\bibnamefont {Shi}},\ }\href
  {http://stacks.iop.org/0256-307X/30/i=2/a=027101} {\bibfield  {journal}
  {\bibinfo  {journal} {Chin. Phys. Lett.}\ }\textbf {\bibinfo {volume} {30}},\
  \bibinfo {pages} {027101} (\bibinfo {year} {2013})}\BibitemShut {NoStop}%
\bibitem [{\citenamefont {Vazifeh}\ and\ \citenamefont
  {Franz}(2013)}]{Vazifeh2013PRL}%
  \BibitemOpen
  \bibfield  {author} {\bibinfo {author} {\bibfnamefont {M.~M.}\ \bibnamefont
  {Vazifeh}}\ and\ \bibinfo {author} {\bibfnamefont {M.}~\bibnamefont
  {Franz}},\ }\href {\doibase 10.1103/PhysRevLett.111.027201} {\bibfield
  {journal} {\bibinfo  {journal} {Phys. Rev. Lett.}\ }\textbf {\bibinfo
  {volume} {111}},\ \bibinfo {pages} {027201} (\bibinfo {year}
  {2013})}\BibitemShut {NoStop}%
\bibitem [{\citenamefont {Landsteiner}(2014)}]{Landsteiner2014PRB}%
  \BibitemOpen
  \bibfield  {author} {\bibinfo {author} {\bibfnamefont {K.}~\bibnamefont
  {Landsteiner}},\ }\href {\doibase 10.1103/PhysRevB.89.075124} {\bibfield
  {journal} {\bibinfo  {journal} {Phys. Rev. B}\ }\textbf {\bibinfo {volume}
  {89}},\ \bibinfo {pages} {075124} (\bibinfo {year} {2014})}\BibitemShut
  {NoStop}%
\bibitem [{\citenamefont {Chang}\ and\ \citenamefont
  {Yang}(2015)}]{Chang2015PRBcme1}%
  \BibitemOpen
  \bibfield  {author} {\bibinfo {author} {\bibfnamefont {M.-C.}\ \bibnamefont
  {Chang}}\ and\ \bibinfo {author} {\bibfnamefont {M.-F.}\ \bibnamefont
  {Yang}},\ }\href {\doibase 10.1103/PhysRevB.91.115203} {\bibfield  {journal}
  {\bibinfo  {journal} {Phys. Rev. B}\ }\textbf {\bibinfo {volume} {91}},\
  \bibinfo {pages} {115203} (\bibinfo {year} {2015})}\BibitemShut {NoStop}%
\bibitem [{\citenamefont {O'Brien}\ \emph {et~al.}(2017)\citenamefont
  {O'Brien}, \citenamefont {Beenakker},\ and\ \citenamefont
  {Adagideli}}]{OBrien2017PRL}%
  \BibitemOpen
  \bibfield  {author} {\bibinfo {author} {\bibfnamefont {T.~E.}\ \bibnamefont
  {O'Brien}}, \bibinfo {author} {\bibfnamefont {C.~W.~J.}\ \bibnamefont
  {Beenakker}}, \ and\ \bibinfo {author} {\bibfnamefont {i.~d.~I.}\
  \bibnamefont {Adagideli}},\ }\href {\doibase 10.1103/PhysRevLett.118.207701}
  {\bibfield  {journal} {\bibinfo  {journal} {Phys. Rev. Lett.}\ }\textbf
  {\bibinfo {volume} {118}},\ \bibinfo {pages} {207701} (\bibinfo {year}
  {2017})}\BibitemShut {NoStop}%
\bibitem [{\citenamefont {Kharzeev}\ and\ \citenamefont
  {Warringa}(2009)}]{KharzeevPRD2009}%
  \BibitemOpen
  \bibfield  {author} {\bibinfo {author} {\bibfnamefont {D.~E.}\ \bibnamefont
  {Kharzeev}}\ and\ \bibinfo {author} {\bibfnamefont {H.~J.}\ \bibnamefont
  {Warringa}},\ }\href {\doibase 10.1103/PhysRevD.80.034028} {\bibfield
  {journal} {\bibinfo  {journal} {Phys. Rev. D}\ }\textbf {\bibinfo {volume}
  {80}},\ \bibinfo {pages} {034028} (\bibinfo {year} {2009})}\BibitemShut
  {NoStop}%
\bibitem [{\citenamefont {Son}\ and\ \citenamefont
  {Yamamoto}(2013)}]{SonDT2013PRD}%
  \BibitemOpen
  \bibfield  {author} {\bibinfo {author} {\bibfnamefont {D.~T.}\ \bibnamefont
  {Son}}\ and\ \bibinfo {author} {\bibfnamefont {N.}~\bibnamefont {Yamamoto}},\
  }\href {\doibase 10.1103/PhysRevD.87.085016} {\bibfield  {journal} {\bibinfo
  {journal} {Phys. Rev. D}\ }\textbf {\bibinfo {volume} {87}},\ \bibinfo
  {pages} {085016} (\bibinfo {year} {2013})}\BibitemShut {NoStop}%
\bibitem [{\citenamefont {Ma}\ and\ \citenamefont {Pesin}(2015)}]{Ma2015PRB}%
  \BibitemOpen
  \bibfield  {author} {\bibinfo {author} {\bibfnamefont {J.}~\bibnamefont
  {Ma}}\ and\ \bibinfo {author} {\bibfnamefont {D.~A.}\ \bibnamefont {Pesin}},\
  }\href {\doibase 10.1103/PhysRevB.92.235205} {\bibfield  {journal} {\bibinfo
  {journal} {Phys. Rev. B}\ }\textbf {\bibinfo {volume} {92}},\ \bibinfo
  {pages} {235205} (\bibinfo {year} {2015})}\BibitemShut {NoStop}%
\bibitem [{\citenamefont {Zhong}\ \emph {et~al.}(2016)\citenamefont {Zhong},
  \citenamefont {Moore},\ and\ \citenamefont {Souza}}]{ZhongSD2016PRLGME}%
  \BibitemOpen
  \bibfield  {author} {\bibinfo {author} {\bibfnamefont {S.}~\bibnamefont
  {Zhong}}, \bibinfo {author} {\bibfnamefont {J.~E.}\ \bibnamefont {Moore}}, \
  and\ \bibinfo {author} {\bibfnamefont {I.}~\bibnamefont {Souza}},\ }\href
  {\doibase 10.1103/PhysRevLett.116.077201} {\bibfield  {journal} {\bibinfo
  {journal} {Phys. Rev. Lett.}\ }\textbf {\bibinfo {volume} {116}},\ \bibinfo
  {pages} {077201} (\bibinfo {year} {2016})}\BibitemShut {NoStop}%
\bibitem [{\citenamefont {Wan}\ \emph {et~al.}(2011)\citenamefont {Wan},
  \citenamefont {Turner}, \citenamefont {Vishwanath},\ and\ \citenamefont
  {Savrasov}}]{WanPRB2011Weyl}%
  \BibitemOpen
  \bibfield  {author} {\bibinfo {author} {\bibfnamefont {X.}~\bibnamefont
  {Wan}}, \bibinfo {author} {\bibfnamefont {A.~M.}\ \bibnamefont {Turner}},
  \bibinfo {author} {\bibfnamefont {A.}~\bibnamefont {Vishwanath}}, \ and\
  \bibinfo {author} {\bibfnamefont {S.~Y.}\ \bibnamefont {Savrasov}},\ }\href
  {\doibase 10.1103/PhysRevB.83.205101} {\bibfield  {journal} {\bibinfo
  {journal} {Phys. Rev. B}\ }\textbf {\bibinfo {volume} {83}},\ \bibinfo
  {pages} {205101} (\bibinfo {year} {2011})}\BibitemShut {NoStop}%
\bibitem [{\citenamefont {Son}\ and\ \citenamefont
  {Spivak}(2013)}]{SonSpivakPRB2013}%
  \BibitemOpen
  \bibfield  {author} {\bibinfo {author} {\bibfnamefont {D.~T.}\ \bibnamefont
  {Son}}\ and\ \bibinfo {author} {\bibfnamefont {B.~Z.}\ \bibnamefont
  {Spivak}},\ }\href {\doibase 10.1103/PhysRevB.88.104412} {\bibfield
  {journal} {\bibinfo  {journal} {Phys. Rev. B}\ }\textbf {\bibinfo {volume}
  {88}},\ \bibinfo {pages} {104412} (\bibinfo {year} {2013})}\BibitemShut
  {NoStop}%
\bibitem [{\citenamefont {Burkov}(2014)}]{Burkov2014PRL}%
  \BibitemOpen
  \bibfield  {author} {\bibinfo {author} {\bibfnamefont {A.~A.}\ \bibnamefont
  {Burkov}},\ }\href {\doibase 10.1103/PhysRevLett.113.247203} {\bibfield
  {journal} {\bibinfo  {journal} {Phys. Rev. Lett.}\ }\textbf {\bibinfo
  {volume} {113}},\ \bibinfo {pages} {247203} (\bibinfo {year}
  {2014})}\BibitemShut {NoStop}%
\bibitem [{\citenamefont {Gorbar}\ \emph {et~al.}(2014)\citenamefont {Gorbar},
  \citenamefont {Miransky},\ and\ \citenamefont {Shovkovy}}]{GorbarPRB2014}%
  \BibitemOpen
  \bibfield  {author} {\bibinfo {author} {\bibfnamefont {E.~V.}\ \bibnamefont
  {Gorbar}}, \bibinfo {author} {\bibfnamefont {V.~A.}\ \bibnamefont
  {Miransky}}, \ and\ \bibinfo {author} {\bibfnamefont {I.~A.}\ \bibnamefont
  {Shovkovy}},\ }\href {\doibase 10.1103/PhysRevB.89.085126} {\bibfield
  {journal} {\bibinfo  {journal} {Phys. Rev. B}\ }\textbf {\bibinfo {volume}
  {89}},\ \bibinfo {pages} {085126} (\bibinfo {year} {2014})}\BibitemShut
  {NoStop}%
\bibitem [{\citenamefont {Gao}\ \emph {et~al.}(2017)\citenamefont {Gao},
  \citenamefont {Yang},\ and\ \citenamefont {Niu}}]{Gao2017PRB}%
  \BibitemOpen
  \bibfield  {author} {\bibinfo {author} {\bibfnamefont {Y.}~\bibnamefont
  {Gao}}, \bibinfo {author} {\bibfnamefont {S.~A.}\ \bibnamefont {Yang}}, \
  and\ \bibinfo {author} {\bibfnamefont {Q.}~\bibnamefont {Niu}},\ }\href
  {\doibase 10.1103/PhysRevB.95.165135} {\bibfield  {journal} {\bibinfo
  {journal} {Phys. Rev. B}\ }\textbf {\bibinfo {volume} {95}},\ \bibinfo
  {pages} {165135} (\bibinfo {year} {2017})}\BibitemShut {NoStop}%
\bibitem [{\citenamefont {Dai}\ \emph {et~al.}(2017)\citenamefont {Dai},
  \citenamefont {Du},\ and\ \citenamefont {Lu}}]{Dai2017PRL}%
  \BibitemOpen
  \bibfield  {author} {\bibinfo {author} {\bibfnamefont {X.}~\bibnamefont
  {Dai}}, \bibinfo {author} {\bibfnamefont {Z.~Z.}\ \bibnamefont {Du}}, \ and\
  \bibinfo {author} {\bibfnamefont {H.-Z.}\ \bibnamefont {Lu}},\ }\href
  {\doibase 10.1103/PhysRevLett.119.166601} {\bibfield  {journal} {\bibinfo
  {journal} {Phys. Rev. Lett.}\ }\textbf {\bibinfo {volume} {119}},\ \bibinfo
  {pages} {166601} (\bibinfo {year} {2017})}\BibitemShut {NoStop}%
\bibitem [{\citenamefont {Kim}\ \emph {et~al.}(2013)\citenamefont {Kim},
  \citenamefont {Kim}, \citenamefont {Wang}, \citenamefont {Sasaki},
  \citenamefont {Satoh}, \citenamefont {Ohnishi}, \citenamefont {Kitaura},
  \citenamefont {Yang},\ and\ \citenamefont {Li}}]{Kim2013PRL}%
  \BibitemOpen
  \bibfield  {author} {\bibinfo {author} {\bibfnamefont {H.-J.}\ \bibnamefont
  {Kim}}, \bibinfo {author} {\bibfnamefont {K.-S.}\ \bibnamefont {Kim}},
  \bibinfo {author} {\bibfnamefont {J.-F.}\ \bibnamefont {Wang}}, \bibinfo
  {author} {\bibfnamefont {M.}~\bibnamefont {Sasaki}}, \bibinfo {author}
  {\bibfnamefont {N.}~\bibnamefont {Satoh}}, \bibinfo {author} {\bibfnamefont
  {A.}~\bibnamefont {Ohnishi}}, \bibinfo {author} {\bibfnamefont
  {M.}~\bibnamefont {Kitaura}}, \bibinfo {author} {\bibfnamefont
  {M.}~\bibnamefont {Yang}}, \ and\ \bibinfo {author} {\bibfnamefont
  {L.}~\bibnamefont {Li}},\ }\href {\doibase 10.1103/PhysRevLett.111.246603}
  {\bibfield  {journal} {\bibinfo  {journal} {Phys. Rev. Lett.}\ }\textbf
  {\bibinfo {volume} {111}},\ \bibinfo {pages} {246603} (\bibinfo {year}
  {2013})}\BibitemShut {NoStop}%
\bibitem [{\citenamefont {Huang}\ \emph {et~al.}(2015)\citenamefont {Huang},
  \citenamefont {Zhao}, \citenamefont {Long}, \citenamefont {Wang},
  \citenamefont {Chen}, \citenamefont {Yang}, \citenamefont {Liang},
  \citenamefont {Xue}, \citenamefont {Weng}, \citenamefont {Fang},
  \citenamefont {Dai},\ and\ \citenamefont {Chen}}]{Huang2015PRX}%
  \BibitemOpen
  \bibfield  {author} {\bibinfo {author} {\bibfnamefont {X.}~\bibnamefont
  {Huang}}, \bibinfo {author} {\bibfnamefont {L.}~\bibnamefont {Zhao}},
  \bibinfo {author} {\bibfnamefont {Y.}~\bibnamefont {Long}}, \bibinfo {author}
  {\bibfnamefont {P.}~\bibnamefont {Wang}}, \bibinfo {author} {\bibfnamefont
  {D.}~\bibnamefont {Chen}}, \bibinfo {author} {\bibfnamefont {Z.}~\bibnamefont
  {Yang}}, \bibinfo {author} {\bibfnamefont {H.}~\bibnamefont {Liang}},
  \bibinfo {author} {\bibfnamefont {M.}~\bibnamefont {Xue}}, \bibinfo {author}
  {\bibfnamefont {H.}~\bibnamefont {Weng}}, \bibinfo {author} {\bibfnamefont
  {Z.}~\bibnamefont {Fang}}, \bibinfo {author} {\bibfnamefont {X.}~\bibnamefont
  {Dai}}, \ and\ \bibinfo {author} {\bibfnamefont {G.}~\bibnamefont {Chen}},\
  }\href {\doibase 10.1103/PhysRevX.5.031023} {\bibfield  {journal} {\bibinfo
  {journal} {Phys. Rev. X}\ }\textbf {\bibinfo {volume} {5}},\ \bibinfo {pages}
  {031023} (\bibinfo {year} {2015})}\BibitemShut {NoStop}%
\bibitem [{\citenamefont {Xiong}\ \emph {et~al.}(2015)\citenamefont {Xiong},
  \citenamefont {Kushwaha}, \citenamefont {Liang}, \citenamefont {Krizan},
  \citenamefont {Hirschberger}, \citenamefont {Wang}, \citenamefont {Cava},\
  and\ \citenamefont {Ong}}]{Xiong2015Science}%
  \BibitemOpen
  \bibfield  {author} {\bibinfo {author} {\bibfnamefont {J.}~\bibnamefont
  {Xiong}}, \bibinfo {author} {\bibfnamefont {S.~K.}\ \bibnamefont {Kushwaha}},
  \bibinfo {author} {\bibfnamefont {T.}~\bibnamefont {Liang}}, \bibinfo
  {author} {\bibfnamefont {J.~W.}\ \bibnamefont {Krizan}}, \bibinfo {author}
  {\bibfnamefont {M.}~\bibnamefont {Hirschberger}}, \bibinfo {author}
  {\bibfnamefont {W.}~\bibnamefont {Wang}}, \bibinfo {author} {\bibfnamefont
  {R.~J.}\ \bibnamefont {Cava}}, \ and\ \bibinfo {author} {\bibfnamefont
  {N.~P.}\ \bibnamefont {Ong}},\ }\href {\doibase 10.1126/science.aac6089}
  {\bibfield  {journal} {\bibinfo  {journal} {Science}\ }\textbf {\bibinfo
  {volume} {350}},\ \bibinfo {pages} {413} (\bibinfo {year}
  {2015})}\BibitemShut {NoStop}%
 %
  \bibitem [{\citenamefont {Li}\ \emph {et~al.}(2015)\citenamefont {Li},
  \citenamefont {Wang}, \citenamefont {Liu}, \citenamefont {Wang},
  \citenamefont {Liao},\ and\ \citenamefont {Yu}}]{li2015NC}%
  \BibitemOpen
  \bibfield  {author} {\bibinfo {author} {\bibfnamefont {C.-Z.}\ \bibnamefont
  {Li}}, \bibinfo {author} {\bibfnamefont {L.-X.}\ \bibnamefont {Wang}},
  \bibinfo {author} {\bibfnamefont {H.}~\bibnamefont {Liu}}, \bibinfo {author}
  {\bibfnamefont {J.}~\bibnamefont {Wang}}, \bibinfo {author} {\bibfnamefont
  {Z.-M.}\ \bibnamefont {Liao}}, \ and\ \bibinfo {author} {\bibfnamefont
  {D.-P.}\ \bibnamefont {Yu}},\ }\href {http://dx.doi.org/10.1038/ncomms10137}
  {\bibfield  {journal} {\bibinfo  {journal} {Nat. Commun.}\ }\textbf {\bibinfo
  {volume} {6}},\ \bibinfo {pages} {10137} (\bibinfo {year}
  {2015})}\BibitemShut {NoStop}%
%
\bibitem [{\citenamefont {Li}\ \emph {et~al.}(2016{\natexlab{a}})\citenamefont
  {Li}, \citenamefont {He}, \citenamefont {Lu}, \citenamefont {Zhang},
  \citenamefont {Liu}, \citenamefont {Ma}, \citenamefont {Fan}, \citenamefont
  {Shen},\ and\ \citenamefont {Wang}}]{LiH2016NC}%
  \BibitemOpen
  \bibfield  {author} {\bibinfo {author} {\bibfnamefont {H.}~\bibnamefont
  {Li}}, \bibinfo {author} {\bibfnamefont {H.}~\bibnamefont {He}}, \bibinfo
  {author} {\bibfnamefont {H.-Z.}\ \bibnamefont {Lu}}, \bibinfo {author}
  {\bibfnamefont {H.}~\bibnamefont {Zhang}}, \bibinfo {author} {\bibfnamefont
  {H.}~\bibnamefont {Liu}}, \bibinfo {author} {\bibfnamefont {R.}~\bibnamefont
  {Ma}}, \bibinfo {author} {\bibfnamefont {Z.}~\bibnamefont {Fan}}, \bibinfo
  {author} {\bibfnamefont {S.-Q.}\ \bibnamefont {Shen}}, \ and\ \bibinfo
  {author} {\bibfnamefont {J.}~\bibnamefont {Wang}},\ }\href
  {http://dx.doi.org/10.1038/ncomms10301} {\bibfield  {journal} {\bibinfo
  {journal} {Nat. Commun.}\ }\textbf {\bibinfo {volume} {7}},\ \bibinfo {pages}
  {10301} (\bibinfo {year} {2016}{\natexlab{a}})}\BibitemShut {NoStop}%
\bibitem [{\citenamefont {Li}\ \emph {et~al.}(2016{\natexlab{b}})\citenamefont
  {Li}, \citenamefont {Kharzeev}, \citenamefont {Zhang}, \citenamefont {Huang},
  \citenamefont {Pletikosic}, \citenamefont {Fedorov}, \citenamefont {Zhong},
  \citenamefont {Schneeloch}, \citenamefont {Gu},\ and\ \citenamefont
  {Valla}}]{LiQ2016NP}%
  \BibitemOpen
  \bibfield  {author} {\bibinfo {author} {\bibfnamefont {Q.}~\bibnamefont
  {Li}}, \bibinfo {author} {\bibfnamefont {D.~E.}\ \bibnamefont {Kharzeev}},
  \bibinfo {author} {\bibfnamefont {C.}~\bibnamefont {Zhang}}, \bibinfo
  {author} {\bibfnamefont {Y.}~\bibnamefont {Huang}}, \bibinfo {author}
  {\bibfnamefont {I.}~\bibnamefont {Pletikosic}}, \bibinfo {author}
  {\bibfnamefont {A.~V.}\ \bibnamefont {Fedorov}}, \bibinfo {author}
  {\bibfnamefont {R.~D.}\ \bibnamefont {Zhong}}, \bibinfo {author}
  {\bibfnamefont {J.~A.}\ \bibnamefont {Schneeloch}}, \bibinfo {author}
  {\bibfnamefont {G.~D.}\ \bibnamefont {Gu}}, \ and\ \bibinfo {author}
  {\bibfnamefont {T.}~\bibnamefont {Valla}},\ }\href
  {http://dx.doi.org/10.1038/nphys3648} {\bibfield  {journal} {\bibinfo
  {journal} {Nat. Phys.}\ }\textbf {\bibinfo {volume} {12}},\ \bibinfo {pages}
  {550} (\bibinfo {year} {2016}{\natexlab{b}})}\BibitemShut {NoStop}%
\bibitem [{for()}]{forster1995hydroF}%
  \BibitemOpen
  \href@noop {} {\bibinfo  {journal} {D. Foster, \emph{Hydrodynamic
  Fluctuations, Broken Symmetry, and Correlation Functions},
  (Benjamin-Cummings, Reading, 1975)}\ }\BibitemShut {NoStop}%
\bibitem [{\citenamefont {Giuliani}\ and\ \citenamefont
  {Vignale}(2005)}]{giuliani2005qtel}%
  \BibitemOpen
\bibfield  {journal} {  }\bibfield  {author} {\bibinfo {author} {\bibfnamefont
  {G.}~\bibnamefont {Giuliani}}\ and\ \bibinfo {author} {\bibfnamefont
  {G.}~\bibnamefont {Vignale}},\ }\href@noop {} {\emph {\bibinfo {title}
  {Quantum theory of the electron liquid}}}\ (\bibinfo  {publisher} {Cambridge
  University Press},\ \bibinfo {address} {Cambridge, UK},\ \bibinfo {year}
  {2005})\BibitemShut {NoStop}%
\bibitem [{\citenamefont {Mahan}(2007)}]{MahanMPP3}%
  \BibitemOpen
  \bibfield  {author} {\bibinfo {author} {\bibfnamefont {G.~D.}\ \bibnamefont
  {Mahan}},\ }\href@noop {} {\emph {\bibinfo {title} {{Many-Particle
  Physics}}}},\ \bibinfo {edition} {3rd}\ ed.\ (\bibinfo  {publisher}
  {Springer},\ \bibinfo {address} {New York, N.Y.},\ \bibinfo {year}
  {2007})\BibitemShut {NoStop}%
\bibitem [{\citenamefont {Nagaosa}\ \emph {et~al.}(2010)\citenamefont
  {Nagaosa}, \citenamefont {Sinova}, \citenamefont {Onoda}, \citenamefont
  {MacDonald},\ and\ \citenamefont {Ong}}]{Nagaosa2010RMP}%
  \BibitemOpen
  \bibfield  {author} {\bibinfo {author} {\bibfnamefont {N.}~\bibnamefont
  {Nagaosa}}, \bibinfo {author} {\bibfnamefont {J.}~\bibnamefont {Sinova}},
  \bibinfo {author} {\bibfnamefont {S.}~\bibnamefont {Onoda}}, \bibinfo
  {author} {\bibfnamefont {A.~H.}\ \bibnamefont {MacDonald}}, \ and\ \bibinfo
  {author} {\bibfnamefont {N.~P.}\ \bibnamefont {Ong}},\ }\href {\doibase
  10.1103/RevModPhys.82.1539} {\bibfield  {journal} {\bibinfo  {journal} {Rev.
  Mod. Phys.}\ }\textbf {\bibinfo {volume} {82}},\ \bibinfo {pages} {1539}
  (\bibinfo {year} {2010})}\BibitemShut {NoStop}%
\bibitem [{\citenamefont {Landau}\ and\ \citenamefont
  {Lifshitz}(1984)}]{LandauECM}%
  \BibitemOpen
  \bibfield  {author} {\bibinfo {author} {\bibfnamefont {L.~D.}\ \bibnamefont
  {Landau}}\ and\ \bibinfo {author} {\bibfnamefont {E.~M.}\ \bibnamefont
  {Lifshitz}},\ }\href@noop {} {\emph {\bibinfo {title} {{Electrodynamics of
  Continuous Media}}}},\ \bibinfo {edition} {2nd}\ ed.\ (\bibinfo  {publisher}
  {Pergamon},\ \bibinfo {address} {New York, N.Y.},\ \bibinfo {year}
  {1984})\BibitemShut {NoStop}%
\bibitem [{\citenamefont {Passarino}\ and\ \citenamefont
  {Veltman}(1979)}]{PASSARINO1979NPB}%
  \BibitemOpen
  \bibfield  {author} {\bibinfo {author} {\bibfnamefont {G.}~\bibnamefont
  {Passarino}}\ and\ \bibinfo {author} {\bibfnamefont {M.}~\bibnamefont
  {Veltman}},\ }\href {\doibase http://dx.doi.org/10.1016/0550-3213(79)90234-7}
  {\bibfield  {journal} {\bibinfo  {journal} {Nucl. Phys. B}\ }\textbf
  {\bibinfo {volume} {160}},\ \bibinfo {pages} {151 } (\bibinfo {year}
  {1979})}\BibitemShut {NoStop}%
\bibitem [{\citenamefont {Chang}\ \emph {et~al.}(2015)\citenamefont {Chang},
  \citenamefont {Zhou}, \citenamefont {Wang}, \citenamefont {Shan},\ and\
  \citenamefont {Xiao}}]{Zhang2015PRB}%
  \BibitemOpen
  \bibfield  {author} {\bibinfo {author} {\bibfnamefont {H.-R.}\ \bibnamefont
  {Chang}}, \bibinfo {author} {\bibfnamefont {J.}~\bibnamefont {Zhou}},
  \bibinfo {author} {\bibfnamefont {S.-X.}\ \bibnamefont {Wang}}, \bibinfo
  {author} {\bibfnamefont {W.-Y.}\ \bibnamefont {Shan}}, \ and\ \bibinfo
  {author} {\bibfnamefont {D.}~\bibnamefont {Xiao}},\ }\href {\doibase
  10.1103/PhysRevB.92.241103} {\bibfield  {journal} {\bibinfo  {journal} {Phys.
  Rev. B}\ }\textbf {\bibinfo {volume} {92}},\ \bibinfo {pages} {241103}
  (\bibinfo {year} {2015})}\BibitemShut {NoStop}%
%
\bibitem{GFb} In this paper, we mainly consider the case, in which $\boldsymbol{b}=0$ or $|\boldsymbol{b}|$ is sufficiently great
  so that Weyl fermions near the two Weyl nodes with opposite chirality behave independently in the
  presence of external fields.
 %
\bibitem [{\citenamefont {Abrikosov}\ and\ \citenamefont
  {Beneslavskii}(1971)}]{Abrikosov1971}%
  \BibitemOpen
  \bibfield  {author} {\bibinfo {author} {\bibfnamefont {A.~A.}\ \bibnamefont
  {Abrikosov}}\ and\ \bibinfo {author} {\bibfnamefont {S.~D.}\ \bibnamefont
  {Beneslavskii}},\ }\href@noop {} {\bibfield  {journal} {\bibinfo  {journal}
  {Sov. Phys. JETP}\ }\textbf {\bibinfo {volume} {32}},\ \bibinfo {pages} {699}
  (\bibinfo {year} {1971})}\BibitemShut {NoStop}%
 %
  \bibitem [{\citenamefont {Lv}\ and\ \citenamefont
  {Zhang}(2013)}]{LvZhangWeyl2013}%
  \BibitemOpen
  \bibfield  {author} {\bibinfo {author} {\bibfnamefont {M.}~\bibnamefont
  {Lv}}\ and\ \bibinfo {author} {\bibfnamefont {S.-C.}\ \bibnamefont {Zhang}},\
  }\href {\doibase 10.1142/S0217979213501774} {\bibfield  {journal} {\bibinfo
  {journal} {Int. J. Mod. Phys. B}\ }\textbf {\bibinfo {volume} {27}},\
  \bibinfo {pages} {1350177} (\bibinfo {year} {2013})}\BibitemShut {NoStop}%
 %
  \bibitem [{\citenamefont {Zhou}\ \emph {et~al.}(2015)\citenamefont {Zhou},
  \citenamefont {Chang},\ and\ \citenamefont {Xiao}}]{ZhouJH2015PRB}%
  \BibitemOpen
\bibfield  {journal} {  }\bibfield  {author} {\bibinfo {author} {\bibfnamefont
  {J.}~\bibnamefont {Zhou}}, \bibinfo {author} {\bibfnamefont {H.-R.}\
  \bibnamefont {Chang}}, \ and\ \bibinfo {author} {\bibfnamefont
  {D.}~\bibnamefont {Xiao}},\ }\href {\doibase 10.1103/PhysRevB.91.035114}
  {\bibfield  {journal} {\bibinfo  {journal} {Phys. Rev. B}\ }\textbf {\bibinfo
  {volume} {91}},\ \bibinfo {pages} {035114} (\bibinfo {year}
  {2015})}\BibitemShut {NoStop}%
%
\bibitem [{\citenamefont {Hofmann}\ and\ \citenamefont
  {Das~Sarma}(2015)}]{Hofmann2015PRB}%
  \BibitemOpen
  \bibfield  {author} {\bibinfo {author} {\bibfnamefont {J.}~\bibnamefont
  {Hofmann}}\ and\ \bibinfo {author} {\bibfnamefont {S.}~\bibnamefont
  {Das~Sarma}},\ }\href {\doibase 10.1103/PhysRevB.91.241108} {\bibfield
  {journal} {\bibinfo  {journal} {Phys. Rev. B}\ }\textbf {\bibinfo {volume}
  {91}},\ \bibinfo {pages} {241108} (\bibinfo {year} {2015})}\BibitemShut
  {NoStop}%
\bibitem [{\citenamefont {Kharzeev}\ \emph {et~al.}(2015)\citenamefont
  {Kharzeev}, \citenamefont {Pisarski},\ and\ \citenamefont
  {Yee}}]{KharzeevPRL2015}%
  \BibitemOpen
  \bibfield  {author} {\bibinfo {author} {\bibfnamefont {D.~E.}\ \bibnamefont
  {Kharzeev}}, \bibinfo {author} {\bibfnamefont {R.~D.}\ \bibnamefont
  {Pisarski}}, \ and\ \bibinfo {author} {\bibfnamefont {H.-U.}\ \bibnamefont
  {Yee}},\ }\href {\doibase 10.1103/PhysRevLett.115.236402} {\bibfield
  {journal} {\bibinfo  {journal} {Phys. Rev. Lett.}\ }\textbf {\bibinfo
  {volume} {115}},\ \bibinfo {pages} {236402} (\bibinfo {year}
  {2015})}\BibitemShut {NoStop}%
\bibitem [{\citenamefont {Pellegrino}\ \emph {et~al.}(2015)\citenamefont
  {Pellegrino}, \citenamefont {Katsnelson},\ and\ \citenamefont
  {Polini}}]{PellegrinoHeliconWeyl2015}%
  \BibitemOpen
  \bibfield  {author} {\bibinfo {author} {\bibfnamefont {F.~M.~D.}\
  \bibnamefont {Pellegrino}}, \bibinfo {author} {\bibfnamefont {M.~I.}\
  \bibnamefont {Katsnelson}}, \ and\ \bibinfo {author} {\bibfnamefont
  {M.}~\bibnamefont {Polini}},\ }\href {\doibase 10.1103/PhysRevB.92.201407}
  {\bibfield  {journal} {\bibinfo  {journal} {Phys. Rev. B}\ }\textbf {\bibinfo
  {volume} {92}},\ \bibinfo {pages} {201407} (\bibinfo {year}
  {2015})}\BibitemShut {NoStop}%
%
\bibitem [{\citenamefont {Zyuzin}\ and\ \citenamefont
  {Zyuzin}(2015)}]{Zyuzin2015PRB}%
  \BibitemOpen
  \bibfield  {author} {\bibinfo {author} {\bibfnamefont {A.~A.}\ \bibnamefont
  {Zyuzin}}\ and\ \bibinfo {author} {\bibfnamefont {V.~A.}\ \bibnamefont
  {Zyuzin}},\ }\href {\doibase 10.1103/PhysRevB.92.115310} {\bibfield
  {journal} {\bibinfo  {journal} {Phys. Rev. B}\ }\textbf {\bibinfo {volume}
  {92}},\ \bibinfo {pages} {115310} (\bibinfo {year} {2015})}\BibitemShut
  {NoStop}%
%
\bibitem [{\citenamefont {Kotov}\ and\ \citenamefont
  {Lozovik}(2016)}]{KotovPRB2016}%
  \BibitemOpen
  \bibfield  {author} {\bibinfo {author} {\bibfnamefont {O.~V.}\ \bibnamefont
  {Kotov}}\ and\ \bibinfo {author} {\bibfnamefont {Y.~E.}\ \bibnamefont
  {Lozovik}},\ }\href {\doibase 10.1103/PhysRevB.93.235417} {\bibfield
  {journal} {\bibinfo  {journal} {Phys. Rev. B}\ }\textbf {\bibinfo {volume}
  {93}},\ \bibinfo {pages} {235417} (\bibinfo {year} {2016})}\BibitemShut
  {NoStop}%
\bibitem [{\citenamefont {Ferreiros}\ and\ \citenamefont
  {Cortijo}(2016)}]{Ferreiros2016PRB}%
  \BibitemOpen
  \bibfield  {author} {\bibinfo {author} {\bibfnamefont {Y.}~\bibnamefont
  {Ferreiros}}\ and\ \bibinfo {author} {\bibfnamefont {A.}~\bibnamefont
  {Cortijo}},\ }\href {\doibase 10.1103/PhysRevB.93.195154} {\bibfield
  {journal} {\bibinfo  {journal} {Phys. Rev. B}\ }\textbf {\bibinfo {volume}
  {93}},\ \bibinfo {pages} {195154} (\bibinfo {year} {2016})}\BibitemShut
  {NoStop}%
\bibitem [{\citenamefont {Yan}\ \emph {et~al.}(2016)\citenamefont {Yan},
  \citenamefont {Huang},\ and\ \citenamefont {Wang}}]{YanZB2016PRB}%
  \BibitemOpen
  \bibfield  {author} {\bibinfo {author} {\bibfnamefont {Z.}~\bibnamefont
  {Yan}}, \bibinfo {author} {\bibfnamefont {P.-W.}\ \bibnamefont {Huang}}, \
  and\ \bibinfo {author} {\bibfnamefont {Z.}~\bibnamefont {Wang}},\ }\href
  {\doibase 10.1103/PhysRevB.93.085138} {\bibfield  {journal} {\bibinfo
  {journal} {Phys. Rev. B}\ }\textbf {\bibinfo {volume} {93}},\ \bibinfo
  {pages} {085138} (\bibinfo {year} {2016})}\BibitemShut {NoStop}%
\bibitem [{\citenamefont {Song}\ \emph {et~al.}(2016)\citenamefont {Song},
  \citenamefont {Zhao}, \citenamefont {Fang},\ and\ \citenamefont
  {Dai}}]{Song2016PRB}%
  \BibitemOpen
  \bibfield  {author} {\bibinfo {author} {\bibfnamefont {Z.}~\bibnamefont
  {Song}}, \bibinfo {author} {\bibfnamefont {J.}~\bibnamefont {Zhao}}, \bibinfo
  {author} {\bibfnamefont {Z.}~\bibnamefont {Fang}}, \ and\ \bibinfo {author}
  {\bibfnamefont {X.}~\bibnamefont {Dai}},\ }\href {\doibase
  10.1103/PhysRevB.94.214306} {\bibfield  {journal} {\bibinfo  {journal} {Phys.
  Rev. B}\ }\textbf {\bibinfo {volume} {94}},\ \bibinfo {pages} {214306}
  (\bibinfo {year} {2016})}\BibitemShut {NoStop}%
\bibitem [{\citenamefont {Rinkel}\ \emph {et~al.}(2017)\citenamefont {Rinkel},
  \citenamefont {Lopes},\ and\ \citenamefont {Garate}}]{Rinkel2017PRL}%
  \BibitemOpen
  \bibfield  {author} {\bibinfo {author} {\bibfnamefont {P.}~\bibnamefont
  {Rinkel}}, \bibinfo {author} {\bibfnamefont {P.~L.~S.}\ \bibnamefont
  {Lopes}}, \ and\ \bibinfo {author} {\bibfnamefont {I.}~\bibnamefont
  {Garate}},\ }\href {\doibase 10.1103/PhysRevLett.119.107401} {\bibfield
  {journal} {\bibinfo  {journal} {Phys. Rev. Lett.}\ }\textbf {\bibinfo
  {volume} {119}},\ \bibinfo {pages} {107401} (\bibinfo {year}
  {2017})}\BibitemShut {NoStop}%
\bibitem [{\citenamefont {Liu}\ and\ \citenamefont {Shi}(2017)}]{Liu2017PRL}%
  \BibitemOpen
  \bibfield  {author} {\bibinfo {author} {\bibfnamefont {D.}~\bibnamefont
  {Liu}}\ and\ \bibinfo {author} {\bibfnamefont {J.}~\bibnamefont {Shi}},\
  }\href {\doibase 10.1103/PhysRevLett.119.075301} {\bibfield  {journal}
  {\bibinfo  {journal} {Phys. Rev. Lett.}\ }\textbf {\bibinfo {volume} {119}},\
  \bibinfo {pages} {075301} (\bibinfo {year} {2017})}\BibitemShut {NoStop}%
%
\bibitem [{\citenamefont {Araki}\ and\ \citenamefont
  {Nomura}(2016)}]{Araki2016PRB}%
  \BibitemOpen
  \bibfield  {author} {\bibinfo {author} {\bibfnamefont {Y.}~\bibnamefont
  {Araki}}\ and\ \bibinfo {author} {\bibfnamefont {K.}~\bibnamefont {Nomura}},\
  }\href {\doibase 10.1103/PhysRevB.93.094438} {\bibfield  {journal} {\bibinfo
  {journal} {Phys. Rev. B}\ }\textbf {\bibinfo {volume} {93}},\ \bibinfo
  {pages} {094438} (\bibinfo {year} {2016})}\BibitemShut {NoStop}%
%
\bibitem{FitTem} The general finite-temperature correlation functions of 3D Weyl/Dirac semimetals cannot be entirely expressed in terms of elementary functions. However, an elegant approach might be useful to connect the finite-temperature correlation functions to the zero-temperature ones (see Sec. 4.4.4 in Ref. \cite{giuliani2005qtel}).
%
\bibitem [{\citenamefont {Peskin}\ and\ \citenamefont
  {Schroeder}(1995)}]{peskin1995qft}%
  \BibitemOpen
  \bibfield  {author} {\bibinfo {author} {\bibfnamefont {M.}~\bibnamefont
  {Peskin}}\ and\ \bibinfo {author} {\bibfnamefont {D.}~\bibnamefont
  {Schroeder}},\ }\href {https://books.google.com.hk/books?id=i35LALN0GosC}
  {\emph {\bibinfo {title} {An Introduction to Quantum Field Theory}}}\
  (\bibinfo  {publisher} {Addison-Wesley, New York},\ \bibinfo {year}
  {1995})\BibitemShut {NoStop}%
\bibitem [{\citenamefont {Kotov}\ \emph {et~al.}(2012)\citenamefont {Kotov},
  \citenamefont {Uchoa}, \citenamefont {Pereira}, \citenamefont {Guinea},\ and\
  \citenamefont {Castro~Neto}}]{Kotov2012RMP}%
  \BibitemOpen
  \bibfield  {author} {\bibinfo {author} {\bibfnamefont {V.~N.}\ \bibnamefont
  {Kotov}}, \bibinfo {author} {\bibfnamefont {B.}~\bibnamefont {Uchoa}},
  \bibinfo {author} {\bibfnamefont {V.~M.}\ \bibnamefont {Pereira}}, \bibinfo
  {author} {\bibfnamefont {F.}~\bibnamefont {Guinea}}, \ and\ \bibinfo {author}
  {\bibfnamefont {A.~H.}\ \bibnamefont {Castro~Neto}},\ }\href {\doibase
  10.1103/RevModPhys.84.1067} {\bibfield  {journal} {\bibinfo  {journal} {Rev.
  Mod. Phys.}\ }\textbf {\bibinfo {volume} {84}},\ \bibinfo {pages} {1067}
  (\bibinfo {year} {2012})}\BibitemShut {NoStop}%
\bibitem [{\citenamefont {Bjorken}\ and\ \citenamefont
  {Drell}(1965)}]{bjorken1965RQF}%
  \BibitemOpen
  \bibfield  {author} {\bibinfo {author} {\bibfnamefont {J.}~\bibnamefont
  {Bjorken}}\ and\ \bibinfo {author} {\bibfnamefont {S.}~\bibnamefont
  {Drell}},\ }\href {https://books.google.com.hk/books?id=ZczvAAAAMAAJ} {\emph
  {\bibinfo {title} {Relativistic quantum fields}}}\ (\bibinfo  {publisher}
  {McGraw-Hill, New York},\ \bibinfo {year} {1965})\BibitemShut {NoStop}%
\bibitem [{\citenamefont {Stauber}\ \emph {et~al.}(2015)\citenamefont
  {Stauber}, \citenamefont {Noriega-P\'erez},\ and\ \citenamefont
  {Schliemann}}]{Stauber2015PRB}%
  \BibitemOpen
  \bibfield  {author} {\bibinfo {author} {\bibfnamefont {T.}~\bibnamefont
  {Stauber}}, \bibinfo {author} {\bibfnamefont {D.}~\bibnamefont
  {Noriega-P\'erez}}, \ and\ \bibinfo {author} {\bibfnamefont {J.}~\bibnamefont
  {Schliemann}},\ }\href {\doibase 10.1103/PhysRevB.91.115407} {\bibfield
  {journal} {\bibinfo  {journal} {Phys. Rev. B}\ }\textbf {\bibinfo {volume}
  {91}},\ \bibinfo {pages} {115407} (\bibinfo {year} {2015})}\BibitemShut
  {NoStop}%
\bibitem [{\citenamefont {Tabert}\ \emph {et~al.}(2016)\citenamefont {Tabert},
  \citenamefont {Carbotte},\ and\ \citenamefont {Nicol}}]{Tabert2016PRB}%
  \BibitemOpen
  \bibfield  {author} {\bibinfo {author} {\bibfnamefont {C.~J.}\ \bibnamefont
  {Tabert}}, \bibinfo {author} {\bibfnamefont {J.~P.}\ \bibnamefont
  {Carbotte}}, \ and\ \bibinfo {author} {\bibfnamefont {E.~J.}\ \bibnamefont
  {Nicol}},\ }\href {\doibase 10.1103/PhysRevB.93.085426} {\bibfield  {journal}
  {\bibinfo  {journal} {Phys. Rev. B}\ }\textbf {\bibinfo {volume} {93}},\
  \bibinfo {pages} {085426} (\bibinfo {year} {2016})}\BibitemShut {NoStop}%
%
\bibitem [{\citenamefont {Roy}\ \emph {et~al.}(2016)\citenamefont {Roy},
  \citenamefont {Juricic},\ and\ \citenamefont {Das~Sarma}}]{Roy2016SR}%
  \BibitemOpen
  \bibfield  {author} {\bibinfo {author} {\bibfnamefont {B.}~\bibnamefont
  {Roy}}, \bibinfo {author} {\bibfnamefont {V.}~\bibnamefont {Juricic}}, \ and\
  \bibinfo {author} {\bibfnamefont {S.}~\bibnamefont {Das~Sarma}},\ }\href
  {http://dx.doi.org/10.1038/srep32446} {\bibfield  {journal} {\bibinfo
  {journal} {Scientific Reports}\ }\textbf {\bibinfo {volume} {6}},\ \bibinfo
  {pages} {32446} (\bibinfo {year} {2016})}\BibitemShut {NoStop}%
 %
\bibitem [{\citenamefont {Roy}\ \emph {et~al.}(2017)\citenamefont {Roy},
  \citenamefont {Goswami},\ and\ \citenamefont {Juri\ifmmode \check{c}\else
  \v{c}\fi{}i\ifmmode~\acute{c}\else \'{c}\fi{}}}]{roy2017prb}%
  \BibitemOpen
  \bibfield  {author} {\bibinfo {author} {\bibfnamefont {B.}~\bibnamefont
  {Roy}}, \bibinfo {author} {\bibfnamefont {P.}~\bibnamefont {Goswami}}, \ and\
  \bibinfo {author} {\bibfnamefont {V.}~\bibnamefont {Juri\ifmmode
  \check{c}\else \v{c}\fi{}i\ifmmode~\acute{c}\else \'{c}\fi{}}},\ }\href
  {\doibase 10.1103/PhysRevB.95.201102} {\bibfield  {journal} {\bibinfo
  {journal} {Phys. Rev. B}\ }\textbf {\bibinfo {volume} {95}},\ \bibinfo
  {pages} {201102} (\bibinfo {year} {2017})}\BibitemShut {NoStop}%
%
\bibitem [{\citenamefont {Bardeen}(1969)}]{Bardeen1969PR}%
  \BibitemOpen
  \bibfield  {author} {\bibinfo {author} {\bibfnamefont {W.~A.}\ \bibnamefont
  {Bardeen}},\ }\href {\doibase 10.1103/PhysRev.184.1848} {\bibfield  {journal}
  {\bibinfo  {journal} {Phys. Rev.}\ }\textbf {\bibinfo {volume} {184}},\
  \bibinfo {pages} {1848} (\bibinfo {year} {1969})}\BibitemShut {NoStop}%
\bibitem [{\citenamefont {Bardeen}\ and\ \citenamefont
  {Zumino}(1984)}]{BARDEEN1984NPB}%
  \BibitemOpen
  \bibfield  {author} {\bibinfo {author} {\bibfnamefont {W.~A.}\ \bibnamefont
  {Bardeen}}\ and\ \bibinfo {author} {\bibfnamefont {B.}~\bibnamefont
  {Zumino}},\ }\href {\doibase https://doi.org/10.1016/0550-3213(84)90322-5}
  {\bibfield  {journal} {\bibinfo  {journal} {Nucl. Phys. B}\ }\textbf
  {\bibinfo {volume} {244}},\ \bibinfo {pages} {421 } (\bibinfo {year}
  {1984})}\BibitemShut {NoStop}%
\bibitem [{\citenamefont {Gorbar}\ \emph {et~al.}(2017)\citenamefont {Gorbar},
  \citenamefont {Miransky}, \citenamefont {Shovkovy},\ and\ \citenamefont
  {Sukhachov}}]{GorbarPRL2017}%
  \BibitemOpen
  \bibfield  {author} {\bibinfo {author} {\bibfnamefont {E.~V.}\ \bibnamefont
  {Gorbar}}, \bibinfo {author} {\bibfnamefont {V.~A.}\ \bibnamefont
  {Miransky}}, \bibinfo {author} {\bibfnamefont {I.~A.}\ \bibnamefont
  {Shovkovy}}, \ and\ \bibinfo {author} {\bibfnamefont {P.~O.}\ \bibnamefont
  {Sukhachov}},\ }\href {\doibase 10.1103/PhysRevLett.118.127601} {\bibfield
  {journal} {\bibinfo  {journal} {Phys. Rev. Lett.}\ }\textbf {\bibinfo
  {volume} {118}},\ \bibinfo {pages} {127601} (\bibinfo {year}
  {2017})}\BibitemShut {NoStop}%
\bibitem [{\citenamefont {Huang}\ \emph {et~al.}(2017)\citenamefont {Huang},
  \citenamefont {Zhou},\ and\ \citenamefont {Shen}}]{Huang2017PRB}%
  \BibitemOpen
  \bibfield  {author} {\bibinfo {author} {\bibfnamefont {Z.-M.}\ \bibnamefont
  {Huang}}, \bibinfo {author} {\bibfnamefont {J.}~\bibnamefont {Zhou}}, \ and\
  \bibinfo {author} {\bibfnamefont {S.-Q.}\ \bibnamefont {Shen}},\ }\href
  {\doibase 10.1103/PhysRevB.96.085201} {\bibfield  {journal} {\bibinfo
  {journal} {Phys. Rev. B}\ }\textbf {\bibinfo {volume} {96}},\ \bibinfo
  {pages} {085201} (\bibinfo {year} {2017})}\BibitemShut {NoStop}%
\bibitem [{\citenamefont {Koshino}\ and\ \citenamefont
  {Hizbullah}(2016)}]{Koshino2016PRB}%
  \BibitemOpen
  \bibfield  {author} {\bibinfo {author} {\bibfnamefont {M.}~\bibnamefont
  {Koshino}}\ and\ \bibinfo {author} {\bibfnamefont {I.~F.}\ \bibnamefont
  {Hizbullah}},\ }\href {\doibase 10.1103/PhysRevB.93.045201} {\bibfield
  {journal} {\bibinfo  {journal} {Phys. Rev. B}\ }\textbf {\bibinfo {volume}
  {93}},\ \bibinfo {pages} {045201} (\bibinfo {year} {2016})}\BibitemShut
  {NoStop}%
\bibitem [{\citenamefont {Moll}\ \emph {et~al.}(2016)\citenamefont {Moll},
  \citenamefont {Potter}, \citenamefont {Nair}, \citenamefont {Ramshaw},
  \citenamefont {Modic}, \citenamefont {Riggs}, \citenamefont {Zeng},
  \citenamefont {Ghimire}, \citenamefont {Bauer}, \citenamefont {Kealhofer},
  \citenamefont {Ronning},\ and\ \citenamefont {Analytis}}]{moll2016NC}%
  \BibitemOpen
  \bibfield  {author} {\bibinfo {author} {\bibfnamefont {P.~J.~W.}\
  \bibnamefont {Moll}}, \bibinfo {author} {\bibfnamefont {A.~C.}\ \bibnamefont
  {Potter}}, \bibinfo {author} {\bibfnamefont {N.~L.}\ \bibnamefont {Nair}},
  \bibinfo {author} {\bibfnamefont {B.~J.}\ \bibnamefont {Ramshaw}}, \bibinfo
  {author} {\bibfnamefont {K.~A.}\ \bibnamefont {Modic}}, \bibinfo {author}
  {\bibfnamefont {S.}~\bibnamefont {Riggs}}, \bibinfo {author} {\bibfnamefont
  {B.}~\bibnamefont {Zeng}}, \bibinfo {author} {\bibfnamefont {N.~J.}\
  \bibnamefont {Ghimire}}, \bibinfo {author} {\bibfnamefont {E.~D.}\
  \bibnamefont {Bauer}}, \bibinfo {author} {\bibfnamefont {R.}~\bibnamefont
  {Kealhofer}}, \bibinfo {author} {\bibfnamefont {F.}~\bibnamefont {Ronning}},
  \ and\ \bibinfo {author} {\bibfnamefont {J.~G.}\ \bibnamefont {Analytis}},\
  }\href {http://dx.doi.org/10.1038/ncomms12492} {\bibfield  {journal}
  {\bibinfo  {journal} {Nat. Commun.}\ }\textbf {\bibinfo {volume} {7}},\
  \bibinfo {pages} {12492} (\bibinfo {year} {2016})}\BibitemShut {NoStop}%
%
\bibitem{AHEmu} Within the linear model for Weyl
  fermions, the anomalous Hall conductivity keeps unaltered at a finite density, which is consistent with the fact that
  the chiral anomaly is unaffected by a finite chemical potential~\cite{Dolan1974PRD}.
%
\bibitem [{\citenamefont {Dolan}\ and\ \citenamefont
  {Jackiw}(1974)}]{Dolan1974PRD}%
  \BibitemOpen
\bibfield  {journal} {  }\bibfield  {author} {\bibinfo {author} {\bibfnamefont
  {L.}~\bibnamefont {Dolan}}\ and\ \bibinfo {author} {\bibfnamefont
  {R.}~\bibnamefont {Jackiw}},\ }\href {\doibase 10.1103/PhysRevD.9.3320}
  {\bibfield  {journal} {\bibinfo  {journal} {Phys. Rev. D}\ }\textbf {\bibinfo
  {volume} {9}},\ \bibinfo {pages} {3320} (\bibinfo {year} {1974})}\BibitemShut{NoStop}%
%
\bibitem [{\citenamefont {Principi}\ \emph {et~al.}(2009)\citenamefont
  {Principi}, \citenamefont {Polini},\ and\ \citenamefont
  {Vignale}}]{Principi2009PRB}%
  \BibitemOpen
\bibfield  {journal} {  }\bibfield  {author} {\bibinfo {author} {\bibfnamefont
  {A.}~\bibnamefont {Principi}}, \bibinfo {author} {\bibfnamefont
  {M.}~\bibnamefont {Polini}}, \ and\ \bibinfo {author} {\bibfnamefont
  {G.}~\bibnamefont {Vignale}},\ }\href {\doibase 10.1103/PhysRevB.80.075418}
  {\bibfield  {journal} {\bibinfo  {journal} {Phys. Rev. B}\ }\textbf {\bibinfo
  {volume} {80}},\ \bibinfo {pages} {075418} (\bibinfo {year}
  {2009})}\BibitemShut {NoStop}%
\bibitem [{\citenamefont {Scholz}\ and\ \citenamefont
  {Schliemann}(2011)}]{Scholz2011PRB}%
  \BibitemOpen
  \bibfield  {author} {\bibinfo {author} {\bibfnamefont {A.}~\bibnamefont
  {Scholz}}\ and\ \bibinfo {author} {\bibfnamefont {J.}~\bibnamefont
  {Schliemann}},\ }\href {\doibase 10.1103/PhysRevB.83.235409} {\bibfield
  {journal} {\bibinfo  {journal} {Phys. Rev. B}\ }\textbf {\bibinfo {volume}
  {83}},\ \bibinfo {pages} {235409} (\bibinfo {year} {2011})}\BibitemShut
  {NoStop}%
\bibitem [{raa()}]{raab2005MTEM}%
  \BibitemOpen
  \href@noop {} {\bibinfo  {journal} {R. E. Raab and O. L. De Lange,
  \emph{Multipole Theory in Electromagnetism: Classical, Quantum, and Symmetry
  Aspects, with Applications}, (Oxford University Press, Oxford, United
  Kingdom, 2005)}\ }\BibitemShut {NoStop}%
%
\bibitem [{\citenamefont {Thakur}\ \emph {et~al.}()\citenamefont {Thakur},
  \citenamefont {Sadhukhan},\ and\ \citenamefont {Agarwal}}]{Thakur2017}%
  \BibitemOpen
\bibfield  {journal} {  }\bibfield  {author} {\bibinfo {author} {\bibfnamefont
  {A.}~\bibnamefont {Thakur}}, \bibinfo {author} {\bibfnamefont
  {K.}~\bibnamefont {Sadhukhan}}, \ and\ \bibinfo {author} {\bibfnamefont
  {A.}~\bibnamefont {Agarwal}},\ }\href@noop {} {\bibinfo  {journal}
  {arXiv:1706.09201 [cond-mat.mes-hall]}\ }\BibitemShut {NoStop}%
%
\end{thebibliography}
%
\end{CJK*}
\end{document}